\documentclass[%
 reprint,
superscriptaddress,
preprintnumbers,
 amsmath,amssymb,
 aps,
floatfix,
]{revtex4-2}

\usepackage{graphicx}
\usepackage{dcolumn}
\usepackage{bm}
\usepackage{dblfloatfix}
\usepackage[compat=1.1.0]{tikz-feynhand}
\usepackage{physics}
\usepackage{slashed}
\usepackage{hyperref}
\usepackage{subfigure}
\usepackage{booktabs}
\usepackage[normalem]{ulem}

\newcommand\numberthis{\addtocounter{equation}{1}\tag{\theequation}}

\makeatletter
\renewcommand\subsubsection{\@startsection{subsubsection}{3}{\z@}%
                                     {-3.25ex\@plus -1ex \@minus -.2ex}%
                                     {1.5ex \@plus .2ex}%
                                     {\normalfont\normalsize\bfseries\boldmath}}
\makeatother

\begin{document}

\preprint{ADP-23-24/T1233}

\title{Chiral analysis of the nucleon mass and sigma commutator}

\author{S.~Owa}
\affiliation{CSSM and ARC Centre of Excellence for Dark Matter Particle Physics, Department of Physics, University of Adelaide, South Australia 5005, Australia}
\author{D.~B.~Leinweber}
\affiliation{CSSM, Department of Physics, University of Adelaide, South Australia 5005, Australia}
\author{A.~W.~Thomas}
\author{X. G. Wang}
\affiliation{CSSM and ARC Centre of Excellence for Dark Matter Particle Physics, Department of Physics, University of Adelaide, South Australia 5005, Australia}

\date{\today}

%
\begin{abstract}
Schemes for describing the light quark mass dependence of the nucleon mass calculated in lattice QCD are compared. The three schemes in consideration include a fully relativistic and Lorentz covariant scheme, one that is fully relativistic but not Lorentz covariant, and a semirelativistic scheme utilizing the heavy baryon approximation. Calculations of observables involving pseudoscalar meson loop diagrams generate nonanalytic terms proportional to square roots and logarithms of the quark mass. The three schemes all yield the correct model independent leading and next-to-leading nonanalytic terms of the chiral expansion of the baryon mass. Results for the masses of the other members of the octet are also presented. Here, low-energy coefficients of the analytic terms of the expansion for the nucleon and hyperons are constrained by lattice QCD results and are demonstrated to be independent of the renormalization scheme used. The differences in the leading coefficient of the chiral expansions are found to be consistent with strange quark counting. Using the schemes examined herein, we report results for the pion-nucleon sigma commutator based upon recent lattice results from the CLS Collaboration. We find $\sigma_{\pi N}=51.7 \pm 3.2 \pm 1.4$ MeV where the uncertainties are statistical and systematic respectively.
\end{abstract}

\maketitle

\section{\label{sec:intro}Introduction}
The existence of lattice QCD calculations as a function of light quark mass offers important opportunities for gaining insight into hadron structure. There have been many studies of the dependence of nucleon properties on the masses of the light quarks, from its mass~\cite{PACSCSdata,Ottnad:2022axz,Leinweber:1999ig,Young:2002ib,Leinweber:2003dg,LECsProcura,LECsBernard,bernard2005chiral,c1c2c3Meissner,Goeke2006,procura2006nucleon,Armour:2008ke,Hall:2010ai,Ren:2012aj,Bruns:2012eh,Shanahan:2012wh,Alvarez-Ruso:2013fza,Ren:2014vea,Lutz:2018cqo,bali2022leading,Lutz:2023xpi,Copeland:2021qni} to its electromagnetic~\cite{Leinweber:2002qb,Leinweber:2004tc,Leinweber:2006ug,Hall:2012yx,Hall:2013oga,CSSM:2014knt,Hall:2012pk,Hall:2013dva,Bignell:2018acn,Bignell:2020xkf} and axial form factors~\cite{Edwards:2005ym,Horsley:2013ayv,Liang:2016fgy,Berkowitz:2017gql,Lutz:2020dfi}, the properties of its excited states
\cite{%
Mahbub:2010rm,%
Edwards:2011jj,%
Lang:2012db,%
Mahbub:2013ala,%
Engel:2013ig,%
Hall:2013qba,%
Roberts:2013ipa,%
Roberts:2013oea,%
Alexandrou:2014mka,%
Liu:2015ktc,%
Kiratidis:2015vpa,%
Leinweber:2015kyz,%
Stokes:2013fgw,%
Kiratidis:2016hda,%
Liu:2016uzk,%
Wu:2016ixr,%
Lang:2016hnn,%
Wu:2017qve,%
Andersen:2017una,%
Stokes:2019zdd,%
Stokes:2019yiz,%
Virgili:2019shg,%
Khan:2020ahz,%
Morningstar:2021ewk,%
Abell:2021awi,%
Bulava:2022vpq,%
Bulava:2023uma,%
Abell:2023qgj,%
Abell:2023nex%
},
and most recently its generalized parton distributions~\cite{Wang:2010hp,Scapellato:2021uke,He:2022leb,Lin:2023gxz}.  Here we focus on the nucleon mass, and we report results for the other members of the octet. We also present a new result for the pion-nucleon light quark sigma commutator, $\sigma_{\pi N}$, based upon an analysis of the most recent CLS data~\cite{Ottnad:2022axz}.

In studying the nucleon mass, $M_N$, as a function of quark mass, chiral symmetry provides important guidance. First, we know that at leading order $m_\pi^2 \propto m_q$, where this appears to be a good approximation for values of $m_\pi$ as large as 0.8 GeV. For this reason, we will show baryon masses as functions of $m_\pi^2$. Second, terms involving odd powers of $m_\pi$ or $\ln m_\pi$ are nonanalytic in the quark mass, with the leading and next-to-leading nonanalytic terms (LNA and NLNA) being proportional to $m_\pi^3$ and $m_\pi^4 \ln m_\pi$, respectively. These terms arise from pion loops and have coefficients which are, in principle, model independent. 

Unfortunately, the convergence properties of an expansion of $M_N$ in powers of $m_\pi^2$ plus the nonanalytic terms are poor. Indeed, the series is badly divergent outside the so called power counting regime (PCR), which corresponds roughly to $m_\pi$ below 0.2 to 0.3 GeV. Attempts to fit lattice QCD results over a wider range of $m_\pi$ have often led to values of the coefficients of the nonanalytic terms being adjusted to values which are inconsistent with the model independent constraints of chiral symmetry.

Here we have two key aims. First, we examine attempts to describe the nucleon mass from lattice QCD over a wide range of pion mass beyond the PCR. Of particular interest is an examination of relativistic effects in the effective field theory. We compare two relativistic formulations with the heavy baryon approximation to discern these effects.  Finite-range regularization (FRR) is used to re-sum the power-series expansion and address larger pion masses in a careful manner, while preserving the leading and next-to-leading nonanalytic behavior of chiral perturbation theory exactly. 

Three schemes are considered including a fully relativistic and Lorentz covariant scheme which uses a four-dimensional regulator; a fully relativistic scheme, similar to the covariant scheme, but using a three-dimensional regulator; and the semirelativistic heavy baryon (HB) approximation, corresponding to the limit of infinitely heavy baryons but including relativistic meson energies and using the three-dimensional regulator.

We find that the results obtained with these different schemes yield accurate and mutually compatible renormalized residual series coefficients (RSCs) of the lower powers of $m_\pi^2$. This is in contrast to claims in the literature~\cite{LagrangianCopeland}. 

Second, having established the efficacy of the various formulations considered, we use them to tackle the highly topical question of the pion-nucleon sigma commutator.

The structure of this paper is as follows. In Sec.~\ref{sec:theory} we outline the theoretical framework, including a discussion of the nonanalytic behavior required by chiral symmetry, as well as the finite-volume corrections to the lattice QCD results. The degree of compatibility of the various schemes considered is investigated in Sec.~\ref{sec:HBvR} over a large range of $m_\pi^2$ using data generated by the PACS-CS collaboration~\cite{PACSCSdata}. Results are also presented for the other members of the nucleon octet. In Sec.~\ref{sec:analysisCLSdata} we analyze the latest lattice QCD results from the CLS collaboration~\cite{Ottnad:2022axz} and compare the extracted low-energy constants (LECs) with contemporary analyses. This naturally leads to a study of the pion-nucleon sigma commutator, $\sigma_{\pi N}$. In Sec.~\ref{sec:compar of LECs and sigma} and \ref{sec:conclusion} we summarize our key findings and make some concluding remarks.

\section{\label{sec:theory}Theoretical framework}
\subsection{\label{sec:chiEFT}Chiral effective field theory}
We present a chiral perturbation theory ($\chi$PT)-inspired model which gives the same leading and next-to-leading model-independent terms in the chiral expansion. The effective SU(3)$_L\times$SU(3)$_R$ chiral Lagrangian (density) is given by
\begin{equation}
	\mathcal{L} = \mathcal{L}_\phi+\mathcal{L}_{\phi B}+\mathcal{L}_{\phi T}+\mathcal{L}_{\phi BT},
\end{equation}
where $\mathcal{L}$ consists of the free meson Lagrangian $\mathcal{L}_\phi$, meson-octet Lagrangian $\mathcal{L}_{\phi B}$, meson-decuplet Lagrangian $\mathcal{L}_{\phi T}$, and the meson-octet-decuplet Lagrangian $\mathcal{L}_{\phi BT}$. Explicitly, at leading order, they are \cite{LagrangianJenkins,LagrangianLutz,LagrangianLedwig,LagrangianCopeland}
\begin{align*}\label{eq:02}
	\mathcal{L}_\phi &= \frac{f_\phi^2}{4}\Tr\left[ D_\mu U(D^\mu U)^\dagger \right]+\frac{f_\phi^2}{4}\Tr\left[ \chi U^\dagger+U\chi^\dagger \right], \\
	\mathcal{L}_{\phi B} &= \Tr\left[ \bar{B}(i\slashed{D}-M_B)B  \right]-\frac{D}{2}\Tr\left[ \bar{B}\gamma^\mu\gamma_5 \{ u_\mu,B \} \right] \\
	&-\frac{F}{2}\Tr\left[ \bar{B}\gamma^\mu\gamma_5\left[ u_\mu,B \right] \right], \\
	\mathcal{L}_{\phi T} &= \left( \bar{T}_\mu \right)^{ijk}(i\gamma^{\mu\nu\alpha}D_\alpha-M_T\gamma^{\mu\nu} )\left( T_\nu \right)^{ijk}, \\
	\mathcal{L}_{\phi BT} &= -\frac{\mathcal{C}}{2}\left[ \epsilon^{ijk}\left( \bar{T}_\mu \right)^{ilm}\Theta^{\mu\nu}(u_\nu)^{lj}(B)^{mk}+\mathrm{H.c.} \right], \numberthis 
\end{align*}
where $f_\phi$ is the pseudoscalar decay constant, $D$ and $F$ are the meson-octet coupling constants, $\mathcal{C}$ is the meson-octet-decuplet constant, and $M_B$ and $M_T$ are the octet and decuplet baryon masses respectively. In the numerical calculations we use the values $f_\phi=93$ MeV, $D=0.86$ and $F=0.41$ leading to $g_A=D+F=1.27$, and $\mathcal{C}=(6/5)g_A$. Additionally, we have used the definition, $\gamma^{\mu\nu}\equiv\frac{1}{2}\left[ \gamma^\mu,\gamma^\nu \right]=-i\sigma^{\mu\nu}$ and $\gamma^{\mu\nu\alpha} \equiv \frac{1}{2}\left\lbrace \gamma^{\mu\nu},\gamma^\alpha \right\rbrace=i\epsilon^{\mu\nu\alpha\beta}\gamma_\beta\gamma_5$. 

We also include the next-to-leading order (NLO) terms,  corresponding to the Lagrangian~\cite{2ndLagrangianOller}
\begin{align*}\label{eq:03}
    \mathcal{L}_{\phi B}^{(2)} &= b_0\Tr[\chi_+]\Tr[\bar{B}B]+b_D\Tr[\bar{B}\{ \chi_+,B\}] \\
    &+b_F\Tr\left[ \bar{B}[\chi_+,B] \right]+\sum_{i=1}^8 b_iO^{(2)}_i+\cdots, \numberthis
\end{align*}
where $b_i$ are the NLO LECs of $\chi$PT. We refer the readers to Appendix~\ref{app:A} for explicit expressions of the fields and the covariant derivatives of the Lagrangian. 

Relevant to the later sections, we also present the NLO Lagrangian in the SU(2) limit \cite{SU2NLOLagrangianGasser,SU2NLOLagrangianFettes,EOMSFuchs}
\begin{align*}
\label{eq:04}
    \mathcal{L}_{\pi N}^{(2)} &= c_1\Tr[\chi_+]\bar{\Psi}\Psi \\
    &-\frac{c_2}{4 M_N^2}\Tr[u_\mu u_\nu]\left( \bar{\Psi}D^\mu D^\nu\Psi+ \text{H.c.} \right) \\
    &+\frac{c_3}{2}\Tr[u^\mu u_\mu]\bar{\Psi}\Psi +\cdots \, . \numberthis
\end{align*}
Here, $c_i$ are the SU(2) dimension-two LECs, $\Psi$ is the nucleon doublet, and $u$ (equivalently $U$) becomes the $2\times2$ unimodular unitary matrix with only pions.

\subsection{Chiral expansion using FRR}
Our focus here is on the analysis of lattice QCD data over a wide range of $m_\pi^2$. As the naive series expansion is badly divergent for $m_\pi$ beyond 0.2 to 0.3 GeV, we explore the application of FRR, which aims to re-sum the series in a physically motivated way~\cite{Leinweber:2003dg,Young:2002ib}. 

Historically, the FRR approach has had a number of successes, such as the accurate prediction of the strange quark contribution to the magnetic moment~\cite{Leinweber:2004tc} and charge radius~\cite{Leinweber:2006ug} of the proton, some years before experimental measurements confirmed the predictions~\cite{Armstrong:2012bi,HAPPEX:1998epc,G0:2009wvv}. It also led to the prediction of the excess of $\bar{d}$ over $\bar{u}$ quarks in the proton a decade before experimental confirmation~\cite{Thomas:1983fh}. Here, we are particularly interested in examining the historical criticism of the use of the heavy baryon approximation in early analyses of lattice QCD results.

In the analysis of the mass of the baryon as a function of meson mass using FRR, the mass of a baryon, $B$, is written as 
\begin{equation}
\label{eq:05}
    M_B =a_0+\sum_{\phi,\mathcal{B}'} a_{2,\phi} m_\phi^2+a_{4,\phi} m_\phi^4+\Sigma_{B\mathcal{B}'\phi}+\Sigma_{B\phi,\mathrm{tad}} \, ,
\end{equation}
where $\mathcal{B}'$ denotes intermediate octet and decuplet baryons, and $a_{i,\phi}$ are the unrenormalized RSCs, which after renormalization relate to the LECs of the $\chi$PT expansion. Details of the renormalization procedure are presented later in this section. 

The self-energy contributions, $\Sigma$, are defined to be the on-shell matrix elements of the transition operator $\hat{\Sigma}$, that is
\begin{equation}
    \Sigma = \frac{1}{2}\sum_{s}\bar{u}(p,s)\, \hat{\Sigma}\, u(p,s) \, ,
\end{equation}
where $u$ is the Dirac spinor, normalized as $\bar{u}(p,s)\, u(p,s')=\delta_{ss'}$, and the sum is taken over the spin of the external baryon state. 
\begin{figure}[t]
    \centering
	\begin{tikzpicture}[baseline=(o), scale=0.55]
	\setlength{\feynhandlinesize}{1pt}
	\setlength{\feynhanddotsize}{0.2pt}
		\begin{feynhand}
			\vertex (o) at (0,0);
			\vertex (a) at (-4.5,0) {$B$};
			\vertex [dot] (b) at (-2,0) {};
			\vertex [dot] (c) at (2,0) {};
			\vertex (d) at (4.5,0) {$B$};
			\vertex (e) at (0,2.6) {$\phi$};
			\propag[plain] (a) to (b);
			\propag[plain] (c) to (d);
			\propag[sca] (b) to [half left, looseness=1.7] (c);
			\propag[plain] (b) to [edge label'= $B'$] (c);
		\end{feynhand}
	\end{tikzpicture}
	\begin{tikzpicture}[baseline=(o), scale=0.55]
	\setlength{\feynhandlinesize}{1pt}
	\setlength{\feynhanddotsize}{0.2pt}
		\begin{feynhand}
			\vertex (o) at (0,0);
			\vertex (a) at (-4.5,0) {$B$};
			\vertex [dot] (b) at (-2,0) {};
			\vertex [dot] (b2) at (-2,0.2) {};
			\vertex [dot] (c) at (2,0) {};
			\vertex [dot] (c2) at (2,0.2) {};
			\vertex (d) at (4.5,0) {$B$};
			\vertex (e) at (0,2.6) {$\phi$};
			\propag[plain] (a) to (b);
			\propag[plain] (c) to (d);
			\propag[sca] (b) to [half left, looseness=1.7] (c);
			\propag[plain] (b) to [edge label'= $T'$] (c);
			\propag[plain] (b2) to (c2);
		\end{feynhand}
	\end{tikzpicture}
	\begin{tikzpicture}[baseline=(o), scale=0.55]
	\setlength{\feynhandlinesize}{1pt}
	\setlength{\feynhanddotsize}{0.2pt}
		\begin{feynhand}
			\vertex (o) at (0,0);
			\vertex (a) at (-4.5,0) {$B$};
			\vertex [dot] (b) at (0,0) {};
			\vertex [dot] (c) at (0,2.4) {};
			\vertex (d) at (4.5,0) {$B$};
			\vertex (e) at (0,3.0) {$\phi$};
			\propag[plain] (a) to (b);
			\propag[plain] (b) to (d);
			\propag[sca] (c) to [out=180, in=145] (b);
			\propag[sca] (b) to [out=35, in=0] (c);
		\end{feynhand}
	\end{tikzpicture}
	\caption{Contributions to the octet baryon self-energies at $\mathcal{O}(p^4)$ from the $\chi$EFT Lagrangian. These include octet-meson, decuplet-meson and tadpole terms. Note that there are additional diagrams that contribute to the tree-level correction at $\mathcal{O}(p^4)$ in $\chi$PT, they are implicitly incorporated in our model through the series expansion of the baryon mass.}
    \label{fig:01}
\end{figure}
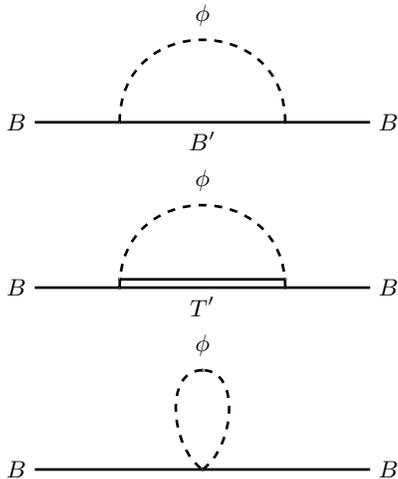

Illustrated in Fig.~\ref{fig:01}, we consider all possible octet-octet-meson, octet-decuplet-meson, and tadpole transitions. Note that the first two diagrams are generated by the leading order Lagrangian, while the tadpole diagram is solely generated by the NLO Lagrangian. All meson loops are included and the strange quark mass is fixed at its physical value. In this way, we do not expand about the SU(3) chiral limit, but rather the broken SU(3) limit, with $m_u=m_d$ (where $m_u$ and $m_d$ represent the light quark masses) and the strange quark mass fixed at its physical value.  We refer to this as the light flavor chiral limit.

In principle, the computation of Eq.~\eqref{eq:05} is circular within the model, as $M_B$ itself appears in the Lagrangian. This means that we require prior knowledge of $M_B$ in the recursive calculation. At leading order approximation, one typically sets $M_B$ to its physical value with no meson mass dependence. For a more sophisticated treatment, one could take some parametrization of $M_B$ in terms of the pion mass from lattice QCD studies. In a preliminary analysis, we computed the renormalized RSCs for the nucleon and the pion-nucleon sigma commutator, and found that there is no significant difference between the results fixing $M_N$ and using some parametrization of $M_N$. In light of this, we fix $M_B$ in the Lagrangian density to their physical values for computational convenience.

The self-energy contributions from the relevant one-loop diagrams, for internal meson momentum $k$ and external baryon momentum $p$, are 
\begin{widetext}
    \begin{align}
	&\hat{\Sigma}_{BB'\phi} = i\left( \frac{C_{BB'\phi}}{f_\phi} \right)^2\int\frac{d^4k}{(2\pi)^4}\slashed{k}\gamma_5\frac{(\slashed{p}-\slashed{k}+M_B)}{D_{B'}}\slashed{k}\gamma_5\frac{1}{D_\phi}, \label{eq:07} \\
	&\hat{\Sigma}_{BT'\phi} = i\left( \frac{C_{BT'\phi}}{f_\phi} \right)^2\int\frac{d^4k}{(2\pi)^4}\Theta^{\mu\nu}k_\nu\frac{(\slashed{p}-\slashed{k}+M_{T'})\Lambda_{\mu\rho}(p-k)}{D_{T'}}\Theta^{\rho\sigma}k_\sigma\frac{1}{D_\phi}, \label{eq:08} \\
	&\hat{\Sigma}_{B\phi,\text{tad}} = i\frac{C_{B\phi,\text{tad}}}{f_\phi^2} \int\frac{d^4k}{(2\pi)^4} \frac{1}{D_\phi}, \label{eq:09}
    \end{align}
\end{widetext}
where $C_{B\mathcal{B}'\phi}$ and $C_{B\phi,\text{tad}}$ are the octet-octet/decuplet-meson and octet-tadpole coupling constants, respectively (given in the Appendix~\ref{app:B}),
$\Theta^{\mu\nu} = g^{\mu\nu}-\gamma^\mu\gamma^\nu$, the spin-3/2 projector is $\Lambda_{\mu\rho}(q) = g_{\mu\rho}-\frac{1}{3}\gamma_\mu\gamma_\rho-\frac{\gamma_\mu q_\rho-\gamma_\rho q_\mu}{3 M_{T'}}-\frac{2q_\mu q_\rho}{3M_{T'}^2}$, and $D_X \equiv q^2-m_X^2+i\varepsilon$ with $q$ and $m_X$ being the momentum and mass of the hadron, respectively. 

We commence with a consideration of a fully relativistic and Lorentz covariant formalism. In this case we introduce a 4-dimensional dipole form factor to regulate the divergent integrals 
\begin{equation}
\label{eq:10}
    w(k) = \left( \frac{\Lambda^2}{k^2-\Lambda^2} \right)^2 \, ,
\end{equation}
where $\Lambda$ is a cutoff scale. Correspondingly, in an alternate relativistic formalism (discussed in further detail in the next section) we use a 3-dimensional dipole regulator
\begin{equation}
\label{eq:11}
    w_{3\text{D}} (\vec{k})= \left( \frac{\Lambda^2}{\vec{k}^2+\Lambda^2} \right)^2 \, .
\end{equation}
This three-dimensional regulator is also used in the heavy baryon case. Closed expressions for the self-energies with these regulators are presented in Appendix~\ref{app:C}.

The essential feature of the FRR approach is that it guarantees the correct, model-independent, LNA and NLNA behavior of the nucleon mass as a function of $m_\pi^2$. Of course, it also generates a nonanalytic term of order $m_\pi^5$, which does depend on the regulator mass, $\Lambda$. However, the choice $\Lambda \in 0.8 - 1.0$ GeV, motivated by considerations of the size of the nucleon~\cite{Thomas:1982kv,Leinweber:1999ig,Thomas:2002sj,Donoghue:1998bs} and analyses examining the renormalization flow of the LECs of the chiral expansion \cite{Hall:2010ai,Hall:2011en,Hall:2012pk}, is in reasonable agreement with the higher-order two-loop calculation of McGovern and Birse~\cite{McGovern:2006fm}, which included an estimate of the effect of nucleon size on the nucleon self-energy within chiral perturbation theory.

The presence of multiple contact interactions in the formal expansion of the chiral Lagrangian density, in Eq.~\eqref{eq:03}, with coefficients typically adjusted to describe pion-nucleon scattering~\cite{Alarcon:2011kh} up to 70 MeV above threshold, leads to tadpole diagrams which generate NLNA terms involving $m_\pi^4 \ln m_\pi$. In the limit where the pion mass is much less than $\Delta_{\Delta N}$ (the $\Delta-N$ mass difference), $m_\pi \ll \Delta_{\Delta N}$, the self-energy term $\Sigma_{N \Delta \pi}$ generates exactly this nonanalytic behavior. The coefficient of the NLNA term arising from $\Sigma_{N \Delta \pi}$ in this limit is approximately $-3.1$ GeV$^{-1}$, which lies within the range for the NLNA coefficient of the tadpole quoted by Frink and Meissner~\cite{Frink:2005ru}, namely $-8.4^{+5.2}_{-4.4}$ GeV$^{-1}$. However, in the physically more relevant case, where $m_\pi$ is comparable to, or larger than $\Delta$, the nonanalytic behavior involves a square root branch point at $m_\pi = \Delta$.

As the tadpole terms generated by the contact interactions will have overlap with a small-$m_\pi$ expansion of the $\Delta \pi$ self-energy term, we examine the effect of including the explicit tadpole contribution of Eq.~\eqref{eq:09}, with a coefficient in the range suggested in Ref.~\cite{Frink:2005ru}, but corrected for the $\Delta \pi$ contribution.

In any regularization scheme, the expressions of observables should preserve the symmetries of the underlying theory. Building from  Ref.~\cite{ModpropDjukanovic}, it can be shown that the regulators presented above can be generated by an alternate chirally invariant Lagrangian. The alternate Lagrangian includes additional terms to the chiral effective theory ($\chi$EFT) Lagrangian, which modifies the hadron propagators to incorporate the regulator. Renormalization can be carried out using the extended on-mass-shell (EOMS) scheme~\cite{EOMSFuchs}, for which one systematically removes the chiral symmetry and power counting violating terms. While we do not strictly follow the EOMS renormalization, our renormalization scheme is tantamount to that of the EOMS scheme up to the desired order in the chiral expansion. 

In this investigation, we make use of lattice QCD data to determine the renormalized RSCs of the chiral expansion. For the lattice data that we consider, the simulations are carried out in $N_f=2+1$ flavours. Given that the strange quark mass is fixed (typically at the physical point), this serves as a good justification to consider the baryon mass expansion only in terms of the light quark mass expressed by $m_\pi^2$. That is, using the Gell-Mann-Oakes-Renner relation with the strange quark mass fixed, at leading order, the squared kaon and eta masses can be written as a function of varying pion mass as
\begin{align}
    m_K^2 &= \frac{1}{2}m_\pi^2 + \left(m_K^2|_{\text{phys}}  - \frac{1}{2}m_\pi^2 |_{\text{phys}}\right), \\
    m_\eta^2 &= \frac{1}{3}m_\pi^2+\frac{4}{3} \left( m_K^2|_{\text{phys}} - \frac{1}{2}m_\pi^2 |_{\text{phys}}\right),
\end{align}
where ``phys" denotes the physical (experimental) value. As a result, the RSCs associated with each meson $a_{i,\phi}$ in Eq.~\eqref{eq:05} are absorbed into one $a_i$ at each order of $m_\pi^2$. 

Given this background, we can now explicitly write the renormalized chiral expansion of the baryon mass as
\begin{align*}
\label{eq:14}
    M_B &= C_0^B+C_2^B\, m_\pi^2+a_4^B\, m_\pi^4 \\
    &+\sum_{\phi,\mathcal{B}'} \left( \tilde{\Sigma}_{B\mathcal{B}'\phi}(m_\pi^2)+\tilde{\Sigma}_{B\phi,\mathrm{tad}}(m_\pi^2) \right) \, \numberthis,
\end{align*}
where $C_i^B$ are the renormalized RSCs unique to each baryon. Here, $C_0^B$ is identified as the mass of the baryon in the light flavor chiral limit. The subtracted self-energies in Eq.~\eqref{eq:14} are defined by
\begin{align*}
\label{eq:tildeSigma}
    \tilde{\Sigma}_{B\mathcal{B}'\phi} &\equiv \Sigma_{B\mathcal{B}'\phi}- \Sigma_{B\mathcal{B}'\phi}(0)-m_\pi^2 \frac{\partial \Sigma_{B\mathcal{B}'\phi} }{\partial m_\pi^2}(0), \\
    \tilde{\Sigma}_{B\phi,\mathrm{tad}} &\equiv \Sigma_{B\phi,\mathrm{tad}}-\Sigma_{B\phi,\mathrm{tad}}(0), \numberthis
\end{align*}
where, for brevity, $\Sigma(0)=\Sigma(m_\pi^2=0)$. The above is adequate in our renormalization scheme because the lowest order nonanalytic term starts at $\mathcal{O}(m_\pi^3)$ and the tadpole terms enter at $\mathcal{O}(m_\pi^4)$. For computational convenience we did not renormalize the coefficients of the terms analytic in $m_q$ beyond $\mathcal{O}(m_\pi^2)$ (i.e., $a_4$ is left unrenormalized).

Evaluation of the NLNA contributions in our FRR scheme reveals that the kaon contributions are small and vary slowly with changes in the pion mass.  As a result we focus on the pion-loop contributions and consider the three SU(2) dimension-two LECs, $c_1$, $c_2$, and $c_3$. These will be constrained by the requirement that our model respects the same nonanalytic behavior as SU(2) $\chi$PT up to and including $\mathcal{O}(m_\pi^4\,\ln m_\pi)$ for the nucleon. 

In order to estimate the tadpole contribution to the mass of the nucleon, we first consider the values of $c_1$, $c_2$, and $c_3$, summarised in Refs.~\cite{c1c2c3Meissner,c1c2c3Frink}, where these LECs were determined based on phenomenological considerations. Although they are determined from different schemes (where the associated higher-order terms differ), the physical pion mass is well within the PCR such that higher order terms do not contribute in a significant manner. The role of these LECs is to set a plausible central value for the nucleon-pion tadpole coefficient. Despite the rather expansive uncertainty range on the LECs from the aforementioned references, we extend this range by considering LECs more recently determined from Ref.~\cite{LECsSiemens:2017}. We consider results with explicit $\Delta$ degrees of freedom because the LECs undergo significant changes with the explicit inclusion of $\Delta$. Thus, with significant error bounds, we evaluate the effects of the tadpole term in the final result. Matching subtleties between SU(2) and SU(3) $\chi$PT are not considered.

In $\chi$PT, the nucleon-pion tadpole self-energy is written as \cite{Fuchs}
\begin{equation}
\label{eq:16}
    \Sigma_{N\pi,\text{tad}} = -i\frac{3}{f_\phi^2}\left( -2c_1+\frac{c_2}{4}+c_3 \right)m_\pi^2\int\frac{d^4k}{(2\pi)^4}\frac{1}{D_\pi} \, .
\end{equation}
In the relativistic theory, the $NN\pi$ self-energy contributes both a LNA contribution proportional to $m_\pi^3$ and a NLNA contribution proportional to $m_\pi^4\,\ln m_\pi$, such that the coefficient of the NLNA term, $k_3$, in the $\chi$PT nucleon mass expansion (discussed later in Sec.~\ref{compar of LECs} Eq.~\eqref{eq:27}) is
\begin{equation}
\label{eq:17}
    k_3 = -\frac{3}{32\pi^2f_\phi^2}\left( -8c_1+c_2+4c_3+\frac{g_A^2}{M_N} \right).
\end{equation}
It is precisely the combination of $c_1$, $c_2$, and $c_3$ in Eq.~\eqref{eq:16} minus the contribution from the $N \Delta \pi$ loop, discussed earlier, that we take as a single parameter $\chi C$ as
\begin{equation}
\label{eq:18}
   \chi C \equiv -2c_1+\frac{c_2}{4}+c_3+\frac{\mathcal{A}}{4} = -1.3_{-1.1}^{+3.3}\ \text{GeV}^{-1},
\end{equation}
from Refs.~\cite{c1c2c3Meissner,c1c2c3Frink} and upper bound appropriately extended to encapsulate values (with explicit $\Delta$) from Ref.~\cite{LECsSiemens:2017}. $\mathcal{A}$ is proportional to the $N\Delta\pi$ NLNA contribution and is given by
\begin{equation}
    \mathcal{A} \equiv \frac{24}{25}g_A^2\frac{M_N^2}{\Delta_{\Delta N}(M_N+\Delta_{\Delta N})^2}\approx 3.1\ \text{GeV}^{-1},
\end{equation}
quoted earlier. The $\Delta-N$ mass difference, $\Delta_{\Delta N}$, is taken at the physical point. In this manner, $k_3$ in our model is congruent to that of $\chi$PT, once we separately include the $N \Delta \pi$ self-energy contribution
\begin{equation}
    k_3 = -\frac{3}{32\pi^2f_\phi^2}\left( 4\chi C-\mathcal{A}+\frac{g_A^2}{M_N} \right).
\end{equation}

\subsection{\label{sec:FVCs} Finite-volume corrections}
In order to fit the chiral expansion to lattice results, one either needs to compute the expansions in finite-volume, or to correct the lattice results to infinite-volume. For the latter case, it is necessary to calculate the finite-volume corrections (FVCs). Following common practice, we choose to treat the spatial and temporal dimensions differently, such that the temporal integral is performed over infinite-volume. Then, for some FRR integrand, $I(\vec{k},m_\pi^2,\Lambda)$, the FVC is defined as follows
\begin{equation}
    \delta^{\text{FVC}}(m_\pi^2,L,\Lambda) \equiv \int \frac{d^3k}{(2\pi)^3}I(\vec{k},m_\pi^2,\Lambda)-\frac{1}{L^3}\sum_{\vec{k}}I(\vec{k},m_\pi^2,\Lambda),
\end{equation}
where $L$ is the length of the box. The lattice QCD results are corrected to infinite-volume through the addition of $ \delta^{\text{FVC}}(m_\pi^2,L,\Lambda)$ to each finite-volume lattice value. 
	
It has been demonstrated that in an earlier study of Ref.~\cite{Hall:2010ai}, FVCs of self-energy integrals, using the same regulator presented in Eq.~\eqref{eq:11}, have negligible dependence on $\Lambda$ beyond 0.8 GeV. At a box size of $L=2.9$ fm, the FVCs vary by about 1 MeV in the domain $0.8\leq \Lambda \leq 2.4$ GeV.

\section{\label{sec:HBvR}Heavy Baryon versus Relativistic Formalisms}

\subsection{\label{sec:schemes}
The correspondence of the schemes}
In order to compare the application of the HB approximation with relativistic approaches for analyzing lattice QCD data for $M_N(m_\pi^2)$, we consider three schemes:
\begin{enumerate}
    \item Covariant (Cov) - a fully relativistic and Lorentz covariant scheme which uses the four-dimensional dipole regulator of Eq.~\eqref{eq:10}, 
    \item Relativistic (Rel) - a fully relativistic scheme, similar to the covariant scheme, but uses the three-dimensional dipole regulator of Eq.~\eqref{eq:11}, and 
    \item Heavy baryon (HB) - a semirelativistic scheme corresponding to the limit of infinitely heavy baryons but including relativistic meson energies and using the three-dimensional dipole regulator.
\end{enumerate}

One may obtain the self-energy expressions in the HB scheme by performing the $k_0$ integral of Eqs.~\eqref{eq:07}-\eqref{eq:09} and taking the limit $M_{\mathcal{B}}$ and $M_{\mathcal{B'}}$ to infinity in the relativistic scheme. We obtain
\begin{align}
    \Sigma^{(\text{HB})}_{BB'\phi} &= -\frac{1}{2}\left( \frac{C_{BB'\phi}}{f_\phi} \right)^2\int \frac{d^3k}{(2\pi)^3} \frac{k^2}{\omega_k(\omega_k+\delta)}, \label{eq:22} \\ 
    \Sigma^{(\text{HB})}_{BT'\phi} &= -\frac{1}{3}\left( \frac{C_{BT'\phi}}{f_\phi} \right)^2\int \frac{d^3k}{(2\pi)^3} \frac{k^2}{\omega_k(\omega_k+\Delta)}, \label{eq:23} \\
    \Sigma^{(\text{HB})}_{B\phi,\text{tad}} &= \frac{1}{2}\frac{C_{B\phi,\text{tad}}}{f_\phi^2}\int \frac{d^3k}{(2\pi)^3} \frac{1}{\omega_k} \, ,    
\end{align}
where $\delta$ ($\Delta$) is the mass difference between the external octet and the internal octet (decuplet) baryon. The meson energy is $\omega_k = \sqrt{k^2+m_\phi^2}$. 

As the regulator cutoff parameters,  $\Lambda$, are {\em a priori} not the same in different schemes, we can determine a correspondence between the three cutoff scales $\Lambda_{(\text{Cov})}$,  $\Lambda_{(\text{Rel})}$, and  $\Lambda_{(\text{HB})}$. This correspondence allows for a reasonable comparison between the schemes, since it may compensate for the differences in the suppression of large momenta.

The correspondence between the cutoff scales is determined by computing the self-energy contributions in one scheme, and fitting the others to it. Given the abundance of literature pointing towards the optimal range $0.8 \leq \Lambda_{(\text{HB})} \leq 1.0\ \text{GeV}$ for the HB scheme in SU(2), we fix $\Lambda_{(\text{HB})} = 0.8$ GeV and perform a fit of the form
\begin{align*}
&\sum_{\phi,\mathcal{B}'}\tilde{\Sigma}_{B\mathcal{B}'\phi}^{(\text{HB})}(m_\pi^2,\Lambda_{(\text{HB})}) \\
&\approx d_4^{(\text{Cov/Rel})}\, m_\pi^4 +\sum_{\phi,\mathcal{B}'}\tilde{\Sigma}^{(\text{Cov/Rel})}_{B\mathcal{B}'\phi}(m_\pi^2,\Lambda_{(\text{Cov/Rel})}), \numberthis \label{eq:25}
\end{align*}
over the domain $0.0\ \leq m_\pi^2 \leq 0.6\ \text{GeV}^2$, by adjusting $d_4$ and $\Lambda$. 

We note that the subtracted self-energies are void of the leading constant and $m_\pi^2$ terms. Since we left $a_4$ unrenormalized in Eq.~\eqref{eq:14}, we have introduced the $d_4$ term to compensate for the differences at $\mathcal{O}(m_\pi^4)$. Such a difference will simply be absorbed by $a_4$ when performing the fits to lattice data.
\begin{figure}[t]
	\centering
	\hspace*{-1.3mm}\includegraphics[width=0.97\columnwidth]{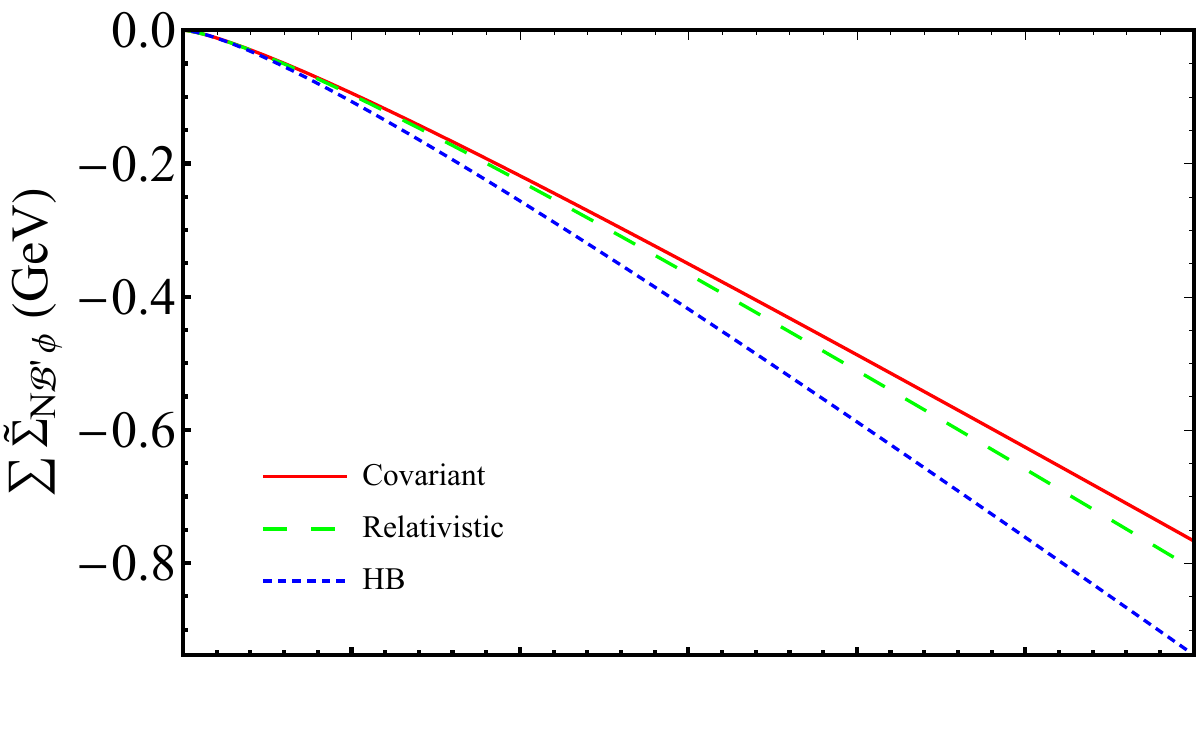} \\[-14pt]
	\includegraphics[width=\columnwidth]{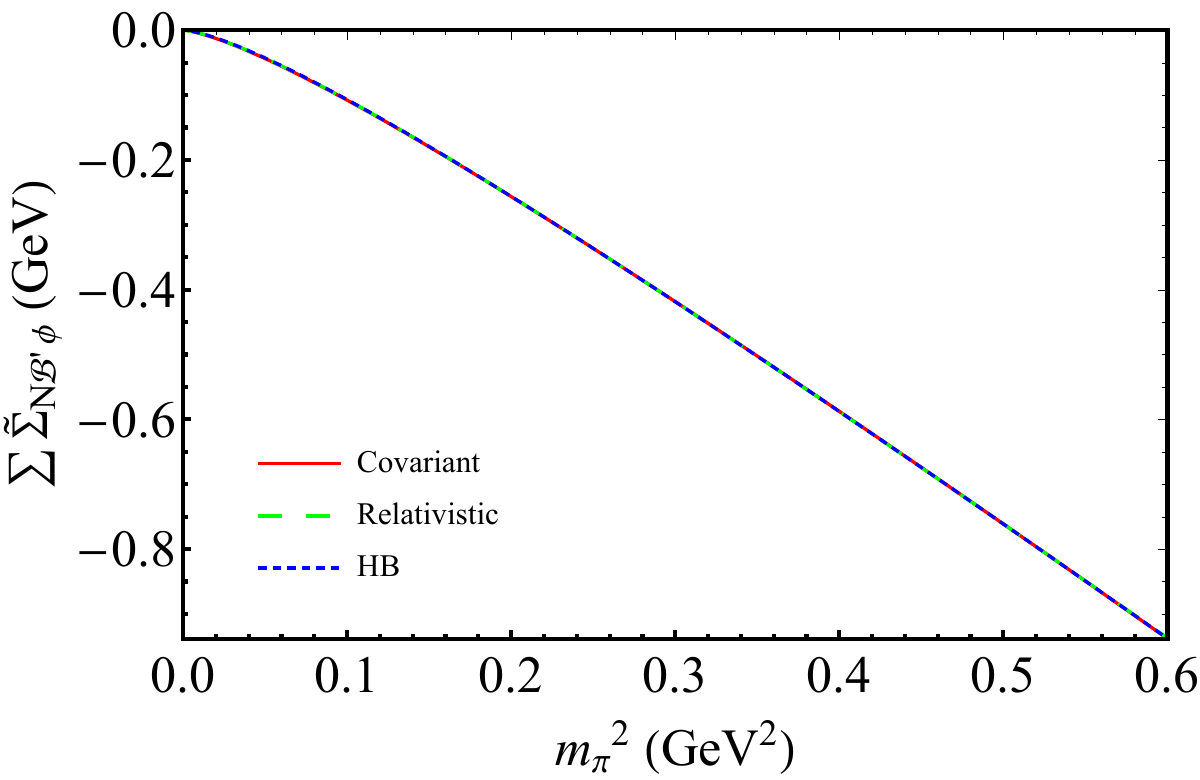}
	\caption{Sum of all FRR renormalized nucleon-baryon-meson self-energy contributions in different schemes. The upper figure compares the schemes at the same cutoff scale $\Lambda_{(\text{Cov})}=\Lambda_{(\text{Rel})}=\Lambda_{(\text{HB})}=0.8$ GeV. The bottom figure shows the fit of the relativistic schemes to the HB scheme at $\Lambda_{(\text{HB})}=0.8$ GeV. The fit parameters are shown in Table~\ref{tab:01}.}
	\label{fig:02}
\end{figure}
\begin{table}[b]
	\caption{Regulator mass and difference in unrenormalized coefficient of $m_\pi^4$ determined by fitting the corresponding self-energies calculated in the HB scheme (Eq.~\eqref{eq:25}), using $\Lambda_{(\text{HB})} \, = \, 0.8$ GeV. }
	\label{tab:01}
	\begin{ruledtabular}
		\begin{tabular}{cccc}
			Scheme & Baryon & $\Lambda$ (GeV) & $d_4$  
			($10^{-2}$ GeV$^{-3}$ )\\[2pt]
			\hline \\[-8pt]
			Covariant & Nucleon & 1.05 & 0.69  \\
			& Lambda  & 1.15 & 8.82  \\
			& Sigma  & 0.93 & -0.01 \\
			& Xi  & 1.04 & 2.74  \\[2pt]
			\hline \\[-8pt]
			Relativistic & Nucleon & 1.01 & 6.78  \\
			& Lambda  & 1.06 & 9.79  \\
			& Sigma  & 0.91 & 2.38 \\
			& Xi  & 0.98 & 3.20
		\end{tabular}
	\end{ruledtabular}
\end{table}

In Fig.~\ref{fig:02} we show the sum of all FRR renormalized nucleon-baryon-meson self-energy contributions for each scheme. As we see in the top half, the relativistic schemes do differ at the 10-20\% level when the same cutoff mass, 0.8 GeV, is used in all of them. However, as shown in the bottom half of Fig.~\ref{fig:02}, if we allow the values of the cutoff mass to vary with scheme, one finds excellent agreement between all three. The variation in $\Lambda$ is indeed somewhat subtle. While not presented here, qualitatively, the plots for $\sum_{\phi,\mathcal{B}'}\tilde{\Sigma}_{\Lambda\mathcal{B}'\phi}$, $\sum_{\phi,\mathcal{B}'}\tilde{\Sigma}_{\Sigma\mathcal{B}'\phi}$, and $\sum_{\phi,\mathcal{B}'}\tilde{\Sigma}_{\Xi\mathcal{B}'\phi}$ as a function of $m_\pi^2$ are virtually the same to that of Fig.~\ref{fig:02}.

While we have considered one value of $\Lambda$ for the sum over intermediate baryons in this comparison, one may conduct a comparison on a diagram-by-diagram basis. In such a case, one could theoretically have different regulator cutoffs for each baryon-baryon-meson contribution. This is because the regulator is associated with the induced pseudoscalar form factor of the baryon in phenomenological models. Lattice QCD calculations have demonstrated that strange quarks are more localized, which suggests that regulators, and thus the cutoff, might be expected to vary for each baryon, depending on the number of strange quarks present. However, with the level of agreement seen in Table~\ref{tab:01} between the octet baryons and the schemes, it seems sufficient to adopt the standard approach and only have one regulator parameter. Therefore, we set $\Lambda_{ (\text{Cov}) } = \Lambda_{ (\text{Rel}) } = 1.0$ GeV in the subsequent fits.

\subsection{\label{sec:fit strategy} Fit Strategy}

The fitting procedure is as follows. Firstly, we apply the FVCs of Sec.~\ref{sec:FVCs} to the lattice results, taking them from finite to infinite-volume. 

Then, we fit the infinite-volume results available at several quark masses and lattice spacings to a function containing both the physics of chiral nonanalytic behavior and a term linear in the square of the lattice spacing to address finite lattice spacing corrections at $\mathcal{O}(a^2)$.  For the nucleon, we have
\begin{align*}
\label{eq:26}
    M_N =\ &C_0 + C_2\, m_\pi^2 + a_4\, m_\pi^4 + \tilde{\Sigma}_{NN\pi} + \tilde{\Sigma}_{N\Lambda K} + \tilde{\Sigma}_{N\Sigma K}\\
    &+ \tilde{\Sigma}_{NN\eta} + \tilde{\Sigma}_{N\Delta\pi} + \tilde{\Sigma}_{N\Sigma^*K} + \tilde{\Sigma}_{N\pi,\mathrm{tad}} + D_a\, a^2 \, , 
    \numberthis
\end{align*}
where, specifically for the nucleon, we have dropped the superscript $N$ on the RSCs. We note the PACS-CS results are obtained with a nonperturbatively improved Wilson-Clover fermion action such that the leading lattice artifact is the same as for the CLS results~\cite{Ottnad:2022axz}, and addressed at $\mathcal{O}(a^2)$ by the final term in Eq.~\eqref{eq:26}. 

The fit parameters include $C_0$, $C_2$, $a_4$ and $D_a$.  With the parameters constrained by lattice QCD results, one can then use Eq.~\eqref{eq:26} to interpolate/extrapolate to any value of $m_\pi^2$ or lattice spacing approaching the continuum limit.  For example, one can expose the lattice spacing dependence of the lattice results by using Eq.~\eqref{eq:26} without the final term to access the physical quark masses and then plot the results as a function of $a^2$.

As a final step, we also explore the addition of the physical baryon mass to the data set and refit to obtain our best estimates of the renormalized RSCs.

In order to check the degree of model dependence associated with the three schemes under consideration, we compare our renormalized RSCs, $C_0$ and $C_2$ to the linear combinations of SU(2) LECs reported using $\chi$PT. 

Finally, given the large uncertainty in the tadpole coefficient of Eq.~\eqref{eq:18}, we consider the upper and lower bounds of this coefficient as a source of systematic uncertainty.

\subsection{\label{fits to lat} Fit to PACS-CS lattice QCD data}

Recent advances in computational capabilities and techniques have led to lattice QCD calculations near physical values of the light quark masses. Our aim in this section is compare analyses of the historical PACS-CS lattice QCD results~\cite{PACSCSdata}, which extend over a wide range of pion mass, using the three schemes described. Note that we choose to exclude two data points from the PACS-CS data set -- one near the physical point with $m_\pi L\approx 2$, and one simulated with a different strange quark mass.

\begin{figure}[t]
    \centering
        \includegraphics[width=\columnwidth]{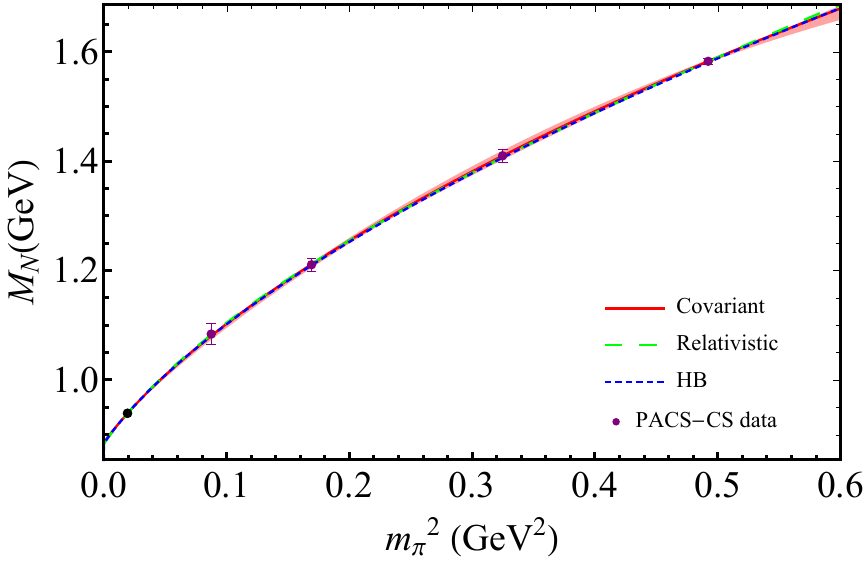}
    \caption{Fits to the PACS-CS nucleon mass data in the three schemes under consideration, covariant and relativistic formalisms at $\Lambda_{(\text{Cov})}=1.0$ GeV, and the heavy baryon scheme at $\Lambda_{(\text{HB})}=0.8$ GeV. The physical nucleon mass (black point) is combined with the lattice results in constraining the fit parameters. The data points are finite-volume corrected but there has been no correction applied for the finite lattice spacing as described in the text. The uncertainties on individual data points are statistical, while the narrow extrapolation band incorporates both statistical and systematic uncertainties.}
    \label{fig:03}
\end{figure}

In the case of the PACS-CS data set~\cite{PACSCSdata}, we cannot readily calculate the $\mathcal{O}(a^2)$ leading lattice artifact correction, because the lattice spacing is kept constant in the PACS-CS scheme. As an initial fit, we set $D_a=0$ in Eq.~\eqref{eq:26}. The results of this initial fit are shown in Fig.~\ref{fig:03} and Table~\ref{tab:02}.
\begin{table}[b]
    \caption{Fits to the nucleon RSCs obtained from the PACS-CS nucleon mass lattice QCD results of Ref.~\cite{PACSCSdata}, combined with the physical mass. The errors associated with the fit parameters are statistical followed by systematic, as described in the text. The regulator cutoff for relativistic and HB formalisms are $\Lambda_{(\text{Cov})}=\Lambda_{(\text{Rel})}=1.0$ GeV and $\Lambda_{(\text{HB})}=0.8$ GeV, respectively.} 
    \label{tab:02}
    \begin{ruledtabular}
        \begin{tabular}{cccl}
         {Scheme} & {$C_0$ (GeV)} & {$C_2$ (GeV$^{-1}$)} & {$a_4$ (GeV$^{-3}$)} \\[2pt]
        \hline \\[-8pt]
        {Covariant} & 0.885(1)(7) & 3.59(5)(52) & -0.17(10)(68) \\ 
        {Relativistic} & 0.882(1)(8) & 3.69(7)(62) & \phantom{-}0.25(13)(112) \\ 
		{HB} & 0.885(1)(5) & 3.44(5)(42)  & -0.29(9)(64) \\ 
        \end{tabular}
    \end{ruledtabular}
\end{table}

We note that the uncertainty associated with the physical point is several orders of magnitude smaller than that of the lattice data, which means that the goodness of fit is primarily determined by the physical point. Therefore, to have some reasonable gauge of the goodness of fit, we exclude the physical point in the $\chi^2_{\text{dof}}$ calculation.

As seen in Fig.~\ref{fig:03}, the fits to the PACS-CS data are virtually indistinguishable. The $\chi^2_{\text{dof}}$ is approximately 0.26 for all three schemes. Not only are the fits to the data over values of $m_\pi^2 \in 0.02-0.50$ GeV$^2$ using the three schemes extremely close but, as we see in Table~\ref{tab:02}, the renormalized RSCs $C_0$ and $C_2$ are also completely consistent. It is also necessary to examine if the value of $C_2$ is consistent with the input used in the coefficient of the tadpole contribution of Eq.~\eqref{eq:18}. This is due to the fact that, in the nucleon mass expansion, the renormalized RSC of the $m_\pi^2$ is related to one of the tadpole coefficients, specifically $C_2=-4c_1$. In obtaining the combination of Eq.~\eqref{eq:18}, Refs.~~\cite{c1c2c3Meissner,c1c2c3Frink} quote $c_1=-0.9^{+0.5}_{-0.2}$ GeV$^{-1}$. Taking the average value of $C_2$ in Table~\ref{tab:02}, we find $c_1=-0.89(1)$ GeV$^{-1}$ (negligible difference when added in quadrature), which is completely consistent with the input.

In summary, contrary to claims in the literature, the use of HB theory with FRR does allow one to determine model independent LECs using lattice data over a wide range of $m_\pi^2$.

The PACS-CS data set further provides the masses for the other octet baryon (hyperon) masses. Since we restricted the study of the nucleon to dimension-two LECs in SU(2), for the hyperons, we use a form analogous to Eq.~\eqref{eq:26},
\begin{align*}
	M_\Lambda &= C_0^\Lambda + C_2^\Lambda\, m_\pi^2 + a_4^\Lambda\, m_\pi^4 + \tilde{\Sigma}_{\Lambda\Sigma\pi} + \tilde{\Sigma}_{\Lambda NK} + \tilde{\Sigma}_{\Lambda\Xi K}\\
	&+ \tilde{\Sigma}_{\Lambda\Lambda\eta} + \tilde{\Sigma}_{\Lambda\Sigma^*\pi} + \tilde{\Sigma}_{\Lambda\Xi^*K}, \\
	M_\Sigma &= C_0^\Sigma + C_2^\Sigma\, m_\pi^2 + a_4^\Sigma\, m_\pi^4 + \tilde{\Sigma}_{\Sigma\Sigma\pi} + \tilde{\Sigma}_{\Sigma\Lambda\pi} + \tilde{\Sigma}_{\Sigma NK}\\
	&+ \tilde{\Sigma}_{\Sigma\Xi K} + \tilde{\Sigma}_{\Sigma\Sigma\eta} + \tilde{\Sigma}_{\Sigma\Sigma^*\pi} + \tilde{\Sigma}_{\Sigma\Delta K} + \tilde{\Sigma}_{\Sigma\Xi^* K} + \tilde{\Sigma}_{\Sigma\Sigma^*\eta}, \\
	M_\Xi &= C_0^\Xi + C_2^\Xi\, m_\pi^2 + a_4^\Xi\, m_\pi^4 + \tilde{\Sigma}_{\Xi\Xi\pi} + \tilde{\Sigma}_{\Xi\Lambda K} + \tilde{\Sigma}_{\Xi\Sigma K}\\
	&+ \tilde{\Sigma}_{\Xi\Xi\eta} + \tilde{\Sigma}_{\Xi\Xi^*\pi} + \tilde{\Sigma}_{\Xi\Sigma^*K} + \tilde{\Sigma}_{\Xi\Omega K} + \tilde{\Sigma}_{\Xi\Xi^*\eta},
	\numberthis
\end{align*}
again omitting the $\mathcal{O}(a^2)$ lattice artifact correction. We reiterate that $C_i^B$ and $a_i^B$ are coefficients of the light flavor chiral expansion of the baryon mass and are unique to each member of the octet.

\begin{figure}[t]
	\centering
	\includegraphics[width=\columnwidth]{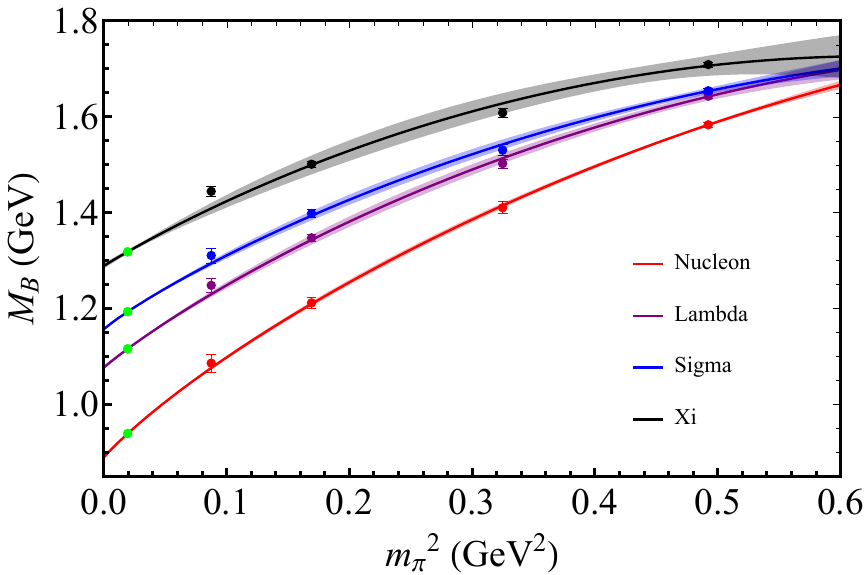}
	\caption{Fits to the PACS-CS octet baryon mass results in the covariant formalism at $\Lambda_{(\text{Cov})}=1.0$ GeV, with the inclusion of the physical points (green dots). The data points are finite-volume corrected. The extrapolation bands and errors on individual data points are purely statistical.
	}
	\label{fig:04}
\end{figure}

In Fig.~\ref{fig:04} we see that the baryon mass expansion for the other octet baryons describes the data fairly well. While we only show the covariant fit, the other fits are indistinguishable, similar to those shown in Fig.~\ref{fig:03}. While not significant for the nucleon, the fact that the strange quark mass used in the PACS-CS simulations is a little large \cite{Menadue:2011pd} is manifest in Fig.~\ref{fig:04}. There the extrapolation curves dip down to pass through the experimental values in a manner that leaves the lattice QCD results at the smallest pion mass positioned above the curves.
\begin{table}[b]
	\caption{Fits to the RSCs (where $B = \Lambda,\ \Sigma,\ \text{and }\Xi$) obtained from the PACS-CS octet baryon mass lattice QCD results of~\cite{PACSCSdata}, combined with the physical masses. The errors associated with the fit parameters are statistical. The regulator cutoff masses for the relativistic and HB formalisms are $\Lambda_{(\text{Cov})}=\Lambda_{(\text{Rel})}=1.0$ GeV and  $\Lambda_{(\text{HB})}=0.8$ GeV, respectively.}
	\label{tab:03}
	\begin{ruledtabular}
		\begin{tabular}{ccccc}
			Baryon & Scheme & $C_0^B$ (GeV)& $C_2^B$ (GeV$^{-1}$)& $a_4^B$ (GeV$^{-3}$) \\[2pt]
			\hline \\[-8pt]
			Lambda & Covariant & 1.077(3) & 2.21(11) & -0.78(22) \\
			& Relativistic & 1.076(3) & 2.24(11) & -0.71(21) \\
			& HB & 1.076(3) & 2.26(10) & -0.76(21) \\[2pt]
			\hline \\[-8pt]
			Sigma & Covariant & 1.157(3) & 2.11(10) & -0.70(21) \\
			& Relativistic & 1.157(3) & 2.11(10) & -0.65(20) \\
			& HB & 1.157(3) & 2.08(10) & -0.74(21) \\[2pt]
			\hline \\[-8pt]
			Xi & Covariant & 1.289(6) & 1.64(22) & -0.97(45) \\
			& Relativistic & 1.288(6) & 1.65(22) & -0.95(45) \\
			& HB & 1.289(6) & 1.62(20) & -0.94(42)
		\end{tabular}
	\end{ruledtabular}
\end{table}

In Table~\ref{tab:03} we see a similar trend to that found for the nucleon. With a cutoff mass chosen to give a good fit over the large range of $m_\pi^2$ provided by PACS-CS, there is very little difference in the renormalized RSCs $C_0$ and $C_2$. Since $a_4$ is not renormalized, $a_4$ in itself does not represent a physical quantity. However, it is worth noting that variation in $a_4$ between different schemes in Table~\ref{tab:03} is less than the variation found in Table~\ref{tab:02}. This is largely attributed to the presence of the tadpole term in Eq.~\eqref{eq:26} which was not included in the fitting form of Eq.~\eqref{eq:25}.

Looking at the differences between the light flavor chiral limit masses, $C_0^B$ and $C_0$ (recall that we have dropped the superscript $N$ for the nucleon), a simple relationship akin to the constituent quark model emerges. $C_0$ subtracted from the average of $C_0^\Lambda$ and $C_0^\Sigma$ is approximately 200 MeV, and $C_0$ subtracted from $C_0^\Xi$ is approximately 400 MeV. One can reasonably interpret the differences in these masses arising from the different number of strange quarks. 

\section{\label{sec:analysisCLSdata}Analysis of the latest CLS Lattice QCD results for the nucleon}
Here, we focus on the most recent nucleon mass lattice QCD results from the CLS collaboration~\cite{Ottnad:2022axz}. This data set is chosen because it presents a range of accurate lattice QCD results up to $m_\pi^2 = 0.14$ GeV$^2$, with the physical mass scale set using the modern gradient flow method~\cite{Luscher:2010iy}.

\begin{figure}[t]
    \centering
        \includegraphics[width=1.05\columnwidth]{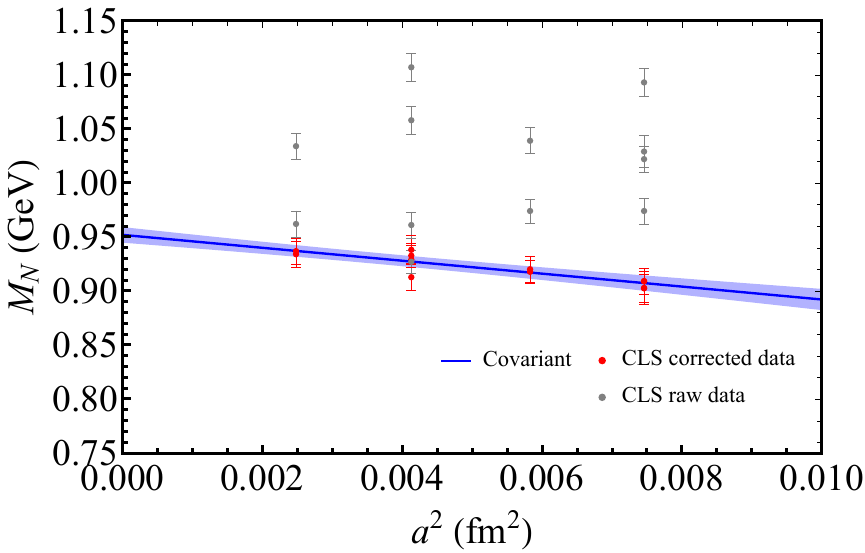}
        \includegraphics[width=\columnwidth]{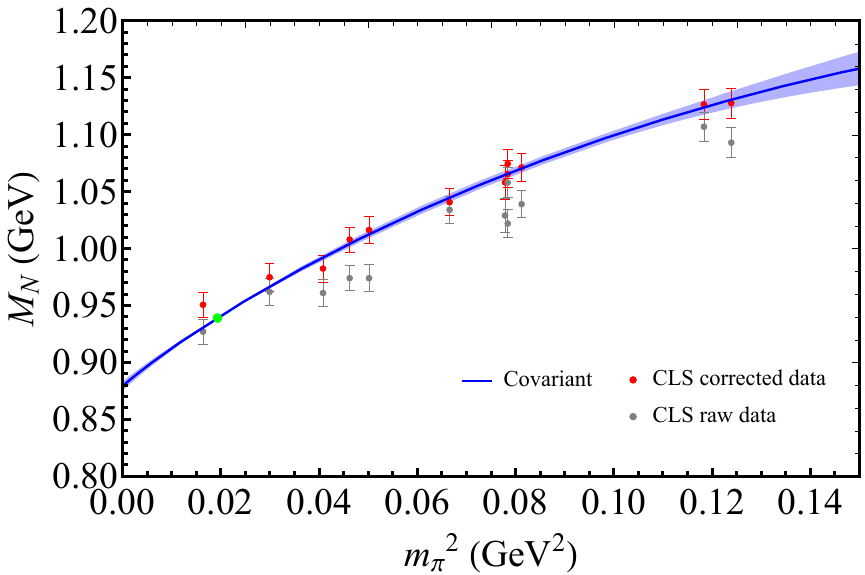}  
    \caption{Fits to the CLS data set in the covariant formalism at $\Lambda_{(\text{Cov})}=1.0$ GeV. The corrected data refers to the data where finite-volume and $\mathcal{O}(a^2)$ corrections are made. The top panel shows the uncorrected data at various pion masses along with a fit to the corrected data evaluated at $m_\pi^2=m_\pi^2|_{\text{phys}}$, where $D_a \approx -6$ GeV fm$^{-2}$. The bottom panel shows our fit to the corrected CLS results with the inclusion of the physical point (green dot). The uncertainties on individual data points are statistical while the extrapolation band incorporates both statistical and systematic uncertainties associated with the tadpole contribution. }
    \label{fig:05}
\end{figure}

By employing the fit strategy described in Sec.~\ref{sec:fit strategy} on the CLS data set, we find $D_a\approx -6$ GeV fm$^{-2}$ in all the schemes. The results for the fits are shown in Fig.~\ref{fig:05} and the renormalized RSCs are provided in Table~\ref{tab:04}. In performing the fits, we have omitted all data with $L<2.5$ fm in order to avoid large FVCs. 

In the same way from the previous section, we perform a consistency check on $C_2$. We find, from Table~\ref{tab:04}, $c_1 = -0.98(6)$ GeV$^{-1}$ which again is consistent with the input.

The $\chi^2_{\text{dof}}$ in all schemes are comparable at approximately 0.62. Again, all the fits are within one standard deviation of one another, and the differences are small. 
\begin{table}[b]
    \caption{Fits to the nucleon RSCs obtained from the corrected CLS nucleon mass lattice QCD results \cite{Ottnad:2022axz}, combined with the physical mass. The errors associated with the fit parameters are statistical followed by systematic, as described in the text. The regulator cutoffs for the covariant, relativistic, and HB formalisms are $\Lambda_{(\text{Cov})}=\Lambda_{(\text{Rel})}=1.0$ GeV and $\Lambda_{(\text{HB})}=0.8$ GeV, respectively.}
    \label{tab:04}
    \begin{ruledtabular}
        \begin{tabular}{cccc}
         Scheme & $C_0$ (GeV) & $C_2$ (GeV$^{-1}$)& $a_4$ (GeV$^{-3}$) \\[2pt] 
        \hline \\[-8pt]
        Covariant & 0.879(4)(2) & 3.92(24)(24) & -3.2(20)(22) \\ 
        Relativistic  & 0.878(4)(2) & 3.97(24)(27) & -2.7(20)(29) \\ 
        HB  & 0.879(4)(2) & 3.85(24)(20) & -4.0(20)(17) \\ 
        \end{tabular}
    \end{ruledtabular}
\end{table}

In relation to the previous section, we also performed an analysis for the combined set of PACS-CS and CLS nucleon lattice data. For this particular case, we allow the coefficient of the tadpole, $\chi C$, to also be a fit parameter, constrained by Eq.~\eqref{eq:18}. This is reasonable, as the wide range of $m_\pi^2$ covered by the PACS-CS data acts to restrict $\chi C$ in a meaningful manner. We use the CLS data to correct for the PACS-CS leading lattice artifact (noting that $D_a$ was set to zero for the PACS-CS data in the previous section). Since the variation in the lattice spacing for the PACS-CS data is small, the $\mathcal{O}(a^2)$ correction simply shifts the data points by a constant amount. We find, for the renormalized RSCs and the coefficient of the tadpole,  $C_0 = 0.884(4)$ GeV and $C_2 = 3.68(28)$ GeV$^{-1}$, and $\chi C = -1.8(10)$ GeV$^{-1}$, respectively. The $\chi^2_{\text{dof}}$ in this case is approximately 0.56. These values are consistent with the those found in the fits presented earlier.

\section{\label{sec:compar of LECs and sigma} Comparison of the LECs and the \texorpdfstring{$\bm{\sigma}$}{σ}-term}

\subsection{\label{compar of LECs} Comparison of the LECs}
Here, we wish to make a comparison between the $\chi$EFT renormalized RSCs of the nucleon extracted using our three FRR schemes and those LECs obtained in earlier $\chi$PT work. In $\chi$PT, the chiral expansion of the nucleon mass in SU(2) is written as \cite{SvenSteininger_1998,SU2chiPTmassexpBecher,EOMSFuchs,SU2chiPTmassexpSchindler}
\begin{equation}\label{eq:27}
    M_N = m + k_1\, m_\pi^2+ k_2\, m_\pi^3+ k_3\, m_\pi^4\ln \frac{m_\pi}{\mu}+ \cdots \, ,
\end{equation}
where $m$ is the renormalized nucleon mass in the chiral limit, $\mu$ is the renormalization parameter, and the coefficients $k_i$ are linear combinations of the renormalized LECs. Explicitly, the $k_i$ are
\begin{align*}\label{eq:28}
    &k_1 = -4c_1, \mkern20mu k_2 = -\frac{3g_A^2}{32\pi f_\phi^2}, \numberthis
\end{align*}
and $k_3$ is provided in Eq.~\eqref{eq:17}. $g_A$ and $f_\phi$ are the chiral limit values of the axial-vector coupling constant and pion decay constant, respectively.

We note that the values of $k_3$ found in the relativistic schemes are different from that found in the HB scheme. Clearly, in the limit $M_N\to \infty$, the term proportional to $g_A^2/M_N$ vanishes, and 
\begin{equation}
    \lim_{M_N\to \infty} \mathcal{A} = \frac{24}{25}g_A^2\frac{1}{\Delta_{\Delta N}}.
\end{equation}
The total discrepancy in the coefficient of the NLNA term from the $NN\pi$ and the $N\Delta\pi$ self-energies amounts to 10\% between the relativistic and HB formalisms. This difference, in comparison with the uncertainty associated with the tadpole coefficient, $\chi C$, is very small. 
\begin{table}[b]
	\caption{Comparison of the renormalized RSCs obtained using FRR in the covariant scheme to analyze the CLS data with those $\chi$PT LECs reported in Refs.~\cite{LECsProcura} and \cite{LECsBernard}. Note in $\chi$PT notation $C_0 = m$ and $C_2=k_1$.} 
	\label{tab:05}
	\begin{ruledtabular}
		\begin{tabular}{lll}
			Scheme & $C_0$ (GeV) & $C_2$ (GeV$^{-1}$) \\[2pt]
			\hline \\[-8pt]
			FRR result (This work) & 0.879(4)(2) & 3.92(24)(24) \\ 
			Procura IR \cite{LECsProcura} & 0.883(3) & 3.72(16) \\ 
			Bernard CR \cite{LECsBernard} & 0.88 & 3.6 (fixed)   
		\end{tabular}
	\end{ruledtabular}
\end{table}

A comparison is shown in Table~\ref{tab:05}, where we present our (covariant) FRR result from the CLS data set Table~\ref{tab:04} with earlier $\chi$PT works from Refs.~\cite{LECsProcura} and \cite{LECsBernard}. These two works include the physical point constraint, but use a different regularization scheme, namely, infrared regularization (IR) and cutoff regularization (CR). The values for $k_3$ vary greatly with $\chi C$. With $\chi C$ in the range of Eq.~\eqref{eq:18}, $k_3 = -7.4$ to 11.9 GeV$^{-3}$, while the values of Procura {\em et al.}~\cite{LECsProcura} and Bernard {\em et al.}~\cite{LECsBernard} yield  $k_3=1.36\pm 3.29\ \text{GeV}^{-3}$ and $k_3=3.82\ \text{GeV}^{-3}$, respectively. 

By way of comparison, we also show the results obtained from the CLS collaboration, by fitting a $\chi$PT-inspired 
form~\cite{Ottnad:2022axz}
\begin{equation}
    M_N(m_\pi,a,L) = \mathring{M}_N+B m_\pi^2+C m_\pi^3+D a^2+E\frac{m_\pi^3}{m_\pi L}e^{-m_\pi L} \, .
\end{equation}
$\mathring{M}_N$ is the nucleon mass in the chiral limit and the letters $B$-$E$ represent fit parameters including the LNA term. The lattice artifact correction and FVC are accounted for by the terms proportional to $D$ and $E$, respectively. Note that in our $\chi$EFT the LNA term is model- and scheme-independent and the coefficient is {\em not} a fit parameter. Nevertheless, with the above form, we obtain (including the physical point) $\mathring{M}_N=0.880(6)$ GeV, $B=3.75(46)$ GeV$^{-1}$, $C=-4.8(13)$ GeV$^{-2}$, $D=-6.0(14)$ GeV fm$^{-2}$, and $E=82(99)$ GeV$^{-2}$. We remark that $C$ in this fit is roughly 14\% smaller than the correct, model independent value, resulting in a shallow slope of the extrapolation curve near the physical point. Without the physical point constraint in the fit, one would find $C$ to be approximately 2 to 3 times smaller than the $\chi$EFT value. Any extrapolation relying on such a value is unphysical and extracting any physically meaningful observable is difficult. Alternatively, one may fix $C$ to $k_2$ of $\chi$EFT (see Eq.~\eqref{eq:28}), and in that case we find $\mathring{M}_N=0.875(1)$ GeV and $B=4.07(5)$ GeV$^{-1}$ producing reasonable agreement with the values quoted in Table~\ref{tab:04}.

\subsection{\label{sigma terms} The \texorpdfstring{$\bm{\sigma}$}{σ}-term}

The pion-nucleon sigma term $\sigma_{\pi N}$ is defined as
\begin{equation}
    \sigma_{\pi N} = \hat{m}\bra{N}\bar{u}u+\bar{d}d\ket{N} \, ,
\end{equation}
where $\hat{m}=(m_u+m_d)/2$. From the quark mass dependence of the baryon and using the Feynman-Hellmann theorem, one can compute $\sigma_{\pi N}$ by
\begin{equation}\label{eq:32}
    \sigma_{\pi N} = m_\pi^2\frac{\partial M_N}{\partial m_\pi^2} \, ,
\end{equation}
at the physical pion mass, where, in our calculation, we use the charged pion mass. The results for the application of Eq.~\eqref{eq:32} to each of the schemes summarized in Table~\ref{tab:04} are presented in Table~\ref{tab:06}. The first set of uncertainties are statistical, while the second set are systematic.
\begin{table}[b]
	\caption{Pion-nucleon sigma term at the physical point computed from the CLS data set using the fits for each scheme as summarized in
		Table~\ref{tab:04}.  
		The first set of uncertainties are statistical and the second set are systematic.}
	\label{tab:06}
	\begin{ruledtabular}
		\begin{tabular}{cc}
			{Scheme} & {$\sigma_{\pi N}$ (MeV)} \\[2pt]
			\hline \\[-5pt]
            Covariant & $51.7\pm 3.1 \pm 1.4$ \\[5pt] 
			Relativistic & $52.2\pm 3.2 \pm 1.4 $  \\[5pt] 
			HB & $51.7\pm 3.2 \pm 1.4 $ \\[5pt]
		\end{tabular}
	\end{ruledtabular}
\end{table}

\begin{figure}[t]
	\centering
	\includegraphics[width=\columnwidth]{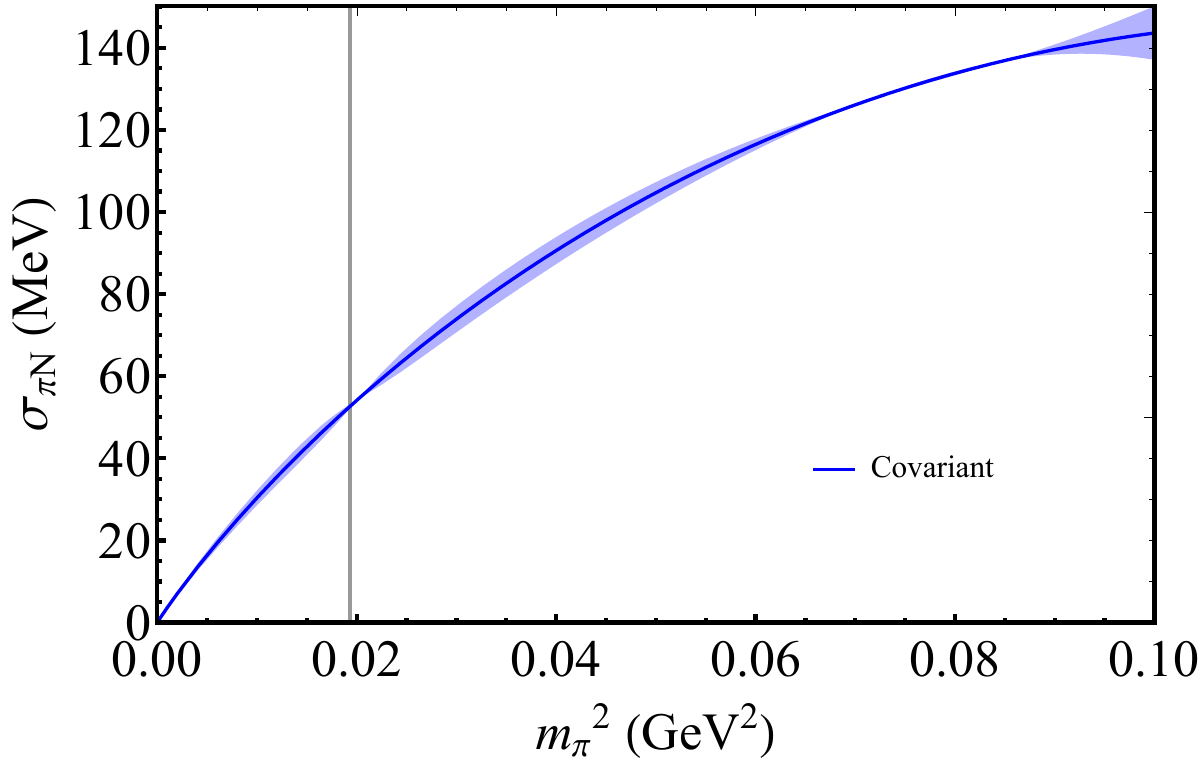}
	\caption{$\sigma_{\pi N}$ as a function of $m_\pi^2$ from the CLS data set fits of Table~\ref{tab:04} in the covariant formalism at $\Lambda_{(\text{Cov})}=1.0$ GeV. The vertical line represents the physical value of charged $m_\pi^2$. The extrapolation band incorporates both statistical and systematic uncertainties.}
	\label{fig:06}
\end{figure}

We also show in Fig.~\ref{fig:06} a plot of the $m_\pi^2$ dependence of $\sigma_{\pi N}$. Again, we show only the result in the covariant scheme, but the curves essentially overlap for the other two schemes, as in Fig.~\ref{fig:03}. Here, the gradient of the curve starts to flatten at around $m_\pi^2=0.1$ GeV$^2$, where the data points become sparse. With points available at larger $m_\pi^2$, as in the PACS-CS data set, we tend to see the curve flattening further out.

In view of the interest in the sigma commutator it is worthwhile to explore the sources of systematic error in some detail. 

In Ref.~\cite{Hoferichter:2023}, it was suggested that using the neutral pion mass, rather than that of the charged pion, could decrease the value of $\sigma_{\pi N}$ by up to 3 MeV. In our analysis, the changes in $\sigma_{\pi N}$ across all schemes used in Table~\ref{tab:04} are consistent, with a maximum reduction of 2 MeV.

The explicit tadpole makes a relatively small contribution to $\sigma_{\pi N}$ of order a few MeV. However, the systematic error associated with it is potentially significant because of the large uncertainty in the corresponding coefficient. This error is determined by refitting to the CLS data using fixed values of $\chi C$ at the upper and lower bound given in Eq.~\eqref{eq:18} and recalculating $\sigma_{\pi N}$ from the respective fits. In this process we observed a strong correlation between the coefficients $k_1$ and $k_3$ which acts to reduce the effect of the systematic uncertainty. Ultimately, the fitted renormalized RSCs are constrained by lattice QCD data, such that a variation in one parameter is mitigated by the other parameters. The net result is a systematic error in $\sigma_{\pi N}$ of approximately $\pm 0.7$ MeV in all schemes. 

Further systematic uncertainties in $\sigma_{\pi N}$ may derive from the choice of regulator and the corresponding cutoff parameter. References~\cite{Young:2002ib} and 
\cite{Leinweber:2003dg} demonstrated that when lattice data was fit using different regulator functions the renormalized RSCs showed little variation. To test that here, we repeated the fits to the CLS data with three choices of form factor, a monopole, a dipole, and a sharp cutoff. The first two showed no model dependence while the more extreme sharp cutoff gave rise to a 2\% variation in the renormalized RSCs. However, even in the latter case the change in $\sigma_{\pi N}$ was just 0.2 MeV. Because the fit is constrained by accurate lattice data, the 2\% change in our $C_2$ was compensated by a change in $a_4$.

One may also explore the light quark mass (or $m_\pi^2$) dependence of the regulator mass parameter, $\Lambda$.  We note that the pion cloud has no direct effect on the axial form factor of the nucleon, which is closely related to the pion-nucleon form factor. Thus the variation with pion mass is expected to be slow. Within the MIT bag model the bag radius, which is inversely proportional to the mass parameter $\Lambda$ in the axial form factor, decreases by just 2\% when $m_\pi^2$  varies between 0.02 and 0.50 GeV$^2$. As a conservative estimate of the effect of the value of $\Lambda$ varying with pion mass, we have reanalyzed the CLS data allowing for a 5\% increase over this mass range. This variation gives rise to a difference of just 0.3 MeV in the value of $\sigma_{\pi N}$.

The remaining input in the fits explored in this work are the coefficients of the pion coupling to the octet and decuplet baryons. A change of $\pm10$\% in the coefficient $\mathcal{C}$ resulted in a variation of 0.1 MeV in $\sigma_{\pi N}$. A 5\% change in $f_\phi$ gave a variation of 0.3 MeV, and finally the more drastic variation of $F$ and $D$ by $\pm 10$\% yielded a change in $\sigma_{\pi N}$ of just 0.3 MeV.

To obtain the final systematic error based on all the sources detailed above, we evaluate the sigma commutator at the average of the values obtained with the charged and neutral pion mass, with an uncertainty of $\pm 1$ MeV. The systematic errors arising from all sources are then combined in quadrature, leading to the values shown in Table~\ref{tab:06}.
\begin{table}[t]
    \caption{Comparison with other determinations of the pion-nucleon sigma commutator.}
    \label{tab:07}
    \begin{ruledtabular}
        \begin{tabular}{llc}
        {Method} & {$\sigma_{\pi N}$ (MeV)} & Ref. \\
        \hline \\[-5pt]
            Data ($\pi N$ scattering) & 59.1(3.5) &\cite{Hoferichter:2023} \\[3pt]
            & 58(5) & \cite{RuizdeElvira:2017stg} \\[3pt]
            & 59(7) &\cite{PhysRevD.85.051503} \\[3pt]
            Direct lattice calculation & 45.9(7.4)(2.8) & \cite{Yang:2015uis} \\[3pt]
            & 41.6(3.8) & \cite{Alexandrou:2019brg} \\[3pt]
            Feynman-Hellmann + lattice data & 45(6) & \cite{Shanahan:2012wh} \\[3pt]
            & 52(3)(8) & \cite{Alvarez-Ruso:2013fza} \\[3pt]
            & 55(1)(4) & \cite{Ren:2014vea} \\[3pt]
            & 44(3)(3) & \cite{Copeland:2021qni} \\[3pt]
            This work from CLS lattice data \cite{Ottnad:2022axz} & $51.7\pm 3.2 \pm 1.4$ & \\[3pt]
        \end{tabular}
    \end{ruledtabular}
\end{table}
In Table~\ref{tab:07} we compare the value 
of $\sigma_{\pi N}$ calculated here with the values extracted by other methods. Deducing the value from pion-nucleon scattering data~\cite{PhysRevD.85.051503} may involve a number of complications~\cite{Ericson:1987uf}. As indicated, the preferred value seems to be around 58 MeV with an uncertainty of order 5 MeV~\cite{RuizdeElvira:2017stg}. This differs by a surprising amount from direct lattice calculations but is within the uncertainties of most applications of the Feynman-Hellmann theorem. The value found in our analysis, namely 
$51.7\pm 3.2 \pm 1.4$ MeV, is compatible with the value extracted from pion-nucleon scattering.

\section{\label{sec:conclusion} Conclusion}
We have presented a detailed study of the mass of the nucleon as a function of pion mass using finite range regularization to evaluate the self-energy integrals. The heavy baryon approximation is compared with a covariant scheme and a relativistic scheme. All three methods produce essentially identical fits to the lattice QCD results from the PACS-CS collaboration~\cite{PACSCSdata} over a wide range of pion mass, well beyond the power counting regime. The renormalized residual series coefficients extracted from those fits are independent of the schemes used and agree well with low-energy constants found in earlier studies using $\chi$PT. A similar degree of scheme independence is found when the schemes are applied to the other members of the nucleon octet. This is contrary to the claims of literature where the heavy baryon approximation is deemed unsuitable.

These schemes were then applied to an analysis of the most recent data on the nucleon mass from the CLS Collaboration~\cite{Ottnad:2022axz}. Of particular interest is the result that the values of pion-nucleon sigma commutator, $\sigma_{\pi N}$, extracted using all three methods including the uncertainty in the coefficient of the (NLNA) $m_\pi^4 \ln m_\pi$ term in the chiral expansion, are completely consistent. The result is $\sigma_{\pi N}=51.7 \pm 3.2 \pm 1.4$ MeV. As illustrated in Table~\ref{tab:07}, this is in reasonable agreement with the value deduced from pion-nucleon scattering data, as well as most applications of the Feynman-Hellmann theorem to lattice data.

\begin{acknowledgments}
This research was undertaken with the assistance of resources from the National Computational
Infrastructure (NCI), provided through the National Computational Merit Allocation Scheme.
This work was supported by the University of Adelaide and by the Australian Research Council through Discovery Projects DP190102215 and DP210103706 (DBL) and DP230101791 (AWT), as well as through the Australian Research Council Centre of Excellence for Dark Matter Particle Physics (CE200100008).
\end{acknowledgments}

\appendix

\section{\label{app:A} Effective SU(3)\texorpdfstring{${}_L\times$}{\_Lx}SU(3)\texorpdfstring{${}_R$}{\_R} chiral Lagrangian}
Beginning in the mesonic sector, $U$ is the 3$\times$3 unimodular unitary matrix which contains the meson fields $\phi$. In particular,
\begin{equation}
    U = \exp \left( i\frac{\phi}{f_\phi} \right)=u^2,
\end{equation}
where
\begin{equation}
    \phi = \sqrt{2}\begin{bmatrix} \frac{1}{\sqrt{2}}\pi^0+\frac{1}{\sqrt{6}}\eta & \pi^+ & K^+ \\ \pi^- & -\frac{1}{\sqrt{2}}\pi^0+\frac{1}{\sqrt{6}}\eta & K^0 \\ K^- & \bar{K}^0 & -\frac{2}{\sqrt{6}}\eta \end{bmatrix},
\end{equation}
with $u$ the unitary square root of $U$. The investigations of the paper assumes the absence of an external field, so that the covariant derivative $D_\mu U\to \partial_\mu U$. The explicit chiral symmetry breaking is accounted for in the matrix $\chi=2B_0\mathcal{M}$ where $\mathcal{M}=\text{diag}(m_u,m_d,m_s)$ and the parameter $B_0$ is related to the singlet quark condensate $3F_0^2B_0=-\left\langle 0| \Bar{q}q |0\right\rangle_{0}$ with $F_0$ the pseudoscalar decay constant in the chiral limit. 

In the octet baryon sector, each baryon field is a four-component Dirac field, and the matrix $B$ follows a similar structure to the meson matrix
\begin{equation}
    B = \begin{bmatrix} \frac{1}{\sqrt{2}}\Sigma^0+\frac{1}{\sqrt{6}}\Lambda & \Sigma^+ & p \\ \Sigma^- & -\frac{1}{\sqrt{2}}\Sigma^0+\frac{1}{\sqrt{6}}\Lambda & n \\ \Xi^- & \Xi^0 & -\frac{2}{\sqrt{6}}\Lambda \end{bmatrix}.
\end{equation}
The covariant derivative of the octet baryon is (again, in the absence of external fields) $D_\mu B=\partial_\mu+[\Gamma_\mu,B]$, where $\Gamma_\mu \equiv \frac{1}{2}(u^\dagger \partial_\mu u+u\partial_\mu u^\dagger)$. With a similar structure to the $\Gamma_\mu$, the baryon fields couple to the meson fields by the so called chiral vielbein  
\begin{equation}
    u_\mu \equiv i(u^\dagger \partial_\mu u-u\partial_\mu u^\dagger),
\end{equation}
which transforms as an axial-vector under parity transformation. In the NLO Lagrangian, additional chirally invariant structures are possible, and in particular we have $\chi_+ = u^\dagger \chi u^\dagger+u\chi^\dagger u.$

In the decuplet baryon sector, the spin-3/2 decuplet fields are defined by the symmetric flavour tensor $T^{ijk}$ with
\begin{align*}
    T^{111}&=\Delta^{++}, &T^{112}&=\frac{1}{\sqrt{3}}\Delta^+,  \\
    T^{122}&=\frac{1}{\sqrt{3}}\Delta^0, &T^{222}&=\Delta^-,  \\
    T^{113}&=\frac{1}{\sqrt{3}}\Sigma^{*+}, &T^{123}&=\frac{1}{\sqrt{6}}\Sigma^{*0}, &T^{223}&=\frac{1}{\sqrt{3}}\Sigma^{*-}, \\
    T^{133}&=\frac{1}{\sqrt{3}}\Xi^{*0}, &T^{233}&=\frac{1}{\sqrt{3}}\Xi^{*-}, &T^{333}&=\Omega^-. \numberthis
\end{align*}
The covariant derivative is defined by $D_\mu \left( T_\nu \right)^{ijk} = \partial_\mu\left( T_\nu \right)^{ijk}+\left( \Gamma_\mu \right)^i_l \left( T_\nu \right)^{ljk}+\left( \Gamma_\mu \right)^j_l \left( T_\nu \right)^{ilk}+\left( \Gamma_\mu \right)^k_l \left( T_\nu \right)^{ijl}$.  

The propagators of the meson and octet baryon Lagrangian are the usual spin-0 and spin-1/2 Feynman propagators respectively, while the spin-3/2 propagator is a little bit more involved. Here, we simply quote the result \cite{chiPTtextbookScherer}
\begin{align*}
    S^{\mu\nu}_F(p) &= -\frac{\slashed{p}+M_T}{p^2-M_T^2+i\varepsilon}\bigg[g^{\mu\nu}-\frac{1}{3}\gamma^\mu\gamma^\nu+\frac{(p^\mu\gamma^\nu-\gamma^\mu p^\nu)}{3M_T} \\
    & -\frac{2}{3M_T^2}p^\mu p^\nu \bigg]+\frac{1}{3M_T^2}\frac{1+A}{1+2A} \bigg[ \bigg( \frac{A}{1+2A}M_T \\
    &-\frac{1+A}{2(A+2A)}\slashed{p}\bigg)\gamma^\mu\gamma^\nu-\gamma^\mu p^\nu-\frac{A}{1+2A}p^\mu \gamma^\nu \bigg]. \numberthis
\end{align*}
In addition, for the leading-order meson-octet-decuplet Lagnragian, the tensor $\Theta^{\mu\nu}$ is defined as
\begin{equation}
    \Theta^{\mu\nu} \equiv g^{\mu\nu}+\frac{3A+1}{2}\gamma^\mu\gamma^\nu,
\end{equation}
where the choice of the off-shell parameter $A=-1$ simplifies both the propagator and the above tensor, as was done in Eq. \eqref{eq:08}.
\begin{table*}[t]
    \caption{Squared coupling constants $C^2_{BB'\phi}$ from two insertions of octet-octet-meson transition.}
    \label{tab:08}
    \begin{ruledtabular}
        \begin{tabular}{llllllll}
        \multicolumn{2}{l}{$N$} & \multicolumn{2}{l}{$\Lambda$} & \multicolumn{2}{l}{$\Sigma$} & \multicolumn{2}{l}{$\Xi$} \\[2pt]
        \hline \\[-8pt]
        $N\pi$ & $\frac{3}{4}(D+F)^2$ & $NK$ & $\frac{1}{6}(D+3F)^2$ & $NK$ & $\frac{1}{2}(D-F)^2$ & $\Lambda K$ & $\frac{1}{12}(D-3F)^2$  \\[5pt]
        $N\eta$ & $\frac{1}{12}(D-3F)^2$ & $\Lambda \eta$ & $\frac{1}{3}D^2$ & $\Lambda\pi$ & $\frac{1}{3}D^2$ & $\Sigma K$ & $\frac{3}{4}(D+F)^2$ \\[5pt]
        $\Lambda K$ & $\frac{1}{12}(D+3F)^2$ & $\Sigma \pi$ & $D^2$ & $\Sigma \pi$ & $2 F^2$ & $\Xi \pi$ & $\frac{3}{4}(D-F)^2$ \\[5pt]
        $\Sigma K$ & $\frac{3}{4}(D-F)^2$ & $\Xi K$ & $\frac{1}{6}(D-3F)^2$ & $\Xi K$ & $\frac{1}{2}(D+F)^2$ & $\Xi \eta$ & $\frac{1}{12}(D+3F)^2$ \\[5pt]
        \end{tabular}
    \end{ruledtabular}
\end{table*}
\begin{table*}[t]
    \caption{Squared coupling constants $C^2_{BT'\phi}$ from two insertions of octet-decuplet-meson transition.}
    \label{tab:09}
    \begin{ruledtabular}
        \begin{tabular}{llllllll}
        \multicolumn{2}{l}{$N$} & \multicolumn{2}{l}{$\Lambda$} & \multicolumn{2}{l}{$\Sigma$} & \multicolumn{2}{l}{$\Xi$} \\[2pt]
        \hline \\[-8pt]
        $\Delta\pi$ & $\mathcal{C}^2$ & $\Sigma^* \pi$ & $\frac{3}{4}\mathcal{C}^2$ & $\Delta K$ & $\frac{2}{3}\mathcal{C}^2$ & $\Sigma^* K$ & $\frac{1}{4}\mathcal{C}^2$  \\[5pt]
        $\Sigma^* K$ & $\frac{1}{4}\mathcal{C}^2$ & $\Xi^* K$ & $\frac{1}{2}\mathcal{C}^2$ & $\Sigma^* \pi$ & $\frac{1}{6}\mathcal{C}^2$ & $\Xi^* \pi$ & $\frac{1}{4}\mathcal{C}^2$  \\[5pt]
        & & & & $\Sigma^* \eta$ & $\frac{1}{4}\mathcal{C}^2$ & $\Xi^* \eta$ & $\frac{1}{4}\mathcal{C}^2$  \\[5pt]
        & & & & $\Xi^* K$ & $\frac{1}{6}\mathcal{C}^2$ & $\Omega K$ & $\frac{1}{2}\mathcal{C}^2$ \\[5pt]
        \end{tabular}
    \end{ruledtabular}
\end{table*}

\section{\label{app:B} Coupling constants of Lagrangian}
Here, we present the coupling constants of the one-loop transitions given in the self-energy expressions Eqs. \eqref{eq:07}-\eqref{eq:09} determined from the effective Lagrangians Eqs.~\eqref{eq:02} and \eqref{eq:04}. These are presented in Tables~\ref{tab:08} and \ref{tab:09}. For the tadpole couplings, we primarily focus on the SU(2) limit, such that we only have 
\begin{equation}
    C_{N\pi,\text{tad}} = -3 \left( -2c_1+\frac{c_2}{4}+c_3 \right)m_\pi^2.
\end{equation}


\section{\label{app:C}Explicit self-energy contributions}
In this section we explicitly show the closed form self-energy contributions in all schemes.
We write the internal octet baryon mass as $M_{B'}=M_B+\delta$ where $\delta$ is the mass difference between the internal and external octet baryons, and we divide the baryon-baryon-meson self-energy into 2 different parts: terms that produce analytic terms in $m_\phi^2$ and terms that produce nonanalytic terms in $m_\phi^2$. Thus, we have, $\Sigma_{B\mathcal{B}'\phi}=\Sigma_{B\mathcal{B}'\phi}^{(1)}+\Sigma_{B\mathcal{B}'\phi}^{(2)}$.

In FRR renormalization, the first part $\Sigma_{B\mathcal{B}'\phi}^{(1)}$, containing only the analytic terms, are not generally of interest since they will be subtracted off, see for example, Eq.~\eqref{eq:tildeSigma}. In fact, if one were to renormalize the chiral expansion of the baryon mass to all orders, everything in $\Sigma_{B\mathcal{B}'\phi}^{(1)}$ will be subtracted. Here, we write down all the parts explicitly in all formalisms. 

\onecolumngrid
\begin{align*}
        \mkern-65mu\Sigma_{BB'\phi}^{(\text{Cov},1)} = &-\frac{C_{BB'\phi}^2 \Lambda ^6 \left(2 M_B+\delta \right)}{96 \pi ^2f_{\phi }^2 M_B ( \Lambda ^2-\delta ^2 )  \left(\Lambda ^2-m_{\phi }^2\right)^3 \left(\left(2 M_B+\delta \right)^2-\Lambda ^2\right)^2} \Bigg\lbrace \Lambda^2 \bigg[ 3 \delta ^3 \left(2 M_B+\delta \right)^4-\delta  \Lambda ^2 \left(2 M_B+\delta \right)^2 \left(7 \delta  M_B+9 M_B^2+6 \delta ^2\right) \\
        &+\Lambda ^4 \left(11 \delta ^2 M_B+13 \delta  M_B^2-2 M_B^3+3 \delta ^3\right)+2 \Lambda ^6 M_B \bigg] \\
        &-m_{\phi }^2\Lambda ^2 \bigg[ \delta  \left(2 M_B+\delta \right)^2 \left(7 \delta  M_B+6 M_B^2+3 \delta ^2\right)+ 2 \Lambda ^2 \left(-4 \delta ^2 M_B+\delta  M_B^2+4 M_B^3-3 \delta ^3\right)+ \Lambda ^4 \left(M_B+3 \delta \right)\bigg] \\
        &+m_{\phi }^4 M_B \bigg[ \delta  \left(2 M_B+\delta \right)^2 \left(3 M_B+2 \delta \right)+ \Lambda ^2 \left(13 \delta  M_B+10 M_B^2+5 \delta ^2\right)-\Lambda^4 \bigg] \Bigg\rbrace \\
        &+ \frac{C_{BB'\phi}^2 \Lambda ^8 \left(2 M_B+\delta \right)^2 \arctan\left( \frac{\sqrt{\left(\delta ^2-\Lambda ^2\right) \left(\Lambda ^2-\left(2 M_B+\delta \right)^2\right)}}{\delta  \left(2 M_B+\delta \right)+\Lambda ^2} \right)}{32 \pi ^2f_{\phi }^2 M_B^3  \left(\Lambda ^2-\delta ^2\right)^{3/2} \left(\Lambda ^2-m_{\phi }^2\right)^4  \left(\left(2 M_B+\delta \right)^2-\Lambda ^2\right)^{5/2}} \Bigg\lbrace \delta ^6 \left(2 M_B+\delta \right)^6-2 \delta ^4 \Lambda ^2 \left(2 M_B+\delta \right)^4 \left(3 \delta  M_B+3 M_B^2+2 \delta ^2\right) \\
        &+6 \delta ^2 \Lambda ^4 \left(2 M_B+\delta \right)^2 \left(3 \delta ^3 M_B+4 \delta ^2 M_B^2+2 \delta  M_B^3+M_B^4+\delta ^4\right)+2 \Lambda ^6 \left(-9 \delta ^5 M_B-15 \delta ^4 M_B^2-10 \delta ^3 M_B^3+6 \delta  M_B^5+2 M_B^6-2 \delta ^6\right) \\
        &+\Lambda ^8 \left(2 \delta ^3 M_B-4 \delta  M_B^3-2 M_B^4+\delta ^4\right)-2 m_\phi^2 \bigg[ \delta ^4 \left(2 M_B+\delta \right)^4 \left(3 \delta  M_B+3 M_B^2+\delta ^2\right) \\
        &-2 \delta ^2 \Lambda ^2 \left(2 M_B+\delta \right)^2 \left(3 \delta  M_B+3 M_B^2+\delta ^2\right) \left(3 \delta  M_B+3 M_B^2+2 \delta ^2\right) \\
        &+6 \Lambda ^4 \left(3 \delta  M_B+3 M_B^2+\delta ^2\right) \left(3 \delta ^3 M_B+4 \delta ^2 M_B^2+2 \delta  M_B^3+M_B^4+\delta ^4\right)-2 \Lambda ^6 \left(\delta  M_B+M_B^2+\delta ^2\right) \left(5 \delta  M_B+5 M_B^2+2 \delta ^2\right) \\
        &+\Lambda ^8 \left(\delta  M_B+M_B^2+\delta ^2\right) \bigg] + m_\phi^4 \bigg[ \delta ^2 \left(2 M_B+\delta \right)^2 \left(6 \delta ^3 M_B+12 \delta ^2 M_B^2+12 \delta  M_B^3+6 M_B^4+\delta ^4\right) \\
        &-2 \Lambda ^2 \left(3 \delta  M_B+3 M_B^2+2 \delta ^2\right) \left(6 \delta ^3 M_B+12 \delta ^2 M_B^2+12 \delta  M_B^3+6 M_B^4+\delta ^4\right)+6 \Lambda ^4 \left(5 \delta ^3 M_B+10 \delta ^2 M_B^2+10 \delta  M_B^3+5 M_B^4+\delta ^4\right) \\
        &-2 \Lambda ^6 \left(5 \delta  M_B+5 M_B^2+2 \delta ^2\right)+\Lambda ^8 \bigg] +4m_\phi^6 M_B^3 \left(M_B+\delta \right)^3 \Bigg\rbrace, \numberthis
\end{align*}
\begin{align*}
        \mkern-55mu\Sigma_{BB'\phi}^{(\text{Cov},2)} &= \frac{C_{BB'\phi}^2 \Lambda ^8 \left(2 M_B+\delta \right)}{64 \pi ^2f_{\phi }^2 M_B  \left(\Lambda ^2-m_{\phi }^2\right)^4} \Bigg\lbrace  4 \delta  m_{\phi }^2 \ln \left(\frac{m_{\phi}}{\Lambda}\right)+\frac{2(m_\phi^2-\delta^2)(\delta+2M_B)}{M_B^2} \bigg[ (\delta\left(2 M_B+\delta \right)-m_{\phi }^2)\ln \left(\frac{m_{\phi}}{\Lambda}\right) \\
        &-\sqrt{(m_{\phi }^2-\delta^2 ) ((2 M_B+\delta)^2 -m_{\phi }^2)} \arctan \left(\frac{\sqrt{(m_{\phi }^2-\delta^2 ) ((2 M_B+\delta)^2 -m_{\phi }^2)}}{\delta  \left(2 M_B+\delta \right)+m_{\phi }^2}\right) \bigg] \Bigg\rbrace. \numberthis
\end{align*}
In a similar fashion, we write $M_{T'}=M_B+\Delta$ for the internal decuplet mass with $\Delta$ the mass difference between the external octet and internal decuplet baryon. Again, we divide the self-energies in two different terms.
\begin{align*}
        \mkern-65mu\Sigma_{BT'\phi}^{(\text{Cov},1)} &= \frac{C_{BT'\phi}^2\Lambda ^6}{384 \pi f_{\phi }^2 M_B \left(M_B+\Delta \right)^2\left(\Lambda ^2-\Delta ^2\right)  \left(\Lambda ^2-m_{\phi }^2\right)^3} \Bigg\lbrace \\
        &\Lambda^2 \bigg[ 2 \Lambda ^6-2 \Delta ^6+4 \Delta  M_B \left(-3 \Delta ^4+\Delta ^2 \Lambda ^2+\Lambda ^4\right)+M_B^2 \left(-24 \Delta ^4+13 \Delta ^2 \Lambda ^2+3 \Lambda ^4\right)-4 M_B^3 \left(4 \Delta ^3-3 \Delta  \Lambda ^2\right) \bigg] \\
        &+\Lambda ^2 m_{\phi }^2 \bigg[ 6 \left(\Delta ^4-\Lambda ^4\right)+4 M_B \left(5 \Delta ^3-3 \Delta  \Lambda ^2\right)+M_B^2 \left(25 \Delta ^2-9 \Lambda ^2\right)+8 \Delta  M_B^3 \bigg] \\
        &-2 m_{\phi }^4 \bigg[ 3 \Delta ^2 \Lambda ^2-3 \Lambda ^4+2 \Delta  \Lambda ^2 M_B+4 \Lambda ^2 M_B^2+2 \Delta  M_B^3 \bigg] \Bigg\rbrace \\
        &+ \frac{C_{BT'\phi}^2\Lambda^8}{192 \pi ^2 f_{\phi }^2M_B^3  \left(M_B+\Delta \right)^2} \Bigg\lbrace \ln \left(\frac{M_B+\Delta }{\Lambda }\right)+ \frac{\arctan \left(\frac{\sqrt{\left(\Delta ^2-\Lambda ^2\right) \left(\Lambda ^2-\left(2 M_B+\Delta \right)^2\right)}}{\Delta  \left(2 M_B+\Delta \right)+\Lambda ^2}\right)}{\left(\Lambda ^2-\Delta ^2\right)^{3/2} \left(\Lambda^2 -m_{\phi}^2\right)^4\sqrt{\left(2 M_B+\Delta \right)^2-\Lambda ^2}} \bigg[ \\
        &-\Delta ^6 \left(2 M_B+\Delta \right)^6+2 \Delta ^4 \Lambda ^2 \left(2 M_B+\Delta \right)^4 \left(3 \Delta  M_B+3 M_B^2+\Delta ^2\right)-4 \Lambda ^6 M_B^3 \left(M_B+\Delta \right)^3 \\
        &+\Lambda ^8 \left(2 \Delta ^3 M_B-4 \Delta  M_B^3-2 M_B^4+\Delta ^4\right)-2 \Lambda ^{10} \left(\Delta  M_B+M_B^2+\Delta ^2\right)+\Lambda ^{12} -2 m_{\phi }^2 \Big( -\Delta ^4 \left(2 M_B+\Delta \right)^4 \left(3 \Delta  M_B+3 M_B^2+2 \Delta ^2\right) \\
        &+2 \Delta ^2 \Lambda ^2 \left(2 M_B+\Delta \right)^2 \left(3 \Delta  M_B+3 M_B^2+\Delta ^2\right) \left(3 \Delta  M_B+3 M_B^2+2 \Delta ^2\right) \\
        &-\Lambda ^4 \left(3 \Delta  M_B+3 M_B^2+2 \Delta ^2\right) \left(6 \Delta ^3 M_B+12 \Delta ^2 M_B^2+12 \Delta  M_B^3+6 M_B^4+\Delta ^4\right)+2 \Lambda ^6 \left(2 \Delta ^3 M_B-4 \Delta  M_B^3-2 M_B^4+\Delta ^4\right) \\
        &-4 \Lambda ^8 \left(\Delta  M_B+M_B^2+\Delta ^2\right)+2 \Lambda ^{10} \Big)+6 m_\phi^4 \Big( -\Delta ^2 \left(2 M_B+\Delta \right)^2 \left(3 \Delta ^3 M_B+4 \Delta ^2 M_B^2+2 \Delta  M_B^3+M_B^4+\Delta ^4\right) \\
        &+2 \Lambda ^2 \left(3 \Delta  M_B+3 M_B^2+\Delta ^2\right) \left(3 \Delta ^3 M_B+4 \Delta ^2 M_B^2+2 \Delta  M_B^3+M_B^4+\Delta ^4\right)-\Lambda ^4 M_B \left(M_B+\Delta \right) \left(7 \Delta  M_B+7 M_B^2+3 \Delta ^2\right) \\
        &-2 \Lambda ^6 \left(\Delta  M_B+M_B^2+\Delta ^2\right)+\Lambda^8 \Big) -2 m_\phi^6 \Big( -9 \Delta ^5 M_B-15 \Delta ^4 M_B^2-10 \Delta ^3 M_B^3+6 \Delta  M_B^5+2 M_B^6-2 \Delta ^6 \\
        &+6 \Lambda ^2 \left(3 \Delta ^3 M_B+4 \Delta ^2 M_B^2+2 \Delta  M_B^3+M_B^4+\Delta ^4\right)-3 \Lambda ^4 \left(3 \Delta  M_B+3 M_B^2+2 \Delta ^2\right)+2 \Lambda ^6 \Big) \bigg] \Bigg\rbrace, \numberthis
\end{align*}
\begin{align*}
        \mkern-55mu\Sigma_{BT'\phi}^{(\text{Cov},2)} &= \frac{C_{BT'\phi}^2 \Lambda^8}{192 \pi ^2 f_{\phi }^2 M_B \left(M_B+\Delta \right)^2 \left(\Lambda^2 -m_{\phi}^2\right)^4} \Bigg\lbrace 2 m_{\phi }^2 \ln \left(\frac{m_{\phi}}{\Lambda}\right) \bigg[ \Delta  \left(2 M_B+\Delta \right)^3+m_{\phi }^2 \left(-6 \Delta  M_B-7 M_B^2-2 \Delta ^2\right)+m_{\phi }^4 \bigg] \\
        &+\frac{1}{M_B^2}\left(m_{\phi}^2-\Delta^2 \right) \left((2 M_B+\Delta)^2 -m_{\phi }^2\right)^2 \bigg[ \ln \left(\frac{m_{\phi}}{\Lambda}\right) \left(\Delta  \left(2 M_B+\Delta \right)-m_{\phi }^2\right) \\
        &-\sqrt{(m_{\phi}^2-\Delta^2 ) ((2 M_B+\Delta)^2 -m_{\phi }^2)}\arctan \left( \frac{\sqrt{(m_{\phi}^2-\Delta^2)((2 M_B+\Delta)^2 -m_{\phi}^2)}}{\Delta \left(2 M_B+\Delta \right)+m_{\phi}^2} \right) \bigg] \Bigg\rbrace. \numberthis
\end{align*}
The explicit nucleon-tadpole self-energy contribution in the covariant formalism is
\begin{equation}
    \Sigma_{N\pi,\text{tad}}^{(\text{Cov})} = -\frac{\chi C\, \Lambda^4 m_\pi^2}{32\pi^2f_\pi^2 \left( \Lambda^2-m_\pi^2 \right)^4} \left[ 2\Lambda^6+3\Lambda^4 m_\pi^2-6\Lambda^2 m_\pi^4+m_\pi^6+12\Lambda^4 m_\pi^2\ln\left( \frac{m_\pi}{\Lambda} \right) \right].
\end{equation}
Here, we present the self-energies in the relativistic scheme
\begin{align*}
        \mkern-55mu\Sigma_{BB'\phi}^{(\text{Rel},1)} &= \frac{C_{BB'\phi}^2 \Lambda ^4 \left(\delta +2 M_B\right)}{192 \pi^2 f_{\phi}^2 \left(\Lambda^2-m^2_{\phi }\right)^3 \left((M_B+\delta)^2 -\Lambda^2\right)^3
         \left( ( m^2_{\phi}-\delta^2 ) (
        	m^2_{\phi }-(2 M_B+\delta)^2 )+4 \Lambda^2 M_B^2 \right)^3} \Bigg\lbrace \\
        &+M_B^{11} \left(M_B+\delta \right) \left(m_{\phi }^2-\Lambda ^2\right){}^3 \left(-10 \Lambda ^2 \left(M_B+\delta \right){}^2+3 \left(M_B+\delta \right){}^4-8 \Lambda ^4\right) \\
        &+ M_B \left(M_B+\delta \right) \left(m_{\phi}^2-\Lambda^2 \right) \left((M_B+\delta)^2 -m_{\phi}^2\right)^5  \bigg[ 14 \Lambda ^6 m_{\phi }^2-8 \Lambda ^4 m_{\phi }^4-2 \left(10 \Lambda ^6-9 \Lambda ^4 m_{\phi }^2+5 \Lambda ^2 m_{\phi }^4\right) \left(\delta
        +M_B\right){}^2 \\
        &+\left(5 \Lambda ^4-2 \Lambda ^2 m_{\phi }^2+3 m_{\phi }^4\right) \left(\delta +M_B\right){}^4 \bigg] + M_B^9 \left(M_B+\delta \right) \left(m_{\phi}^2-\Lambda^2 \right) \bigg[ 2 \Lambda ^4 \left(-56 \Lambda ^6+128 \Lambda ^4 m_{\phi }^2-95 \Lambda ^2 m_{\phi }^4+20 m_{\phi }^6\right) \\
        &+ 2 \left(-12 \Lambda ^8+61 \Lambda ^6 m_{\phi }^2-65 \Lambda ^4 m_{\phi }^4+25 \Lambda ^2 m_{\phi }^6\right) \left(\delta
        +M_B\right){}^2+\left(68 \Lambda ^6-175 \Lambda ^4 m_{\phi }^2+104 \Lambda ^2 m_{\phi }^4-15 m_{\phi }^6\right) \left(\delta
        +M_B\right){}^4 \\
        &+\left(-7 \Lambda ^4+22 \Lambda ^2 m_{\phi }^2-9 m_{\phi }^4\right) \left(\delta +M_B\right)^6 \bigg] + 2 M_B^7 \left(M_B+\delta \right) \left(m_{\phi}^2-\Lambda^2 \right) \bigg[ \\
        &-4 \Lambda ^4 \left(88 \Lambda ^8-204 \Lambda ^6 m_{\phi }^2+170 \Lambda ^4 m_{\phi }^4-61 \Lambda ^2 m_{\phi }^6+10 m_{\phi }^8\right) +2 \Lambda ^2 \left(232 \Lambda ^8-346 \Lambda ^6 m_{\phi }^2+81 \Lambda ^4 m_{\phi }^4+82 \Lambda ^2 m_{\phi }^6-25 m_{\phi }^8\right)
        \left(\delta +M_B\right){}^2 \\
        &+3 \left(-68 \Lambda ^8+40 \Lambda ^6 m_{\phi }^2+37 \Lambda ^4 m_{\phi }^4-38 \Lambda ^2 m_{\phi }^6+5 m_{\phi }^8\right) \left(\delta
        +M_B\right){}^4+2 \left(9 \Lambda ^6+35 \Lambda ^4 m_{\phi }^2-23 \Lambda ^2 m_{\phi }^4+3 m_{\phi }^6\right) \left(\delta +M_B\right){}^6 \\
        &-\left(\Lambda ^4+14 \Lambda ^2 m_{\phi }^2-3 m_{\phi }^4\right) \left(\delta +M_B\right){}^8 \bigg] + \Lambda ^2 \left(m_{\phi }^2-\left(M_B+\delta \right){}^2\right){}^6 \bigg[ 8 \Lambda ^8+m_{\phi }^4 \left(\delta +M_B\right){}^4+10 \Lambda ^6 \left(m_{\phi }^2+\left(\delta +M_B\right){}^2\right) \\
        &+17 \Lambda ^2 m_{\phi }^2 \left(\delta +M_B\right){}^2 \left(m_{\phi }^2+\left(\delta +M_B\right){}^2\right)-3 \Lambda ^4 \left(m_{\phi }^4+19
        m_{\phi }^2 \left(\delta +M_B\right){}^2+\left(\delta +M_B\right){}^4\right) \bigg] + M_B^{10} \bigg[ -16 \Lambda ^{12}-6 m_{\phi }^6 \left(\delta +M_B\right){}^6 \\
        &+48 \Lambda ^{10} \left(m_{\phi }^2+\left(\delta +M_B\right){}^2\right)+13 \Lambda ^2 m_{\phi }^4 \left(\delta +M_B\right){}^4 \left(m_{\phi
        }^2+\left(\delta +M_B\right){}^2\right)-4 \Lambda ^8 \left(7 m_{\phi }^4+36 m_{\phi }^2 \left(\delta +M_B\right){}^2+7 \left(\delta
        +M_B\right){}^4\right) \\
        &-3 \Lambda ^4 m_{\phi }^2 \left(\delta +M_B\right){}^2 \left(11 m_{\phi }^4+8 m_{\phi }^2 \left(\delta +M_B\right){}^2+11 \left(\delta
        +M_B\right){}^4\right) \\
        &+\Lambda ^6 \left(m_{\phi }^2+\left(\delta +M_B\right){}^2\right) \left(11 m_{\phi }^4+73 m_{\phi }^2 \left(\delta
        +M_B\right){}^2+11 \left(\delta +M_B\right){}^4\right) \bigg]-M_B^2 \left((M_B+\delta)^2 -m^2_{\phi }\right){}^4 \bigg[ \\
        &-96 \Lambda ^{12}+6 m_{\phi }^6 \left(\delta +M_B\right){}^6-72 \Lambda ^{10} \left(m_{\phi }^2+\left(\delta +M_B\right){}^2\right)-9 \Lambda ^2
        m_{\phi }^4 \left(\delta +M_B\right){}^4 \left(m_{\phi }^2+\left(\delta +M_B\right){}^2\right) \\
        &-\Lambda ^6 \left(m_{\phi }^2+\left(\delta +M_B\right){}^2\right) \left(23 m_{\phi }^4+557 m_{\phi }^2 \left(\delta +M_B\right){}^2+23
        \left(\delta +M_B\right){}^4\right)+4 \Lambda ^8 \left(29 m_{\phi }^4+194 m_{\phi }^2 \left(\delta +M_B\right){}^2+29 \left(\delta
        +M_B\right){}^4\right) \\
        &+\Lambda ^4 m_{\phi }^2 \left(\delta +M_B\right){}^2 \left(101 m_{\phi }^4+248 m_{\phi }^2 \left(\delta +M_B\right){}^2+101 \left(\delta
        +M_B\right){}^4\right) \bigg]-M_B^3 \left(M_B+\delta \right) \left(m^2_{\phi }-\Lambda^2 \right) \left((M_B+\delta)^2 -m^2_{\phi }\right){}^3 \bigg[ \\ 
        &15 m_{\phi }^6 \left(\delta +M_B\right){}^4+9 m_{\phi }^4 \left(\delta +M_B\right){}^6+24 \Lambda ^8 \left(-7 m_{\phi }^2+10 \left(\delta
        +M_B\right){}^2\right) \\
        &-2 \Lambda ^2 m_{\phi }^2 \left(\delta +M_B\right){}^2 \left(25 m_{\phi }^4+46 m_{\phi }^2 \left(\delta
        +M_B\right){}^2+\left(\delta +M_B\right){}^4\right)-2 \Lambda ^6 \left(-92 m_{\phi }^4+121 m_{\phi }^2 \left(\delta +M_B\right){}^2+91 \left(\delta +M_B\right){}^4\right) \\ 
        &+\Lambda ^4 \left(-40 m_{\phi }^6+124 m_{\phi }^4 \left(\delta +M_B\right){}^2+187 m_{\phi }^2 \left(\delta +M_B\right){}^4+17 \left(\delta
        +M_B\right){}^6\right) \bigg] \\
        &+2 M_B^5 \left(M_B+\delta \right) \left(m_{\phi}^2-\Lambda^2 \right) \left((M_B+\delta)^2 -m^2_{\phi }\right) \bigg[ -192 \Lambda ^{12}+24 \Lambda ^{10} \left(27 m_{\phi }^2+\left(\delta +M_B\right){}^2\right) \\
        &+2 \Lambda ^2 m_{\phi }^2 \left(\delta +M_B\right){}^2 \left(5 m_{\phi }^2+3 \left(\delta +M_B\right){}^2\right) \left(-5 m_{\phi }^4-8 m_{\phi
        }^2 \left(\delta +M_B\right)^2+\left(\delta +M_B\right)^4\right) \\
        &+3 m_{\phi }^4 \left(\delta +M_B\right)^4 \left(5 m_{\phi }^4+2 m_{\phi
        }^2 \left(\delta +M_B\right)^2+\left(\delta +M_B\right){}^4\right)+4 \Lambda ^8 \left(-160 m_{\phi }^4-79 m_{\phi }^2 \left(\delta +M_B\right){}^2+53 \left(\delta +M_B\right){}^4\right) \\
        &+2 \Lambda ^6 \left(121 m_{\phi }^6+54 m_{\phi }^4 \left(\delta +M_B\right){}^2-63 m_{\phi }^2 \left(\delta +M_B\right){}^4-64 \left(\delta
        +M_B\right){}^6\right) \\
        &+\Lambda ^4 \left(-40 m_{\phi }^8+162 m_{\phi }^6 \left(\delta +M_B\right){}^2+117 m_{\phi }^4 \left(\delta +M_B\right){}^4+88 m_{\phi }^2
        \left(\delta +M_B\right){}^6+9 \left(\delta +M_B\right){}^8\right) \bigg] \\
        &+2 M_B^4 \left((M_B+\delta)^2 -m^2_{\phi }\right){}^2 \bigg[ 192 \Lambda ^{14}+48 \Lambda ^{12} \left(m_{\phi }^2+\left(\delta +M_B\right){}^2\right)+12 m_{\phi }^6 \left(\delta +M_B\right){}^6
        \left(m_{\phi }^2+\left(\delta +M_B\right){}^2\right) \\
        &-\Lambda ^2 m_{\phi }^4 \left(\delta +M_B\right){}^4 \left(23 m_{\phi }^4+98 m_{\phi }^2 \left(\delta +M_B\right){}^2+23 \left(\delta
        +M_B\right){}^4\right)-12 \Lambda ^{10} \left(31 m_{\phi }^4+134 m_{\phi }^2 \left(\delta +M_B\right){}^2+31 \left(\delta
        +M_B\right){}^4\right) \\
        &+3 \Lambda ^4 m_{\phi }^2 \left(\delta +M_B\right){}^2 \left(m_{\phi }^2+\left(\delta +M_B\right){}^2\right) \left(39 m_{\phi }^4+122 m_{\phi
        }^2 \left(\delta +M_B\right){}^2+39 \left(\delta +M_B\right){}^4\right) \\
        &+2 \Lambda ^8 \left(m_{\phi }^2+\left(\delta +M_B\right){}^2\right) \left(119 m_{\phi }^4+920 m_{\phi }^2 \left(\delta +M_B\right){}^2+119
        \left(\delta +M_B\right){}^4\right) \\
        &-\Lambda ^6 \left(31 m_{\phi }^8+935 m_{\phi }^6 \left(\delta +M_B\right){}^2+1716 m_{\phi }^4 \left(\delta +M_B\right){}^4+935 m_{\phi }^2
        \left(\delta +M_B\right){}^6+31 \left(\delta +M_B\right){}^8\right) \bigg] 
\end{align*}

\begin{align*}
		&+M_B^8 \bigg[ -320 \Lambda ^{14}+1024 \Lambda ^{12} \left(m_{\phi }^2+\left(\delta +M_B\right){}^2\right)+24 m_{\phi }^6 \left(\delta +M_B\right){}^6
		\left(m_{\phi }^2+\left(\delta +M_B\right){}^2\right) \\
		&-\Lambda ^2 m_{\phi }^4 \left(\delta +M_B\right){}^4 \left(51 m_{\phi }^4+202 m_{\phi }^2 \left(\delta +M_B\right){}^2+51 \left(\delta
		+M_B\right){}^4\right) \\
		&+6 \Lambda ^8 \left(m_{\phi }^2+\left(\delta +M_B\right){}^2\right) \left(55 m_{\phi }^4+422 m_{\phi }^2 \left(\delta
		+M_B\right){}^2+55 \left(\delta +M_B\right){}^4\right)-16 \Lambda ^{10} \left(57 m_{\phi }^4+205 m_{\phi }^2 \left(\delta +M_B\right){}^2+57 \left(\delta +M_B\right){}^4\right)        
        &+\Lambda ^4 m_{\phi }^2 \left(\delta +M_B\right){}^2 \left(m_{\phi }^2+\left(\delta +M_B\right){}^2\right) \left(149 m_{\phi }^4+454 m_{\phi }^2
        \left(\delta +M_B\right){}^2+149 \left(\delta +M_B\right){}^4\right) \\
        &-\Lambda ^6 \left(47 m_{\phi }^8+1055 m_{\phi }^6 \left(\delta +M_B\right){}^2+2052 m_{\phi }^4 \left(\delta +M_B\right){}^4+1055 m_{\phi }^2
        \left(\delta +M_B\right){}^6+47 \left(\delta +M_B\right){}^8\right) \bigg]  \\
        &+ 2 M_B^6 \bigg[ 256 \Lambda ^{16}-256 \Lambda ^{14} \left(m_{\phi }^2+\left(\delta +M_B\right){}^2\right)-6 m_{\phi }^6 \left(\delta +M_B\right){}^6 \left(3
        m_{\phi }^4+2 m_{\phi }^2 \left(\delta +M_B\right){}^2+3 \left(\delta +M_B\right){}^4\right) \\ 
        &-24 \Lambda ^{12} \left(11 m_{\phi }^4+38 m_{\phi }^2 \left(\delta +M_B\right){}^2+11 \left(\delta +M_B\right){}^4\right) \\
        &+\Lambda ^2 m_{\phi }^4
        \left(\delta +M_B\right){}^4 \left(m_{\phi }^2+\left(\delta +M_B\right){}^2\right) \left(37 m_{\phi }^4+126 m_{\phi }^2 \left(\delta
        +M_B\right){}^2+37 \left(\delta +M_B\right){}^4\right) \\
        &+4 \Lambda ^{10} \left(m_{\phi }^2+\left(\delta +M_B\right){}^2\right) \left(159 m_{\phi }^4+428 m_{\phi }^2 \left(\delta +M_B\right){}^2+159
        \left(\delta +M_B\right){}^4\right) \\
        &+3 \Lambda ^6 \left(m_{\phi }^2+\left(\delta +M_B\right){}^2\right) \left(13 m_{\phi }^8+365 m_{\phi }^6 \left(\delta +M_B\right){}^2+188
        m_{\phi }^4 \left(\delta +M_B\right){}^4+365 m_{\phi }^2 \left(\delta +M_B\right){}^6+13 \left(\delta +M_B\right){}^8\right) \\
        &-4 \Lambda ^8 \left(84 m_{\phi }^8+639 m_{\phi }^6 \left(\delta +M_B\right){}^2+614 m_{\phi }^4 \left(\delta +M_B\right){}^4+639 m_{\phi }^2
        \left(\delta +M_B\right){}^6+84 \left(\delta +M_B\right){}^8\right) \\
        &-\Lambda ^4 m_{\phi }^2 \left(\delta +M_B\right){}^2 \left(133 m_{\phi }^8+574 m_{\phi }^6 \left(\delta +M_B\right){}^2+634 m_{\phi }^4
        \left(\delta +M_B\right){}^4+574 m_{\phi }^2 \left(\delta +M_B\right){}^6+133 \left(\delta +M_B\right){}^8\right) \bigg] \Bigg\rbrace \\
        &-\frac{C_{BB'\phi}^2 \Lambda ^3 \left(2 M_B+\delta \right)^2 \arctan\left(\frac{\sqrt{\left(M_B+\delta \right)^2-\Lambda ^2}}{\Lambda}\right)}{64 \pi ^2 f_{\phi }^2 \left( (M_B+\delta)^2 -\Lambda^2 ) \right)^{7/2} \left(  (m_{\phi}^2-\delta^2)  (m_{\phi}^2-(2 M_B+\delta)^2) +4 \Lambda ^2 M_B^2\right)^4} \Bigg\lbrace \\
        &\left(M_B+\delta \right)^7 \left(\delta ^2-m_{\phi }^2\right)^4 \left((2 M_B+\delta)^2 -m_{\phi }^2\right)^3  \\
        &+2 \Lambda ^2 \left(M_B+\delta \right)^5 \left(\delta ^2-m_{\phi }^2\right)^3 \left((2 M_B+\delta)^2 -m_{\phi }^2\right)^2 \bigg[ -3 M_B^2 \left(\delta ^2-3 m_{\phi }^2\right)-9 \delta  M_B \left(\delta^2-m_{\phi }^2 \right)+12 \delta  M_B^3+6 M_B^4-2 \left(\delta ^2-m_{\phi }^2\right)^2 \bigg] \\
        &-4 \Lambda ^4 \left(M_B+\delta \right)^3 \left(\delta ^2-m_{\phi }^2\right)^2 \left((2 M_B+\delta)^2 -m_{\phi }^2\right) \bigg[ M_B^2 \left(13 \delta ^2-11 m_{\phi }^2\right)+11 \delta  M_B \left(\delta ^2-m_{\phi }^2\right)+4 \delta  M_B^3+2 M_B^4+2 \left(\delta ^2-m_{\phi }^2\right)^2 \bigg] \\
        &\times\bigg[  2 M_B^2 \left(\delta ^2+2 m_{\phi }^2\right)-4 \delta  M_B \left(\delta^2 -m_{\phi }^2\right)+12 \delta  M_B^3+6 M_B^4-\left(\delta ^2-m_{\phi }^2\right)^2 \bigg] \\
        &+32 \Lambda ^6 M_B^2 \left(M_B+\delta \right)^3 \left(\delta ^2-m_{\phi }^2\right) \bigg[ 4 \left(\delta ^2-m_{\phi }^2\right)^4 +38 \delta  M_B \left(\delta ^2-m_{\phi }^2\right)^3+2 M_B^2 \left(63 \delta ^2-19 m_{\phi }^2\right) \left(\delta ^2-m_{\phi }^2\right)^2 \\
        &+\delta  M_B^3 \left(\delta^2 -m_{\phi }^2\right) \left(185 \delta ^2-176 m_{\phi }^2\right)+M_B^4\left(141 \delta ^4-203 \delta ^2 m_{\phi }^2+88 m_{\phi }^4\right)+M_B^5 \left(131 \delta ^3-27 \delta  m_{\phi }^2\right)+3 M_B^6 \left(55 \delta ^2-3 m_{\phi }^2\right) \\
        &+104 \delta  M_B^7+26 M_B^8 \bigg]+64 \Lambda ^8 M_B^4 \left(M_B+\delta \right)^3 \bigg[ 12 \left(\delta ^2-m_{\phi }^2\right)^3+31 \delta  M_B \left(\delta ^2-m_{\phi }^2\right)^2-M_B^2 \left(\delta^2 -m_{\phi }^2\right) \left(19 \delta ^2+31 m_{\phi }^2\right) \\
        &+2 M_B^3 \left(50 \delta  m_{\phi }^2-49 \delta ^3\right)+M_B^4 \left(50 m_{\phi }^2-44 \delta ^2\right)+6 \delta  M_B^5+2 M_B^6 \bigg] \\
        &-256 \Lambda ^{10} M_B^5 \left(M_B+\delta \right)^2 \bigg[ M_B^2 \left(22 m_{\phi }^2-20 \delta ^2\right)-22 \delta  M_B \left(\delta^2-m_{\phi }^2 \right)+4 \delta  M_B^3+2 M_B^4-7 \left(\delta ^2-m_{\phi }^2\right)^2 \bigg] \\
        &+512 \Lambda ^{12} M_B^5 \bigg[ 2 m_{\phi }^2 \left(\delta  M_B+M_B^2+\delta ^2\right)-2 \delta ^3 M_B+4 \delta  M_B^3+2 M_B^4-\delta ^4-m_{\phi }^4 \bigg] \Bigg\rbrace, \numberthis
\end{align*}
\begin{align*}
        \mkern-65mu\Sigma_{BB'\phi}^{(\text{Rel},2)} &= \frac{C_{BB'\phi}^2 \Lambda ^3 \left(2 M_B+\delta \right)}{64 \pi^2f_{\phi}^2 \left( (m_{\phi}^2-\delta^2 )(m_{\phi}^2-(2 M_B+\delta)^2 )+4 \Lambda ^2 M_B^2\right)^4} \Bigg( -512 \Lambda ^5  M_B^5 \arctan\left(\frac{\sqrt{(m_{\phi }^2-\delta^2) ( (2 M_B+\delta)^2 -m_{\phi }^2)}}{\delta  \left(2 M_B+\delta \right)+m_{\phi }^2}\right) \\
        &\times \left(2 M_B+\delta \right) \left(m_{\phi }^2-\delta^2 \right)^{3/2} \sqrt{ (2 M_B+\delta)^2 -m_{\phi}^2}+ \frac{1}{\left(m_{\phi }^2-\Lambda ^2\right)^{7/2}}\arctan \left(\frac{\sqrt{m_{\phi }^2-\Lambda ^2}}{\Lambda}\right) \bigg\lbrace \\
        &-m_{\phi }^8 \left(m_{\phi}^2-\delta^2 \right)^4 \left( (2 M_B+\delta)^2 -m_{\phi}^2 \right)^3 \\
        &+2 \Lambda ^2 m_{\phi }^6 \left(m_{\phi}^2-\delta^2 \right)^3 \left( (2 M_B+\delta)^2 -m_{\phi }^2\right)^2  \bigg[ 6 M_B^2 \left(2 \delta ^2-3 m_{\phi }^2\right)+9 \delta  M_B \left(\delta ^2-m_{\phi }^2\right)+4 \delta  M_B^3+2 \left(m_{\phi }^2-\delta ^2\right)^2 \bigg] \\
        &+4 \Lambda ^4 m_{\phi }^4 \left(m_{\phi }^2-\delta^2 \right)^2 \left( m_{\phi }^2-(2 M_B+\delta)^2 \right) \bigg[ M_B^2 \left(4 \delta ^2-10 m_{\phi }^2\right)+4 \delta  M_B \left(\delta ^2-m_{\phi }^2\right)+\left(m_{\phi }^2-\delta ^2\right)^2 \bigg] \\
        &\times \bigg[ M_B^2 \left(20 \delta ^2-18 m_{\phi }^2\right)+11 \delta  M_B \left(\delta ^2-m_{\phi }^2\right)+12 \delta  M_B^3+2 \left(m_{\phi }^2-\delta ^2\right)^2 \bigg] \\
        &-32 \Lambda ^6 M_B^2 m_{\phi }^4 \left(m_{\phi}^2-\delta^2 \right) \bigg[ 4 \left(m_{\phi }^2-\delta ^2\right)^4+38 \delta  M_B \left(\delta ^2-m_{\phi }^2\right)^3-8 M_B^2 \left(m_{\phi }^2-\delta ^2\right)^2 \left(7 m_{\phi }^2-18 \delta ^2\right) \\
        &+\delta  M_B^3 \left(m_{\phi }^2-\delta^2 \right)  \left(263 m_{\phi }^2-272 \delta ^2\right)+2 M_B^4 \left(128 \delta ^4-190 \delta ^2 m_{\phi }^2+75 m_{\phi }^4\right)+\left.M_B^5 \left(96 \delta ^3-60 \delta  m_{\phi }^2\right)\right) \bigg] \\
        &+64 \Lambda ^8 M_B^4 m_{\phi }^4 \bigg[ M_B^3 \left(70 \delta  m_{\phi }^2-68 \delta ^3\right)+31 \delta  M_B \left(m_{\phi }^2-\delta ^2\right)^2+10 M_B^2 \left(m_{\phi}^2-\delta^2 \right) \left(2 \delta ^2+3 m_{\phi }^2\right)-12 \left(m_{\phi }^2-\delta ^2\right)^3 \bigg] \\
        &-256 \Lambda ^{10} M_B^5 m_{\phi }^2 \bigg[ M_B^2 \left(30 \delta  m_{\phi }^2-28 \delta ^3\right)+M_B \left(-28 \delta ^4+34 \delta ^2 m_{\phi }^2-6 m_{\phi}^4 \right)-7 \delta  \left(m_{\phi }^2-\delta ^2\right)^2 \bigg]  \\
        &+512 \Lambda ^{12} M_B^5 \bigg[ M_B^2 \left(6 \delta  m_{\phi }^2-4 \delta ^3\right)-2 M_B \left(2 \delta ^4-3 \delta ^2 m_{\phi }^2+m_{\phi }^4\right)-\delta  \left(m_{\phi }^2-\delta ^2\right)^2 \bigg] \bigg\rbrace \Bigg) , \numberthis
\end{align*}
\begin{align*}
	\mkern-55mu\Sigma_{BT'\phi}^{(\text{Rel},1)}  &= \frac{C_{BT'\phi}^2 \Lambda ^6 M_B^2}{288 \pi ^2 f_{\phi }^2 (m_{\phi }^2-\Lambda^2 )^3\left( M_B+\Delta \right)^2\left( (M_B+\Delta)^2 -\Lambda^2 ) \right)^{3} \left(  (m_{\phi}^2-\Delta^2)  (m_{\phi}^2-(2 M_B+\Delta)^2) +4 \Lambda ^2 M_B^2\right)^3} \Bigg\lbrace \\
	&128 \Lambda ^{14} M_B^6 \left(2 \Delta +3 M_B\right)+ \Lambda ^2 m_{\phi }^2 \left(M_B+\Delta \right){}^3 \left(\Delta ^2-m_{\phi }^2\right) \left((2 M_B+\Delta)^2 -m^2_{\phi }\right){}^2 \bigg[ \\
	&-16 \left(m_{\phi }^2-\Delta ^2\right){}^3 \left(\Delta ^2+m_{\phi }^2\right)+7 \Delta  M_B \left(m_{\phi }^2-\Delta ^2\right){}^2 \left(16 \Delta ^2+m_{\phi }^2\right)M_B^2 +\left(m_{\phi }^2-\Delta ^2\right){}^2 \left(304 \Delta ^2+7 m_{\phi }^2\right)  \\
	&+\left(400 \Delta ^5-830 \Delta ^3 m_{\phi }^2+334 \Delta  m_{\phi }^4\right) M_B^3 +2 \left(128 \Delta ^4-287 \Delta ^2 m_{\phi }^2+63 m_{\phi }^4\right) M_B^4+32 \Delta  \left(2 \Delta ^2-5 m_{\phi }^2\right) M_B^5 \bigg] \\
	&+16 \Lambda ^{12} M_B^3 \bigg[ -6 \left(m_{\phi }^2-\Delta ^2\right){}^3+60 \Delta  M_B \left(m_{\phi }^2-\Delta ^2\right){}^2+69 M_B^2 \left(3 \Delta ^4-4 \Delta ^2 m_{\phi }^2+m_{\phi }^4\right)+6 \Delta  M_B^3 \left(59 \Delta ^2-43 m_{\phi }^2\right) \\
	&+M_B^4 \left(322 \Delta ^2-70 m_{\phi }^2\right)+136 \Delta  M_B^5+10 M_B^6 \bigg] -4 \Lambda ^{10} M_B^2 \bigg[ -12 \Delta  \left(m_{\phi }^2-\Delta ^2\right){}^4-3 M_B \left(m_{\phi }^2-\Delta ^2\right){}^3 \left(27 m_{\phi }^2-17 \Delta ^2\right) \\
	&+6 \Delta  M_B^2 \left(m_{\phi }^2-\Delta ^2\right){}^2 \left(3 \Delta ^2+73 m_{\phi }^2\right)+2 M_B^3 \left(229 \Delta ^6-75 \Delta ^4 m_{\phi }^2-291 \Delta ^2 m_{\phi }^4+137 m_{\phi }^6\right) \\
	&+8 M_B^4 \left(151 \Delta ^5-101 \Delta ^3 m_{\phi }^2+46 \Delta  m_{\phi }^4\right)+8 M_B^5 \left(243 \Delta ^4+50 \Delta ^2 m_{\phi }^2+64 m_{\phi }^4\right)+4 M_B^6 \left(559 \Delta ^3+385 \Delta  m_{\phi }^2\right) \\
	&+4 M_B^7 \left(391 \Delta ^2+155 m_{\phi }^2\right)+480 \Delta  M_B^8+16 M_B^9 \bigg] -\Lambda ^4 \left(M_B+\Delta \right) \left((2 M_B+\Delta)^2 -m^2_{\phi }\right) \bigg[ \\
	&-\left(m_{\phi }^2-\Delta ^2\right){}^5 \left(4 \Delta ^4+61 \Delta ^2 m_{\phi }^2+4 m_{\phi }^4\right)+\Delta  M_B \left(m_{\phi }^2-\Delta ^2\right){}^4 \left(52 \Delta ^4+616 \Delta ^2 m_{\phi }^2-125 m_{\phi }^4\right) \\
	&-2 M_B^2 \left(m_{\phi }^2-\Delta ^2\right){}^3 \left(148 \Delta ^6+1133 \Delta ^4 m_{\phi }^2-991 \Delta ^2 m_{\phi }^4+34 m_{\phi }^6\right) \\
	&+4 \Delta  M_B^3 \left(m_{\phi }^2-\Delta ^2\right){}^2 \left(242 \Delta ^6+705 \Delta ^4 m_{\phi }^2-2040 \Delta ^2 m_{\phi }^4+547 m_{\phi }^6\right) \\
	&+M_B^4 +\left(m^2_{\phi }-\Delta^2 \right)  \left(-2004 \Delta ^8+4283 \Delta ^6 m_{\phi }^2+9439 \Delta ^4 m_{\phi }^4-10391 \Delta ^2 m_{\phi }^6+833 m_{\phi }^8\right) \\
	&+4 \Delta  M_B^5 \left(681 \Delta ^8-4943 \Delta ^6 m_{\phi }^2+4147 \Delta ^4 m_{\phi }^4+1651 \Delta ^2 m_{\phi }^6-1524 m_{\phi }^8\right) \\ 
	&-4 M_B^6 \left(-608 \Delta ^8+6829 \Delta ^6 m_{\phi }^2-7877 \Delta ^4 m_{\phi }^4+1211 \Delta ^2 m_{\phi }^6+397 m_{\phi }^8\right)+8 \Delta  M_B^7 \left(172 \Delta ^6-2557 \Delta ^4 m_{\phi }^2+3074 \Delta ^2 m_{\phi }^4-653 m_{\phi }^6\right) 
\end{align*}
\begin{align*}
	&-16 M_B^8 \left(-28 \Delta ^6+504 \Delta ^4 m_{\phi }^2-573 \Delta ^2 m_{\phi }^4+85 m_{\phi }^6\right)+16 M_B^9 \left(4 \Delta ^5-82 \Delta ^3 m_{\phi }^2+81 \Delta  m_{\phi }^4\right) \bigg] \\ 
	&+2 \Lambda ^8 \bigg[ 2 \Delta  \left(m_{\phi }^2-\Delta ^2\right){}^6+3 M_B \left(m_{\phi }^2-9 \Delta ^2\right) \left(m_{\phi }^2-\Delta ^2\right){}^5+6 \Delta  M_B^2 \left(m_{\phi }^2-\Delta ^2\right){}^4 \left(41 \Delta ^2+5 m_{\phi }^2\right) \\
	&+2 M_B^3 \left(m_{\phi }^2-\Delta ^2\right){}^3 \left(-801 \Delta ^4-257 \Delta ^2 m_{\phi }^2+16 m_{\phi }^4\right) + 12 \Delta  M_B^4 \left(m_{\phi }^2-\Delta ^2\right){}^2 \left(559 \Delta ^4+2 \Delta ^2 m_{\phi }^2-115 m_{\phi }^4\right) \\
	& -3 M_B^5 \left(m^2_{\phi }-\Delta^2 \right) \left(5933 \Delta ^6-4579 \Delta ^4 m_{\phi }^2-1145 \Delta ^2 m_{\phi }^4+263 m_{\phi }^6\right) +8 M_B^6 \left(3792 \Delta ^7-6858 \Delta ^5 m_{\phi }^2+2691 \Delta ^3 m_{\phi }^4+395 \Delta  m_{\phi }^6\right) \\
	&+24 M_B^7 \left(1395 \Delta ^6-2626 \Delta ^4 m_{\phi }^2+1227 \Delta ^2 m_{\phi }^4+52 m_{\phi }^6\right)+24 M_B^8 \left(1000 \Delta ^5-1719 \Delta ^3 m_{\phi }^2+867 \Delta  m_{\phi }^4\right) \\
	&+8 M_B^9 \left(1440 \Delta ^4-1447 \Delta ^2 m_{\phi }^2+707 m_{\phi }^4\right)+384 \Delta  M_B^{10} \left(11 \Delta ^2+m_{\phi }^2\right)+96 M_B^{11} \left(14 \Delta ^2+5 m_{\phi }^2\right)+256 \Delta  M_B^{12} \bigg] \\
	&+\Lambda ^6 \bigg[ 15 \Delta  \left(m_{\phi }^2-\Delta ^2\right){}^6 \left(\Delta ^2+m_{\phi }^2\right)+M_B \left(m_{\phi }^2-\Delta ^2\right){}^5 \left(-227 \Delta ^4-158 \Delta ^2 m_{\phi }^2+28 m_{\phi }^4\right) \\
	&+2 \Delta  M_B^2 \left(m_{\phi }^2-\Delta ^2\right){}^4 \left(727 \Delta ^4-179 \Delta ^2 m_{\phi }^2-260 m_{\phi }^4\right)-2 M_B^3 \left(m_{\phi }^2-\Delta ^2\right){}^3 \left(2485 \Delta ^6-5262 \Delta ^4 m_{\phi }^2-336 \Delta ^2 m_{\phi }^4+173 m_{\phi }^6\right) \\
	&+\Delta  M_B^4 \left(m_{\phi }^2-\Delta ^2\right){}^2 \left(8747 \Delta ^6-55897 \Delta ^4 m_{\phi }^2+25673 \Delta ^2 m_{\phi }^4+1317 m_{\phi }^6\right) \\
	&+M_B^5 \left(m^2_{\phi }-\Delta^2 \right) \left(-2159 \Delta ^8+142822 \Delta ^6 m_{\phi }^2-163232 \Delta ^4 m_{\phi }^4+32010 \Delta ^2 m_{\phi }^6+687 m_{\phi }^8\right) \\
	&+8 M_B^6 \left(-3335 \Delta ^9-22665 \Delta ^7 m_{\phi }^2+52053 \Delta ^5 m_{\phi }^4-28459 \Delta ^3 m_{\phi }^6+2550 \Delta  m_{\phi }^8\right) \\
	&+4 M_B^7 \left(-17087 \Delta ^8-28744 \Delta ^6 m_{\phi }^2+88446 \Delta ^4 m_{\phi }^4-42408 \Delta ^2 m_{\phi }^6+1257 m_{\phi }^8\right) \\
	&-32 M_B^8 \left(2809 \Delta ^7-423 \Delta ^5 m_{\phi }^2-5124 \Delta ^3 m_{\phi }^4+2346 \Delta  m_{\phi }^6\right)-16 M_B^9 \left(4535 \Delta ^6-5148 \Delta ^4 m_{\phi }^2-1242 \Delta ^2 m_{\phi }^4+939 m_{\phi }^6\right) \\
	&-32 M_B^{10} \left(1132 \Delta ^5-1867 \Delta ^3 m_{\phi }^2+423 \Delta  m_{\phi }^4\right)-32 M_B^{11} \left(322 \Delta ^4-570 \Delta ^2 m_{\phi }^2+129 m_{\phi }^4\right)+640 M_B^{12} \left(3 \Delta  m_{\phi }^2-2 \Delta ^3\right) \bigg] \\
	&+3 m_{\phi }^4 \left(M_B+\Delta \right){}^5 \left(m_{\phi }^2-\Delta ^2\right){}^2 \left((2 M_B+\Delta)^2 -m^2_{\phi }\right){}^3  \left(\Delta  \left(M_B+\Delta \right)-m_{\phi }^2\right) \Bigg\rbrace \\
    & -\frac{C_{BT'\phi}^2 \Lambda^3  \arctan\left(\frac{\sqrt{\left(M_B+\Delta \right)^2-\Lambda ^2}}{\Lambda}\right)}{96 \pi ^2 f_{\phi }^2 \left( M_B+\Delta \right)^2\left( (M_B+\Delta)^2 -\Lambda^2 ) \right)^{7/2} \left(  (m_{\phi}^2-\Delta^2)  (m_{\phi}^2-(2 M_B+\Delta)^2) +4 \Lambda ^2 M_B^2\right)^4} \Bigg\lbrace \\
    &\left(M_B+\Delta \right)^7 \left(m_{\phi}^2-\Delta^2 \right)^3 \left((2 M_B+\Delta)^2 -m_{\phi}^2\right)^4 \\
    &-2 \Lambda ^2 \left(M_B+\Delta \right)^5 \left(m_{\phi }^2-\Delta ^2\right)^2 \left( (2 M_B+\Delta)^2 -m_{\phi}^2\right)^3 \bigg[ -3 \left(m_{\phi }^2-\Delta ^2\right)^2+7 \Delta  M_B \left(m_{\phi }^2-\Delta^2 \right)+M_B^2 \left(11 \Delta ^2+7 m_{\phi }^2\right) \\
    &+36 \Delta  M_B^3+18 M_B^4 \bigg] +24 \Lambda ^4 M_B^2 \left(M_B+\Delta \right)^5 \left(\Delta ^2-m_{\phi }^2\right) \left( (2 M_B+\Delta)^2 -m_{\phi}^2\right)^2 \bigg[ 3 M_B^2 \left(9 \Delta ^2-7 m_{\phi }^2\right) +21 \Delta  M_B \left(\Delta ^2-m_{\phi }^2\right) \\
    &+ 4 \left(m_{\phi }^2-\Delta ^2\right)^2 +12 \Delta  M_B^3+6 M_B^4 \bigg] +16 \Lambda ^6 M_B^3 \left(M_B+\Delta \right)^4 \left( (2 M_B+\Delta)^2 -m_{\phi}^2\right)  \bigg[ 35 \left(m_{\phi }^2-\Delta ^2\right)^3 \\
    &-174 \Delta  M_B \left(m_{\phi }^2-\Delta ^2\right)^2 -6 M_B^2 \left(50 \Delta ^4-79 \Delta ^2 m_{\phi }^2+29 m_{\phi }^4\right)+4 M_B^3 \left(63 \Delta  m_{\phi }^2-62 \Delta ^3\right)+6 M_B^4 \left(21 m_{\phi }^2-19 \Delta ^2\right) \\
    &+12 \Delta  M_B^5+4 M_B^6 \bigg] +64 \Lambda ^8 M_B^3 \left(M_B+\Delta \right)^2 \bigg[ -7 \left(m_{\phi }^2-\Delta ^2\right)^4 +70 \Delta  M_B \left(m_{\phi }^2-\Delta ^2\right)^3 +70 M_B^2 \left(m_{\phi }^2-4 \Delta ^2\right) \left(m_{\phi }^2-\Delta ^2\right)^2 \\
    &-4 M_B^3 \left(146 \Delta ^5-251 \Delta ^3 m_{\phi }^2+105 \Delta  m_{\phi }^4\right) +M_B^4\left(-688 \Delta ^4+912 \Delta ^2 m_{\phi }^2-210 m_{\phi }^4\right) +M_B^5 \left(492 \Delta  m_{\phi }^2-436 \Delta ^3\right) \\
    &+4 M_B^6 \left(41 m_{\phi }^2-20 \Delta ^2\right) +56 \Delta  M_B^7+ 14 M_B^8 \bigg] +128 \Lambda ^{10} M_B^3 \bigg[ -\left(m_{\phi }^2-\Delta ^2\right)^4 +10 \Delta  M_B \left(m_{\phi }^2-\Delta ^2\right)^3 \\
    &+ 10 M_B^2 \left(m_{\phi }^2-4 \Delta ^2\right) \left(m_{\phi }^2-\Delta ^2\right)^2 -20 M_B^3 \left(4 \Delta ^5-7 \Delta ^3 m_{\phi }^2+3 \Delta  m_{\phi }^4\right) -10 M_B^4 \left(8 \Delta ^4-12 \Delta ^2 m_{\phi }^2+3 m_{\phi }^4\right) \\
    &+20 M_B^5 \left(3 \Delta  m_{\phi }^2-\Delta ^3\right) +20 M_B^6 \left(2 \Delta ^2+m_{\phi }^2\right) +40 \Delta  M_B^7 +10 M_B^8 \bigg] \Bigg\rbrace , \numberthis
\end{align*}
\begin{align*}
    \mkern-65mu\Sigma_{BT'\phi}^{(\text{Rel},2)} &= \frac{C_{BT'\phi}^2 \Lambda ^5 M_B^2}{96 \pi^2 f_{\phi}^2 \left( M_B+\Delta \right)^2 \left( (m_{\phi}^2-\Delta^2) ( m_{\phi}^2-(2 M_B+\Delta)^2 )+4 \Lambda ^2 M_B^2\right)^4} \Bigg( \\
    & -128 \Lambda ^3 M_B^3 \left(m_{\phi}^2 -\Delta^2 \right)^{3/2} \left( (2 M_B+\Delta)^2 -m_{\phi }^2 \right)^{5/2} \arctan\left(\frac{\sqrt{ (m_{\phi }^2-\Delta^2 ) ((2 M_B+\Delta)^2 -m_{\phi }^2)}}{\Delta  \left(2 M_B+\Delta \right)+m_{\phi }^2}\right) \\
    &-\frac{1}{\left(m_{\phi }^2-\Lambda ^2\right)^{7/2}}\arctan\left(\frac{\sqrt{m_{\phi}^2-\Lambda^2}}{\Lambda}\right) \bigg\lbrace \Delta  m_{\phi }^6 \left(m_{\phi }^2-\Delta ^2\right)^3 \left( (2 M_B+\Delta )^2-m_{\phi}^2 \right)^4 \\
    &+2 \Lambda ^2 m_{\phi }^4 \left(m_{\phi }^2-\Delta ^2\right)^2 \left( (2 M_B+\Delta )^2-m_{\phi}^2 \right)^3 \bigg[ 3 \Delta  \left(m_{\phi }^2-\Delta ^2\right)^2 +M_B \left(12 \Delta ^4-17 \Delta ^2 m_{\phi }^2+5 m_{\phi }^4\right)-6 M_B^2 \left(5 \Delta  m_{\phi }^2-2 \Delta ^3\right) \bigg] \\
    &-24 \Lambda ^4 M_B^2 m_{\phi }^4 \left(m_{\phi }^2-\Delta ^2\right) \left( (2 M_B+\Delta )^2-m_{\phi}^2 \right)^2 \bigg[ 4 \Delta  \left(m_{\phi }^2-\Delta ^2\right)^2-M_B \left(-16 \Delta ^4+11 \Delta ^2 m_{\phi }^2+5 m_{\phi }^4\right) -2 M_B^2 \left(5 \Delta  m_{\phi }^2-8 \Delta ^3\right) \bigg] \\ 
    &+16 \Lambda ^6 M_B^3 m_{\phi }^4 \bigg[ -35 \left(m_{\phi }^2-\Delta ^2\right)^4+314 \Delta  M_B \left(m_{\phi }^2-\Delta ^2\right)^3+2 M_B^2 \left(m_{\phi }^2-\Delta ^2\right)^2 \left(145 m_{\phi }^2-556 \Delta ^2\right)\\
    &-4 M_B^3 \left(484 \Delta ^5-843 \Delta ^3 m_{\phi }^2+359 \Delta  m_{\phi }^4\right)-8 M_B^4 \left(206 \Delta ^4-283 \Delta ^2 m_{\phi }^2+75 m_{\phi }^4\right)+16 M_B^5 \left(35 \Delta  m_{\phi }^2-34 \Delta ^3\right) \bigg] \\
    &-64 \Lambda ^8 M_B^3 m_{\phi }^2 \bigg[ -7 \left(m_{\phi }^2-\Delta ^2\right)^4+70 \Delta  M_B \left(m_{\phi }^2-\Delta ^2\right)^3+70 M_B^2 \left(m_{\phi }^2-4 \Delta ^2\right) \left(m_{\phi }^2-\Delta ^2\right)^2 \\
    &-4 M_B^3 \left(140 \Delta ^5-239 \Delta ^3 m_{\phi }^2+99 \Delta  m_{\phi }^4\right)-2 M_B^4 \left(280 \Delta ^4-372 \Delta ^2 m_{\phi }^2+85 m_{\phi }^4\right)+16 M_B^5 \left(15 \Delta  m_{\phi }^2-14 \Delta ^3\right) \bigg] \\
    &+128 \Lambda ^{10} M_B^3 \bigg[ -\left(m_{\phi }^2-\Delta ^2\right)^4+10 \Delta  M_B \left(m_{\phi }^2-\Delta ^2\right)^3 +10 M_B^2 \left(m_{\phi }^2-4 \Delta ^2\right) \left(m_{\phi }^2-\Delta ^2\right)^2-20 M_B^3 \left(4 \Delta ^5-7 \Delta ^3 m_{\phi }^2+3 \Delta  m_{\phi }^4\right) \\
    &-10 M_B^4 \left(8 \Delta ^4-12 \Delta ^2 m_{\phi }^2+3 m_{\phi }^4\right) +16 M_B^5 \left(3 \Delta  m_{\phi }^2-2 \Delta ^3\right) \bigg] \bigg\rbrace \Bigg), \numberthis
\end{align*}
\begin{equation}
    \mkern-55mu\Sigma^{(\text{Rel})}_{N\pi,\text{tad}} = \frac{\chi C\, \Lambda ^3 m_\pi^2}{256 \pi ^2 f_{\pi }^2 \left(\Lambda ^2-m_{\pi }^2\right)^{7/2}} \left[ \Lambda  \sqrt{\Lambda ^2-m_{\pi }^2} \left(-8 \Lambda ^4-10 \Lambda ^2 m_{\pi }^2+3 m_{\pi }^4\right)+3 m_{\pi }^2 \left(8 \Lambda ^4-4 \Lambda ^2 m_{\pi }^2+m_{\pi }^4\right) \text{artanh} \left(\frac{ \sqrt{\Lambda ^2-m_{\pi }^2}}{\Lambda}\right) \right]. 
\end{equation}

Finally, in the HB scheme
\begin{align*}
        \mkern-55mu\Sigma_{BB'\phi}^{(\text{HB},1)} &= -\frac{C_{BB'\phi}^2\Lambda ^5}{384 \pi ^2 f_{\phi }^2 \left(\delta ^2+\Lambda ^2-m_{\phi }^2\right)^4} \Bigg\lbrace 3 \pi  \left(\Lambda ^6+9 \Lambda ^4 \left(\delta ^2-m_{\phi }^2\right)-9 \Lambda ^2 \left(m_{\phi }^2-\delta ^2\right)^2+\left(m_{\phi }^2-\delta ^2\right)^3\right) \\
        &+\frac{48 \delta  \Lambda  \left(\delta ^2-m_{\phi }^2\right)^2 \left(\delta ^2+\Lambda ^2-m_{\phi }^2\right)}{m_{\phi }^2-\Lambda ^2} \\
        &+ \frac{2 \delta  \Lambda  \left(\delta ^2+\Lambda ^2-m_{\phi }^2\right)^2 \left(-28 \delta ^2 \Lambda ^4+8 \Lambda ^6-\left(21 \delta ^2+67 \Lambda ^2\right) m_{\phi }^4+\left(64 \delta ^2 \Lambda ^2+38 \Lambda ^4\right) m_{\phi }^2+21 m_{\phi }^6\right)}{\left(m_{\phi }^2-\Lambda ^2\right)^3} \Bigg\rbrace, \numberthis
\end{align*}
\begin{align*}
        \mkern-55mu\Sigma_{BB'\phi}^{(\text{HB},2)} &= \frac{C_{BB'\phi}^2\Lambda^8}{64 \pi ^2 f_{\phi }^2 \left(\delta ^2+\Lambda ^2-m_{\phi }^2\right)^4} \Bigg\lbrace 8 \left(m_{\phi}^2-\delta^2\right)^{3/2} \bigg[ 2\arctan\left(\frac{\delta }{\sqrt{m_{\phi }^2-\delta ^2}}\right)- \pi \bigg] \\
        &-\frac{\delta}{\Lambda ^3 \left(\Lambda ^2-m_{\phi }^2\right)^{7/2}}\text{artanh} \left(\frac{\sqrt{\Lambda ^2-m_{\phi }^2}}{\Lambda}\right) \bigg[ -16 \delta ^2 \Lambda ^{10}+8 \Lambda ^8 \left(7 \delta ^2+3 \Lambda ^2\right) m_{\phi }^2+\left(6 \delta ^6 \Lambda ^2+24 \delta ^4 \Lambda ^4-34 \delta ^2 \Lambda ^6-60 \Lambda ^8\right) m_{\phi }^4\\
        &-\left(\delta ^6+27 \delta ^4 \Lambda ^2+39 \delta ^2 \Lambda ^4-35 \Lambda ^6\right) m_{\phi }^6+3 \left(\delta ^4+12 \delta ^2 \Lambda ^2+5 \Lambda ^4\right) m_{\phi }^8-3 \left(\delta ^2+5 \Lambda ^2\right) m_{\phi }^{10}+m_{\phi }^{12} \bigg] \Bigg\rbrace. \numberthis 
\end{align*}
Then observing Eqs. \eqref{eq:22} and \eqref{eq:23}, we note that the expression for the octet-decuplet-meson self-energy will be the same besides the factor of 2/3, and replacing $\delta \to \Delta$ and $C_{BB'\phi}\to C_{BT'\phi}$. We further remark that $\Sigma^{(\text{Rel})}_{N\pi,\text{tad}}=\Sigma^{(\text{HB})}_{N\pi,\text{tad}}$.

\twocolumngrid
\bibliography{bibliography}

\providecommand{\noopsort}[1]{}\providecommand{\singleletter}[1]{#1}%
\begin{thebibliography}{104}%
\makeatletter
\providecommand \@ifxundefined [1]{%
 \@ifx{#1\undefined}
}%
\providecommand \@ifnum [1]{%
 \ifnum #1\expandafter \@firstoftwo
 \else \expandafter \@secondoftwo
 \fi
}%
\providecommand \@ifx [1]{%
 \ifx #1\expandafter \@firstoftwo
 \else \expandafter \@secondoftwo
 \fi
}%
\providecommand \natexlab [1]{#1}%
\providecommand \enquote  [1]{``#1''}%
\providecommand \bibnamefont  [1]{#1}%
\providecommand \bibfnamefont [1]{#1}%
\providecommand \citenamefont [1]{#1}%
\providecommand \href@noop [0]{\@secondoftwo}%
\providecommand \href [0]{\begingroup \@sanitize@url \@href}%
\providecommand \@href[1]{\@@startlink{#1}\@@href}%
\providecommand \@@href[1]{\endgroup#1\@@endlink}%
\providecommand \@sanitize@url [0]{\catcode `\\12\catcode `\$12\catcode
  `\&12\catcode `\#12\catcode `\^12\catcode `\_12\catcode `\%12\relax}%
\providecommand \@@startlink[1]{}%
\providecommand \@@endlink[0]{}%
\providecommand \url  [0]{\begingroup\@sanitize@url \@url }%
\providecommand \@url [1]{\endgroup\@href {#1}{\urlprefix }}%
\providecommand \urlprefix  [0]{URL }%
\providecommand \Eprint [0]{\href }%
\providecommand \doibase [0]{https://doi.org/}%
\providecommand \selectlanguage [0]{\@gobble}%
\providecommand \bibinfo  [0]{\@secondoftwo}%
\providecommand \bibfield  [0]{\@secondoftwo}%
\providecommand \translation [1]{[#1]}%
\providecommand \BibitemOpen [0]{}%
\providecommand \bibitemStop [0]{}%
\providecommand \bibitemNoStop [0]{.\EOS\space}%
\providecommand \EOS [0]{\spacefactor3000\relax}%
\providecommand \BibitemShut  [1]{\csname bibitem#1\endcsname}%
\let\auto@bib@innerbib\@empty
\bibitem [{\citenamefont {Aoki}\ \emph {et~al.}(2009)\citenamefont {Aoki},
  \citenamefont {Ishikawa}, \citenamefont {Ishizuka}, \citenamefont {Izubuchi},
  \citenamefont {Kadoh}, \citenamefont {Kanaya}, \citenamefont {Kuramashi},
  \citenamefont {Namekawa}, \citenamefont {Okawa}, \citenamefont {Taniguchi},
  \citenamefont {Ukawa}, \citenamefont {Ukita},\ and\ \citenamefont
  {Yoshi\'e}}]{PACSCSdata}%
  \BibitemOpen
  \bibfield  {author} {\bibinfo {author} {\bibfnamefont {S.}~\bibnamefont
  {Aoki}}, \bibinfo {author} {\bibfnamefont {K.-I.}\ \bibnamefont {Ishikawa}},
  \bibinfo {author} {\bibfnamefont {N.}~\bibnamefont {Ishizuka}}, \bibinfo
  {author} {\bibfnamefont {T.}~\bibnamefont {Izubuchi}}, \bibinfo {author}
  {\bibfnamefont {D.}~\bibnamefont {Kadoh}}, \bibinfo {author} {\bibfnamefont
  {K.}~\bibnamefont {Kanaya}}, \bibinfo {author} {\bibfnamefont
  {Y.}~\bibnamefont {Kuramashi}}, \bibinfo {author} {\bibfnamefont
  {Y.}~\bibnamefont {Namekawa}}, \bibinfo {author} {\bibfnamefont
  {M.}~\bibnamefont {Okawa}}, \bibinfo {author} {\bibfnamefont
  {Y.}~\bibnamefont {Taniguchi}}, \bibinfo {author} {\bibfnamefont
  {A.}~\bibnamefont {Ukawa}}, \bibinfo {author} {\bibfnamefont
  {N.}~\bibnamefont {Ukita}},\ and\ \bibinfo {author} {\bibfnamefont
  {T.}~\bibnamefont {Yoshi\'e}} (\bibinfo {collaboration} {PACS-CS
  Collaboration}),\ }\bibfield  {title} {\bibinfo {title} {$2+1$ flavor lattice
  {QCD} toward the physical point},\ }\href
  {https://doi.org/10.1103/PhysRevD.79.034503} {\bibfield  {journal} {\bibinfo
  {journal} {Phys. Rev. D}\ }\textbf {\bibinfo {volume} {79}},\ \bibinfo
  {pages} {034503} (\bibinfo {year} {2009})}\BibitemShut {NoStop}%
\bibitem [{\citenamefont {Ottnad}\ \emph {et~al.}(2023)\citenamefont {Ottnad},
  \citenamefont {Djukanovic}, \citenamefont {Meyer}, \citenamefont {von
  Hippel},\ and\ \citenamefont {Wittig}}]{Ottnad:2022axz}%
  \BibitemOpen
  \bibfield  {author} {\bibinfo {author} {\bibfnamefont {K.}~\bibnamefont
  {Ottnad}}, \bibinfo {author} {\bibfnamefont {D.}~\bibnamefont {Djukanovic}},
  \bibinfo {author} {\bibfnamefont {H.~B.}\ \bibnamefont {Meyer}}, \bibinfo
  {author} {\bibfnamefont {G.}~\bibnamefont {von Hippel}},\ and\ \bibinfo
  {author} {\bibfnamefont {H.}~\bibnamefont {Wittig}},\ }\bibfield  {title}
  {\bibinfo {title} {{Mass and isovector matrix elements of the nucleon at
  zero-momentum transfer}},\ }\href {https://doi.org/10.22323/1.430.0117}
  {\bibfield  {journal} {\bibinfo  {journal} {PoS}\ }\textbf {\bibinfo {volume}
  {LATTICE2022}},\ \bibinfo {pages} {117} (\bibinfo {year} {2023})},\ \Eprint
  {https://arxiv.org/abs/2212.09940} {arXiv:2212.09940 [hep-lat]} \BibitemShut
  {NoStop}%
\bibitem [{\citenamefont {Leinweber}\ \emph {et~al.}(2000)\citenamefont
  {Leinweber}, \citenamefont {Thomas}, \citenamefont {Tsushima},\ and\
  \citenamefont {Wright}}]{Leinweber:1999ig}%
  \BibitemOpen
  \bibfield  {author} {\bibinfo {author} {\bibfnamefont {D.~B.}\ \bibnamefont
  {Leinweber}}, \bibinfo {author} {\bibfnamefont {A.~W.}\ \bibnamefont
  {Thomas}}, \bibinfo {author} {\bibfnamefont {K.}~\bibnamefont {Tsushima}},\
  and\ \bibinfo {author} {\bibfnamefont {S.~V.}\ \bibnamefont {Wright}},\
  }\bibfield  {title} {\bibinfo {title} {{Baryon masses from lattice QCD:
  Beyond the perturbative chiral regime}},\ }\href
  {https://doi.org/10.1103/PhysRevD.61.074502} {\bibfield  {journal} {\bibinfo
  {journal} {Phys. Rev. D}\ }\textbf {\bibinfo {volume} {61}},\ \bibinfo
  {pages} {074502} (\bibinfo {year} {2000})},\ \Eprint
  {https://arxiv.org/abs/hep-lat/9906027} {arXiv:hep-lat/9906027} \BibitemShut
  {NoStop}%
\bibitem [{\citenamefont {Young}\ \emph {et~al.}(2003)\citenamefont {Young},
  \citenamefont {Leinweber},\ and\ \citenamefont {Thomas}}]{Young:2002ib}%
  \BibitemOpen
  \bibfield  {author} {\bibinfo {author} {\bibfnamefont {R.~D.}\ \bibnamefont
  {Young}}, \bibinfo {author} {\bibfnamefont {D.~B.}\ \bibnamefont
  {Leinweber}},\ and\ \bibinfo {author} {\bibfnamefont {A.~W.}\ \bibnamefont
  {Thomas}},\ }\bibfield  {title} {\bibinfo {title} {{Convergence of chiral
  effective field theory}},\ }\href
  {https://doi.org/10.1016/S0146-6410(03)00034-6} {\bibfield  {journal}
  {\bibinfo  {journal} {Prog. Part. Nucl. Phys.}\ }\textbf {\bibinfo {volume}
  {50}},\ \bibinfo {pages} {399} (\bibinfo {year} {2003})},\ \Eprint
  {https://arxiv.org/abs/hep-lat/0212031} {arXiv:hep-lat/0212031} \BibitemShut
  {NoStop}%
\bibitem [{\citenamefont {Leinweber}\ \emph {et~al.}(2004)\citenamefont
  {Leinweber}, \citenamefont {Thomas},\ and\ \citenamefont
  {Young}}]{Leinweber:2003dg}%
  \BibitemOpen
  \bibfield  {author} {\bibinfo {author} {\bibfnamefont {D.~B.}\ \bibnamefont
  {Leinweber}}, \bibinfo {author} {\bibfnamefont {A.~W.}\ \bibnamefont
  {Thomas}},\ and\ \bibinfo {author} {\bibfnamefont {R.~D.}\ \bibnamefont
  {Young}},\ }\bibfield  {title} {\bibinfo {title} {{Physical nucleon
  properties from lattice QCD}},\ }\href
  {https://doi.org/10.1103/PhysRevLett.92.242002} {\bibfield  {journal}
  {\bibinfo  {journal} {Phys. Rev. Lett.}\ }\textbf {\bibinfo {volume} {92}},\
  \bibinfo {pages} {242002} (\bibinfo {year} {2004})},\ \Eprint
  {https://arxiv.org/abs/hep-lat/0302020} {arXiv:hep-lat/0302020} \BibitemShut
  {NoStop}%
\bibitem [{\citenamefont {Procura}\ \emph {et~al.}(2004)\citenamefont
  {Procura}, \citenamefont {Hemmert},\ and\ \citenamefont
  {Weise}}]{LECsProcura}%
  \BibitemOpen
  \bibfield  {author} {\bibinfo {author} {\bibfnamefont {M.}~\bibnamefont
  {Procura}}, \bibinfo {author} {\bibfnamefont {T.~R.}\ \bibnamefont
  {Hemmert}},\ and\ \bibinfo {author} {\bibfnamefont {W.}~\bibnamefont
  {Weise}},\ }\bibfield  {title} {\bibinfo {title} {Nucleon mass, sigma term,
  and lattice {QCD}},\ }\href {https://doi.org/10.1103/PhysRevD.69.034505}
  {\bibfield  {journal} {\bibinfo  {journal} {Phys. Rev. D}\ }\textbf {\bibinfo
  {volume} {69}},\ \bibinfo {pages} {034505} (\bibinfo {year}
  {2004})}\BibitemShut {NoStop}%
\bibitem [{\citenamefont {Bernard}\ \emph {et~al.}(2004)\citenamefont
  {Bernard}, \citenamefont {Hemmert},\ and\ \citenamefont
  {Meißner}}]{LECsBernard}%
  \BibitemOpen
  \bibfield  {author} {\bibinfo {author} {\bibfnamefont {V.}~\bibnamefont
  {Bernard}}, \bibinfo {author} {\bibfnamefont {T.~R.}\ \bibnamefont
  {Hemmert}},\ and\ \bibinfo {author} {\bibfnamefont {U.-G.}\ \bibnamefont
  {Meißner}},\ }\bibfield  {title} {\bibinfo {title} {Cutoff schemes in chiral
  perturbation theory and the quark mass expansion of the nucleon mass},\
  }\href {https://doi.org/https://doi.org/10.1016/j.nuclphysa.2003.12.011}
  {\bibfield  {journal} {\bibinfo  {journal} {Nuclear Physics A}\ }\textbf
  {\bibinfo {volume} {732}},\ \bibinfo {pages} {149} (\bibinfo {year}
  {2004})}\BibitemShut {NoStop}%
\bibitem [{\citenamefont {Bernard}\ \emph {et~al.}(2005)\citenamefont
  {Bernard}, \citenamefont {Hemmert},\ and\ \citenamefont
  {Meißner}}]{bernard2005chiral}%
  \BibitemOpen
  \bibfield  {author} {\bibinfo {author} {\bibfnamefont {V.}~\bibnamefont
  {Bernard}}, \bibinfo {author} {\bibfnamefont {T.~R.}\ \bibnamefont
  {Hemmert}},\ and\ \bibinfo {author} {\bibfnamefont {U.-G.}\ \bibnamefont
  {Meißner}},\ }\bibfield  {title} {\bibinfo {title} {Chiral extrapolations
  and the covariant small scale expansion},\ }\href
  {https://doi.org/https://doi.org/10.1016/j.physletb.2005.06.088} {\bibfield
  {journal} {\bibinfo  {journal} {Physics Letters B}\ }\textbf {\bibinfo
  {volume} {622}},\ \bibinfo {pages} {141} (\bibinfo {year}
  {2005})}\BibitemShut {NoStop}%
\bibitem [{\citenamefont {Meißner}(2006)}]{c1c2c3Meissner}%
  \BibitemOpen
  \bibfield  {author} {\bibinfo {author} {\bibfnamefont {U.-G.}\ \bibnamefont
  {Meißner}},\ }\bibfield  {title} {\bibinfo {title} {Quark mass dependence of
  baryon properties: Foundations and applications},\ }\href
  {https://doi.org/https://doi.org/10.1016/j.nuclphysbps.2006.01.025}
  {\bibfield  {journal} {\bibinfo  {journal} {Nuclear Physics B - Proceedings
  Supplements}\ }\textbf {\bibinfo {volume} {153}},\ \bibinfo {pages} {170}
  (\bibinfo {year} {2006})},\ \bibinfo {note} {proceedings of the Workshop on
  Computational Hadron Physics}\BibitemShut {NoStop}%
\bibitem [{\citenamefont {Goeke}\ \emph {et~al.}(2006)\citenamefont {Goeke},
  \citenamefont {Ossmann}, \citenamefont {Schweitzer},\ and\ \citenamefont
  {Silva}}]{Goeke2006}%
  \BibitemOpen
  \bibfield  {author} {\bibinfo {author} {\bibfnamefont {K.}~\bibnamefont
  {Goeke}}, \bibinfo {author} {\bibfnamefont {J.}~\bibnamefont {Ossmann}},
  \bibinfo {author} {\bibfnamefont {P.}~\bibnamefont {Schweitzer}},\ and\
  \bibinfo {author} {\bibfnamefont {A.}~\bibnamefont {Silva}},\ }\bibfield
  {title} {\bibinfo {title} {Pion mass dependence of the nucleon mass in the
  chiral quark soliton model},\ }\href
  {https://doi.org/10.1140/epja/i2005-10229-5} {\bibfield  {journal} {\bibinfo
  {journal} {The European Physical Journal A - Hadrons and Nuclei}\ }\textbf
  {\bibinfo {volume} {27}},\ \bibinfo {pages} {77} (\bibinfo {year}
  {2006})}\BibitemShut {NoStop}%
\bibitem [{\citenamefont {Procura}\ \emph {et~al.}(2006)\citenamefont
  {Procura}, \citenamefont {Musch}, \citenamefont {Wollenweber}, \citenamefont
  {Hemmert},\ and\ \citenamefont {Weise}}]{procura2006nucleon}%
  \BibitemOpen
  \bibfield  {author} {\bibinfo {author} {\bibfnamefont {M.}~\bibnamefont
  {Procura}}, \bibinfo {author} {\bibfnamefont {B.~U.}\ \bibnamefont {Musch}},
  \bibinfo {author} {\bibfnamefont {T.}~\bibnamefont {Wollenweber}}, \bibinfo
  {author} {\bibfnamefont {T.~R.}\ \bibnamefont {Hemmert}},\ and\ \bibinfo
  {author} {\bibfnamefont {W.}~\bibnamefont {Weise}},\ }\bibfield  {title}
  {\bibinfo {title} {Nucleon mass: From lattice {QCD} to the chiral limit},\
  }\href {https://doi.org/10.1103/PhysRevD.73.114510} {\bibfield  {journal}
  {\bibinfo  {journal} {Phys. Rev. D}\ }\textbf {\bibinfo {volume} {73}},\
  \bibinfo {pages} {114510} (\bibinfo {year} {2006})}\BibitemShut {NoStop}%
\bibitem [{\citenamefont {Armour}\ \emph {et~al.}(2010)\citenamefont {Armour},
  \citenamefont {Allton}, \citenamefont {Leinweber}, \citenamefont {Thomas},\
  and\ \citenamefont {Young}}]{Armour:2008ke}%
  \BibitemOpen
  \bibfield  {author} {\bibinfo {author} {\bibfnamefont {W.}~\bibnamefont
  {Armour}}, \bibinfo {author} {\bibfnamefont {C.~R.}\ \bibnamefont {Allton}},
  \bibinfo {author} {\bibfnamefont {D.~B.}\ \bibnamefont {Leinweber}}, \bibinfo
  {author} {\bibfnamefont {A.~W.}\ \bibnamefont {Thomas}},\ and\ \bibinfo
  {author} {\bibfnamefont {R.~D.}\ \bibnamefont {Young}},\ }\bibfield  {title}
  {\bibinfo {title} {{An Analysis of the nucleon spectrum from lattice
  partially-quenched QCD}},\ }\href
  {https://doi.org/10.1016/j.nuclphysa.2010.03.012} {\bibfield  {journal}
  {\bibinfo  {journal} {Nucl. Phys. A}\ }\textbf {\bibinfo {volume} {840}},\
  \bibinfo {pages} {97} (\bibinfo {year} {2010})},\ \Eprint
  {https://arxiv.org/abs/0810.3432} {arXiv:0810.3432 [hep-lat]} \BibitemShut
  {NoStop}%
\bibitem [{\citenamefont {Hall}\ \emph {et~al.}(2010)\citenamefont {Hall},
  \citenamefont {Leinweber},\ and\ \citenamefont {Young}}]{Hall:2010ai}%
  \BibitemOpen
  \bibfield  {author} {\bibinfo {author} {\bibfnamefont {J.~M.~M.}\
  \bibnamefont {Hall}}, \bibinfo {author} {\bibfnamefont {D.~B.}\ \bibnamefont
  {Leinweber}},\ and\ \bibinfo {author} {\bibfnamefont {R.~D.}\ \bibnamefont
  {Young}},\ }\bibfield  {title} {\bibinfo {title} {{Power Counting Regime of
  Chiral Effective Field Theory and Beyond}},\ }\href
  {https://doi.org/10.1103/PhysRevD.82.034010} {\bibfield  {journal} {\bibinfo
  {journal} {Phys. Rev. D}\ }\textbf {\bibinfo {volume} {82}},\ \bibinfo
  {pages} {034010} (\bibinfo {year} {2010})},\ \Eprint
  {https://arxiv.org/abs/1002.4924} {arXiv:1002.4924 [hep-lat]} \BibitemShut
  {NoStop}%
\bibitem [{\citenamefont {Ren}\ \emph {et~al.}(2012)\citenamefont {Ren},
  \citenamefont {Geng}, \citenamefont {Camalich}, \citenamefont {Meng},\ and\
  \citenamefont {Toki}}]{Ren:2012aj}%
  \BibitemOpen
  \bibfield  {author} {\bibinfo {author} {\bibfnamefont {X.-L.}\ \bibnamefont
  {Ren}}, \bibinfo {author} {\bibfnamefont {L.~S.}\ \bibnamefont {Geng}},
  \bibinfo {author} {\bibfnamefont {J.~M.}\ \bibnamefont {Camalich}}, \bibinfo
  {author} {\bibfnamefont {J.}~\bibnamefont {Meng}},\ and\ \bibinfo {author}
  {\bibfnamefont {H.}~\bibnamefont {Toki}},\ }\bibfield  {title} {\bibinfo
  {title} {Octet baryon masses in next-to-next-to-next-to-leading order
  covariant baryon chiral perturbation theory},\ }\href
  {https://doi.org/10.1007/JHEP12(2012)073} {\bibfield  {journal} {\bibinfo
  {journal} {Journal of High Energy Physics}\ }\textbf {\bibinfo {volume}
  {2012}},\ \bibinfo {pages} {73} (\bibinfo {year} {2012})}\BibitemShut
  {NoStop}%
\bibitem [{\citenamefont {Bruns}\ \emph {et~al.}(2013)\citenamefont {Bruns},
  \citenamefont {Greil},\ and\ \citenamefont {Sch\"afer}}]{Bruns:2012eh}%
  \BibitemOpen
  \bibfield  {author} {\bibinfo {author} {\bibfnamefont {P.~C.}\ \bibnamefont
  {Bruns}}, \bibinfo {author} {\bibfnamefont {L.}~\bibnamefont {Greil}},\ and\
  \bibinfo {author} {\bibfnamefont {A.}~\bibnamefont {Sch\"afer}},\ }\bibfield
  {title} {\bibinfo {title} {{Chiral extrapolation of baryon mass ratios}},\
  }\href {https://doi.org/10.1103/PhysRevD.87.054021} {\bibfield  {journal}
  {\bibinfo  {journal} {Phys. Rev. D}\ }\textbf {\bibinfo {volume} {87}},\
  \bibinfo {pages} {054021} (\bibinfo {year} {2013})},\ \Eprint
  {https://arxiv.org/abs/1209.0980} {arXiv:1209.0980 [hep-ph]} \BibitemShut
  {NoStop}%
\bibitem [{\citenamefont {Shanahan}\ \emph {et~al.}(2013)\citenamefont
  {Shanahan}, \citenamefont {Thomas},\ and\ \citenamefont
  {Young}}]{Shanahan:2012wh}%
  \BibitemOpen
  \bibfield  {author} {\bibinfo {author} {\bibfnamefont {P.~E.}\ \bibnamefont
  {Shanahan}}, \bibinfo {author} {\bibfnamefont {A.~W.}\ \bibnamefont
  {Thomas}},\ and\ \bibinfo {author} {\bibfnamefont {R.~D.}\ \bibnamefont
  {Young}},\ }\bibfield  {title} {\bibinfo {title} {{Sigma terms from an SU(3)
  chiral extrapolation}},\ }\href {https://doi.org/10.1103/PhysRevD.87.074503}
  {\bibfield  {journal} {\bibinfo  {journal} {Phys. Rev. D}\ }\textbf {\bibinfo
  {volume} {87}},\ \bibinfo {pages} {074503} (\bibinfo {year} {2013})},\
  \Eprint {https://arxiv.org/abs/1205.5365} {arXiv:1205.5365 [nucl-th]}
  \BibitemShut {NoStop}%
\bibitem [{\citenamefont {Alvarez-Ruso}\ \emph {et~al.}(2013)\citenamefont
  {Alvarez-Ruso}, \citenamefont {Ledwig}, \citenamefont {Martin~Camalich},\
  and\ \citenamefont {Vicente-Vacas}}]{Alvarez-Ruso:2013fza}%
  \BibitemOpen
  \bibfield  {author} {\bibinfo {author} {\bibfnamefont {L.}~\bibnamefont
  {Alvarez-Ruso}}, \bibinfo {author} {\bibfnamefont {T.}~\bibnamefont
  {Ledwig}}, \bibinfo {author} {\bibfnamefont {J.}~\bibnamefont
  {Martin~Camalich}},\ and\ \bibinfo {author} {\bibfnamefont {M.~J.}\
  \bibnamefont {Vicente-Vacas}},\ }\bibfield  {title} {\bibinfo {title}
  {{Nucleon mass and pion-nucleon sigma term from a chiral analysis of lattice
  QCD data}},\ }\href {https://doi.org/10.1103/PhysRevD.88.054507} {\bibfield
  {journal} {\bibinfo  {journal} {Phys. Rev. D}\ }\textbf {\bibinfo {volume}
  {88}},\ \bibinfo {pages} {054507} (\bibinfo {year} {2013})},\ \Eprint
  {https://arxiv.org/abs/1304.0483} {arXiv:1304.0483 [hep-ph]} \BibitemShut
  {NoStop}%
\bibitem [{\citenamefont {Ren}\ \emph {et~al.}(2015)\citenamefont {Ren},
  \citenamefont {Geng},\ and\ \citenamefont {Meng}}]{Ren:2014vea}%
  \BibitemOpen
  \bibfield  {author} {\bibinfo {author} {\bibfnamefont {X.-L.}\ \bibnamefont
  {Ren}}, \bibinfo {author} {\bibfnamefont {L.-S.}\ \bibnamefont {Geng}},\ and\
  \bibinfo {author} {\bibfnamefont {J.}~\bibnamefont {Meng}},\ }\bibfield
  {title} {\bibinfo {title} {{Scalar strangeness content of the nucleon and
  baryon sigma terms}},\ }\href {https://doi.org/10.1103/PhysRevD.91.051502}
  {\bibfield  {journal} {\bibinfo  {journal} {Phys. Rev. D}\ }\textbf {\bibinfo
  {volume} {91}},\ \bibinfo {pages} {051502} (\bibinfo {year} {2015})},\
  \Eprint {https://arxiv.org/abs/1404.4799} {arXiv:1404.4799 [hep-ph]}
  \BibitemShut {NoStop}%
\bibitem [{\citenamefont {Lutz}\ \emph {et~al.}(2018)\citenamefont {Lutz},
  \citenamefont {Heo},\ and\ \citenamefont {Guo}}]{Lutz:2018cqo}%
  \BibitemOpen
  \bibfield  {author} {\bibinfo {author} {\bibfnamefont {M.~F.~M.}\
  \bibnamefont {Lutz}}, \bibinfo {author} {\bibfnamefont {Y.}~\bibnamefont
  {Heo}},\ and\ \bibinfo {author} {\bibfnamefont {X.-Y.}\ \bibnamefont {Guo}},\
  }\bibfield  {title} {\bibinfo {title} {{On the convergence of the chiral
  expansion for the baryon ground-state masses}},\ }\href
  {https://doi.org/10.1016/j.nuclphysa.2018.05.007} {\bibfield  {journal}
  {\bibinfo  {journal} {Nucl. Phys. A}\ }\textbf {\bibinfo {volume} {977}},\
  \bibinfo {pages} {146} (\bibinfo {year} {2018})},\ \Eprint
  {https://arxiv.org/abs/1801.06417} {arXiv:1801.06417 [hep-lat]} \BibitemShut
  {NoStop}%
\bibitem [{\citenamefont {Bali}\ \emph {et~al.}(2022)\citenamefont {Bali},
  \citenamefont {Collins}, \citenamefont {S\"oldner},\ and\ \citenamefont
  {Weish\"aupl}}]{bali2022leading}%
  \BibitemOpen
  \bibfield  {author} {\bibinfo {author} {\bibfnamefont {G.~S.}\ \bibnamefont
  {Bali}}, \bibinfo {author} {\bibfnamefont {S.}~\bibnamefont {Collins}},
  \bibinfo {author} {\bibfnamefont {W.}~\bibnamefont {S\"oldner}},\ and\
  \bibinfo {author} {\bibfnamefont {S.}~\bibnamefont {Weish\"aupl}} (\bibinfo
  {collaboration} {RQCD Collaboration}),\ }\bibfield  {title} {\bibinfo {title}
  {Leading order mesonic and baryonic su(3) low energy constants from
  ${N}_{f}=3$ lattice {QCD}},\ }\href
  {https://doi.org/10.1103/PhysRevD.105.054516} {\bibfield  {journal} {\bibinfo
   {journal} {Phys. Rev. D}\ }\textbf {\bibinfo {volume} {105}},\ \bibinfo
  {pages} {054516} (\bibinfo {year} {2022})}\BibitemShut {NoStop}%
\bibitem [{\citenamefont {Lutz}\ \emph {et~al.}(2023)\citenamefont {Lutz},
  \citenamefont {Heo},\ and\ \citenamefont {Guo}}]{Lutz:2023xpi}%
  \BibitemOpen
  \bibfield  {author} {\bibinfo {author} {\bibfnamefont {M.~F.~M.}\
  \bibnamefont {Lutz}}, \bibinfo {author} {\bibfnamefont {Y.}~\bibnamefont
  {Heo}},\ and\ \bibinfo {author} {\bibfnamefont {X.-Y.}\ \bibnamefont {Guo}},\
  }\bibfield  {title} {\bibinfo {title} {{Low-energy constants in the chiral
  Lagrangian with baryon octet and decuplet fields from Lattice QCD data on CLS
  ensembles}},\ }\href {https://doi.org/10.1140/epjc/s10052-023-11556-1}
  {\bibfield  {journal} {\bibinfo  {journal} {Eur. Phys. J. C}\ }\textbf
  {\bibinfo {volume} {83}},\ \bibinfo {pages} {440} (\bibinfo {year} {2023})},\
  \Eprint {https://arxiv.org/abs/2301.06837} {arXiv:2301.06837 [hep-lat]}
  \BibitemShut {NoStop}%
\bibitem [{\citenamefont {Copeland}\ \emph {et~al.}(2023)\citenamefont
  {Copeland}, \citenamefont {Ji},\ and\ \citenamefont
  {Melnitchouk}}]{Copeland:2021qni}%
  \BibitemOpen
  \bibfield  {author} {\bibinfo {author} {\bibfnamefont {P.~M.}\ \bibnamefont
  {Copeland}}, \bibinfo {author} {\bibfnamefont {C.-R.}\ \bibnamefont {Ji}},\
  and\ \bibinfo {author} {\bibfnamefont {W.}~\bibnamefont {Melnitchouk}},\
  }\bibfield  {title} {\bibinfo {title} {{Octet and decuplet baryon
  \ensuremath{\sigma} terms and mass decompositions}},\ }\href
  {https://doi.org/10.1103/PhysRevD.107.094041} {\bibfield  {journal} {\bibinfo
   {journal} {Phys. Rev. D}\ }\textbf {\bibinfo {volume} {107}},\ \bibinfo
  {pages} {094041} (\bibinfo {year} {2023})},\ \Eprint
  {https://arxiv.org/abs/2112.03198} {arXiv:2112.03198 [nucl-th]} \BibitemShut
  {NoStop}%
\bibitem [{\citenamefont {Leinweber}(2004)}]{Leinweber:2002qb}%
  \BibitemOpen
  \bibfield  {author} {\bibinfo {author} {\bibfnamefont {D.~B.}\ \bibnamefont
  {Leinweber}},\ }\bibfield  {title} {\bibinfo {title} {{Quark contributions to
  baryon magnetic moments in full, quenched and partially quenched QCD}},\
  }\href {https://doi.org/10.1103/PhysRevD.69.014005} {\bibfield  {journal}
  {\bibinfo  {journal} {Phys. Rev. D}\ }\textbf {\bibinfo {volume} {69}},\
  \bibinfo {pages} {014005} (\bibinfo {year} {2004})},\ \Eprint
  {https://arxiv.org/abs/hep-lat/0211017} {arXiv:hep-lat/0211017} \BibitemShut
  {NoStop}%
\bibitem [{\citenamefont {Leinweber}\ \emph {et~al.}(2005)\citenamefont
  {Leinweber}, \citenamefont {Boinepalli}, \citenamefont {Cloet}, \citenamefont
  {Thomas}, \citenamefont {Williams}, \citenamefont {Young}, \citenamefont
  {Zanotti},\ and\ \citenamefont {Zhang}}]{Leinweber:2004tc}%
  \BibitemOpen
  \bibfield  {author} {\bibinfo {author} {\bibfnamefont {D.~B.}\ \bibnamefont
  {Leinweber}}, \bibinfo {author} {\bibfnamefont {S.}~\bibnamefont
  {Boinepalli}}, \bibinfo {author} {\bibfnamefont {I.~C.}\ \bibnamefont
  {Cloet}}, \bibinfo {author} {\bibfnamefont {A.~W.}\ \bibnamefont {Thomas}},
  \bibinfo {author} {\bibfnamefont {A.~G.}\ \bibnamefont {Williams}}, \bibinfo
  {author} {\bibfnamefont {R.~D.}\ \bibnamefont {Young}}, \bibinfo {author}
  {\bibfnamefont {J.~M.}\ \bibnamefont {Zanotti}},\ and\ \bibinfo {author}
  {\bibfnamefont {J.~B.}\ \bibnamefont {Zhang}},\ }\bibfield  {title} {\bibinfo
  {title} {{Precise determination of the strangeness magnetic moment of the
  nucleon}},\ }\href {https://doi.org/10.1103/PhysRevLett.94.212001} {\bibfield
   {journal} {\bibinfo  {journal} {Phys. Rev. Lett.}\ }\textbf {\bibinfo
  {volume} {94}},\ \bibinfo {pages} {212001} (\bibinfo {year} {2005})},\
  \Eprint {https://arxiv.org/abs/hep-lat/0406002} {arXiv:hep-lat/0406002}
  \BibitemShut {NoStop}%
\bibitem [{\citenamefont {Leinweber}\ \emph {et~al.}(2006)\citenamefont
  {Leinweber}, \citenamefont {Boinepalli}, \citenamefont {Thomas},
  \citenamefont {Wang}, \citenamefont {Williams}, \citenamefont {Young},
  \citenamefont {Zanotti},\ and\ \citenamefont {Zhang}}]{Leinweber:2006ug}%
  \BibitemOpen
  \bibfield  {author} {\bibinfo {author} {\bibfnamefont {D.~B.}\ \bibnamefont
  {Leinweber}}, \bibinfo {author} {\bibfnamefont {S.}~\bibnamefont
  {Boinepalli}}, \bibinfo {author} {\bibfnamefont {A.~W.}\ \bibnamefont
  {Thomas}}, \bibinfo {author} {\bibfnamefont {P.}~\bibnamefont {Wang}},
  \bibinfo {author} {\bibfnamefont {A.~G.}\ \bibnamefont {Williams}}, \bibinfo
  {author} {\bibfnamefont {R.~D.}\ \bibnamefont {Young}}, \bibinfo {author}
  {\bibfnamefont {J.~M.}\ \bibnamefont {Zanotti}},\ and\ \bibinfo {author}
  {\bibfnamefont {J.~B.}\ \bibnamefont {Zhang}},\ }\bibfield  {title} {\bibinfo
  {title} {{Strange electric form-factor of the proton}},\ }\href
  {https://doi.org/10.1103/PhysRevLett.97.022001} {\bibfield  {journal}
  {\bibinfo  {journal} {Phys. Rev. Lett.}\ }\textbf {\bibinfo {volume} {97}},\
  \bibinfo {pages} {022001} (\bibinfo {year} {2006})},\ \Eprint
  {https://arxiv.org/abs/hep-lat/0601025} {arXiv:hep-lat/0601025} \BibitemShut
  {NoStop}%
\bibitem [{\citenamefont {Hall}\ \emph
  {et~al.}(2013{\natexlab{a}})\citenamefont {Hall}, \citenamefont {Leinweber},
  \citenamefont {Owen},\ and\ \citenamefont {Young}}]{Hall:2012yx}%
  \BibitemOpen
  \bibfield  {author} {\bibinfo {author} {\bibfnamefont {J.~M.~M.}\
  \bibnamefont {Hall}}, \bibinfo {author} {\bibfnamefont {D.~B.}\ \bibnamefont
  {Leinweber}}, \bibinfo {author} {\bibfnamefont {B.~J.}\ \bibnamefont
  {Owen}},\ and\ \bibinfo {author} {\bibfnamefont {R.~D.}\ \bibnamefont
  {Young}},\ }\bibfield  {title} {\bibinfo {title} {{Finite-volume corrections
  to charge radii}},\ }\href {https://doi.org/10.1016/j.physletb.2013.06.048}
  {\bibfield  {journal} {\bibinfo  {journal} {Phys. Lett. B}\ }\textbf
  {\bibinfo {volume} {725}},\ \bibinfo {pages} {101} (\bibinfo {year}
  {2013}{\natexlab{a}})},\ \Eprint {https://arxiv.org/abs/1210.6124}
  {arXiv:1210.6124 [hep-lat]} \BibitemShut {NoStop}%
\bibitem [{\citenamefont {Hall}\ \emph
  {et~al.}(2013{\natexlab{b}})\citenamefont {Hall}, \citenamefont {Leinweber},\
  and\ \citenamefont {Young}}]{Hall:2013oga}%
  \BibitemOpen
  \bibfield  {author} {\bibinfo {author} {\bibfnamefont {J.~M.~M.}\
  \bibnamefont {Hall}}, \bibinfo {author} {\bibfnamefont {D.~B.}\ \bibnamefont
  {Leinweber}},\ and\ \bibinfo {author} {\bibfnamefont {R.~D.}\ \bibnamefont
  {Young}},\ }\bibfield  {title} {\bibinfo {title} {{Chiral extrapolations for
  nucleon electric charge radii}},\ }\href
  {https://doi.org/10.1103/PhysRevD.88.014504} {\bibfield  {journal} {\bibinfo
  {journal} {Phys. Rev. D}\ }\textbf {\bibinfo {volume} {88}},\ \bibinfo
  {pages} {014504} (\bibinfo {year} {2013}{\natexlab{b}})},\ \Eprint
  {https://arxiv.org/abs/1305.3984} {arXiv:1305.3984 [hep-lat]} \BibitemShut
  {NoStop}%
\bibitem [{\citenamefont {Shanahan}\ \emph {et~al.}(2014)\citenamefont
  {Shanahan}, \citenamefont {Thomas}, \citenamefont {Young}, \citenamefont
  {Zanotti}, \citenamefont {Horsley}, \citenamefont {Nakamura}, \citenamefont
  {Pleiter}, \citenamefont {Rakow}, \citenamefont {Schierholz},\ and\
  \citenamefont {St\"uben}}]{CSSM:2014knt}%
  \BibitemOpen
  \bibfield  {author} {\bibinfo {author} {\bibfnamefont {P.~E.}\ \bibnamefont
  {Shanahan}}, \bibinfo {author} {\bibfnamefont {A.~W.}\ \bibnamefont
  {Thomas}}, \bibinfo {author} {\bibfnamefont {R.~D.}\ \bibnamefont {Young}},
  \bibinfo {author} {\bibfnamefont {J.~M.}\ \bibnamefont {Zanotti}}, \bibinfo
  {author} {\bibfnamefont {R.}~\bibnamefont {Horsley}}, \bibinfo {author}
  {\bibfnamefont {Y.}~\bibnamefont {Nakamura}}, \bibinfo {author}
  {\bibfnamefont {D.}~\bibnamefont {Pleiter}}, \bibinfo {author} {\bibfnamefont
  {P.~E.~L.}\ \bibnamefont {Rakow}}, \bibinfo {author} {\bibfnamefont
  {G.}~\bibnamefont {Schierholz}},\ and\ \bibinfo {author} {\bibfnamefont
  {H.}~\bibnamefont {St\"uben}} (\bibinfo {collaboration} {CSSM,
  QCDSF/UKQCD}),\ }\bibfield  {title} {\bibinfo {title} {{Magnetic form factors
  of the octet baryons from lattice QCD and chiral extrapolation}},\ }\href
  {https://doi.org/10.1103/PhysRevD.89.074511} {\bibfield  {journal} {\bibinfo
  {journal} {Phys. Rev. D}\ }\textbf {\bibinfo {volume} {89}},\ \bibinfo
  {pages} {074511} (\bibinfo {year} {2014})},\ \Eprint
  {https://arxiv.org/abs/1401.5862} {arXiv:1401.5862 [hep-lat]} \BibitemShut
  {NoStop}%
\bibitem [{\citenamefont {Hall}\ \emph {et~al.}(2012)\citenamefont {Hall},
  \citenamefont {Leinweber},\ and\ \citenamefont {Young}}]{Hall:2012pk}%
  \BibitemOpen
  \bibfield  {author} {\bibinfo {author} {\bibfnamefont {J.~M.~M.}\
  \bibnamefont {Hall}}, \bibinfo {author} {\bibfnamefont {D.~B.}\ \bibnamefont
  {Leinweber}},\ and\ \bibinfo {author} {\bibfnamefont {R.~D.}\ \bibnamefont
  {Young}},\ }\bibfield  {title} {\bibinfo {title} {{Chiral extrapolations for
  nucleon magnetic moments}},\ }\href
  {https://doi.org/10.1103/PhysRevD.85.094502} {\bibfield  {journal} {\bibinfo
  {journal} {Phys. Rev. D}\ }\textbf {\bibinfo {volume} {85}},\ \bibinfo
  {pages} {094502} (\bibinfo {year} {2012})},\ \Eprint
  {https://arxiv.org/abs/1201.6114} {arXiv:1201.6114 [hep-lat]} \BibitemShut
  {NoStop}%
\bibitem [{\citenamefont {Hall}\ \emph {et~al.}(2014)\citenamefont {Hall},
  \citenamefont {Leinweber},\ and\ \citenamefont {Young}}]{Hall:2013dva}%
  \BibitemOpen
  \bibfield  {author} {\bibinfo {author} {\bibfnamefont {J.~M.~M.}\
  \bibnamefont {Hall}}, \bibinfo {author} {\bibfnamefont {D.~B.}\ \bibnamefont
  {Leinweber}},\ and\ \bibinfo {author} {\bibfnamefont {R.~D.}\ \bibnamefont
  {Young}},\ }\bibfield  {title} {\bibinfo {title} {{Finite-volume and partial
  quenching effects in the magnetic polarizability of the neutron}},\ }\href
  {https://doi.org/10.1103/PhysRevD.89.054511} {\bibfield  {journal} {\bibinfo
  {journal} {Phys. Rev. D}\ }\textbf {\bibinfo {volume} {89}},\ \bibinfo
  {pages} {054511} (\bibinfo {year} {2014})},\ \Eprint
  {https://arxiv.org/abs/1312.5781} {arXiv:1312.5781 [hep-lat]} \BibitemShut
  {NoStop}%
\bibitem [{\citenamefont {Bignell}\ \emph {et~al.}(2018)\citenamefont
  {Bignell}, \citenamefont {Hall}, \citenamefont {Kamleh}, \citenamefont
  {Leinweber},\ and\ \citenamefont {Burkardt}}]{Bignell:2018acn}%
  \BibitemOpen
  \bibfield  {author} {\bibinfo {author} {\bibfnamefont {R.}~\bibnamefont
  {Bignell}}, \bibinfo {author} {\bibfnamefont {J.}~\bibnamefont {Hall}},
  \bibinfo {author} {\bibfnamefont {W.}~\bibnamefont {Kamleh}}, \bibinfo
  {author} {\bibfnamefont {D.}~\bibnamefont {Leinweber}},\ and\ \bibinfo
  {author} {\bibfnamefont {M.}~\bibnamefont {Burkardt}},\ }\bibfield  {title}
  {\bibinfo {title} {{Neutron magnetic polarizability with Landau mode
  operators}},\ }\href {https://doi.org/10.1103/PhysRevD.98.034504} {\bibfield
  {journal} {\bibinfo  {journal} {Phys. Rev. D}\ }\textbf {\bibinfo {volume}
  {98}},\ \bibinfo {pages} {034504} (\bibinfo {year} {2018})},\ \Eprint
  {https://arxiv.org/abs/1804.06574} {arXiv:1804.06574 [hep-lat]} \BibitemShut
  {NoStop}%
\bibitem [{\citenamefont {Bignell}\ \emph {et~al.}(2020)\citenamefont
  {Bignell}, \citenamefont {Kamleh},\ and\ \citenamefont
  {Leinweber}}]{Bignell:2020xkf}%
  \BibitemOpen
  \bibfield  {author} {\bibinfo {author} {\bibfnamefont {R.}~\bibnamefont
  {Bignell}}, \bibinfo {author} {\bibfnamefont {W.}~\bibnamefont {Kamleh}},\
  and\ \bibinfo {author} {\bibfnamefont {D.}~\bibnamefont {Leinweber}},\
  }\bibfield  {title} {\bibinfo {title} {{Magnetic polarizability of the
  nucleon using a Laplacian mode projection}},\ }\href
  {https://doi.org/10.1103/PhysRevD.101.094502} {\bibfield  {journal} {\bibinfo
   {journal} {Phys. Rev. D}\ }\textbf {\bibinfo {volume} {101}},\ \bibinfo
  {pages} {094502} (\bibinfo {year} {2020})},\ \Eprint
  {https://arxiv.org/abs/2002.07915} {arXiv:2002.07915 [hep-lat]} \BibitemShut
  {NoStop}%
\bibitem [{\citenamefont {Edwards}\ \emph {et~al.}(2006)\citenamefont
  {Edwards}, \citenamefont {Fleming}, \citenamefont {Hagler}, \citenamefont
  {Negele}, \citenamefont {Orginos}, \citenamefont {Pochinsky}, \citenamefont
  {Renner}, \citenamefont {Richards},\ and\ \citenamefont
  {Schroers}}]{Edwards:2005ym}%
  \BibitemOpen
  \bibfield  {author} {\bibinfo {author} {\bibfnamefont {R.~G.}\ \bibnamefont
  {Edwards}}, \bibinfo {author} {\bibfnamefont {G.~T.}\ \bibnamefont
  {Fleming}}, \bibinfo {author} {\bibfnamefont {P.}~\bibnamefont {Hagler}},
  \bibinfo {author} {\bibfnamefont {J.~W.}\ \bibnamefont {Negele}}, \bibinfo
  {author} {\bibfnamefont {K.}~\bibnamefont {Orginos}}, \bibinfo {author}
  {\bibfnamefont {A.~V.}\ \bibnamefont {Pochinsky}}, \bibinfo {author}
  {\bibfnamefont {D.~B.}\ \bibnamefont {Renner}}, \bibinfo {author}
  {\bibfnamefont {D.~G.}\ \bibnamefont {Richards}},\ and\ \bibinfo {author}
  {\bibfnamefont {W.}~\bibnamefont {Schroers}} (\bibinfo {collaboration}
  {LHPC}),\ }\bibfield  {title} {\bibinfo {title} {{The Nucleon axial charge in
  full lattice QCD}},\ }\href {https://doi.org/10.1103/PhysRevLett.96.052001}
  {\bibfield  {journal} {\bibinfo  {journal} {Phys. Rev. Lett.}\ }\textbf
  {\bibinfo {volume} {96}},\ \bibinfo {pages} {052001} (\bibinfo {year}
  {2006})},\ \Eprint {https://arxiv.org/abs/hep-lat/0510062}
  {arXiv:hep-lat/0510062} \BibitemShut {NoStop}%
\bibitem [{\citenamefont {Horsley}\ \emph {et~al.}(2014)\citenamefont
  {Horsley}, \citenamefont {Nakamura}, \citenamefont {Nobile}, \citenamefont
  {Rakow}, \citenamefont {Schierholz},\ and\ \citenamefont
  {Zanotti}}]{Horsley:2013ayv}%
  \BibitemOpen
  \bibfield  {author} {\bibinfo {author} {\bibfnamefont {R.}~\bibnamefont
  {Horsley}}, \bibinfo {author} {\bibfnamefont {Y.}~\bibnamefont {Nakamura}},
  \bibinfo {author} {\bibfnamefont {A.}~\bibnamefont {Nobile}}, \bibinfo
  {author} {\bibfnamefont {P.~E.~L.}\ \bibnamefont {Rakow}}, \bibinfo {author}
  {\bibfnamefont {G.}~\bibnamefont {Schierholz}},\ and\ \bibinfo {author}
  {\bibfnamefont {J.~M.}\ \bibnamefont {Zanotti}},\ }\bibfield  {title}
  {\bibinfo {title} {{Nucleon axial charge and pion decay constant from
  two-flavor lattice QCD}},\ }\href
  {https://doi.org/10.1016/j.physletb.2014.03.002} {\bibfield  {journal}
  {\bibinfo  {journal} {Phys. Lett. B}\ }\textbf {\bibinfo {volume} {732}},\
  \bibinfo {pages} {41} (\bibinfo {year} {2014})},\ \Eprint
  {https://arxiv.org/abs/1302.2233} {arXiv:1302.2233 [hep-lat]} \BibitemShut
  {NoStop}%
\bibitem [{\citenamefont {Liang}\ \emph {et~al.}(2017)\citenamefont {Liang},
  \citenamefont {Yang}, \citenamefont {Liu}, \citenamefont {Alexandru},
  \citenamefont {Draper},\ and\ \citenamefont {Sufian}}]{Liang:2016fgy}%
  \BibitemOpen
  \bibfield  {author} {\bibinfo {author} {\bibfnamefont {J.}~\bibnamefont
  {Liang}}, \bibinfo {author} {\bibfnamefont {Y.-B.}\ \bibnamefont {Yang}},
  \bibinfo {author} {\bibfnamefont {K.-F.}\ \bibnamefont {Liu}}, \bibinfo
  {author} {\bibfnamefont {A.}~\bibnamefont {Alexandru}}, \bibinfo {author}
  {\bibfnamefont {T.}~\bibnamefont {Draper}},\ and\ \bibinfo {author}
  {\bibfnamefont {R.~S.}\ \bibnamefont {Sufian}},\ }\bibfield  {title}
  {\bibinfo {title} {{Lattice Calculation of Nucleon Isovector Axial Charge
  with Improved Currents}},\ }\href
  {https://doi.org/10.1103/PhysRevD.96.034519} {\bibfield  {journal} {\bibinfo
  {journal} {Phys. Rev. D}\ }\textbf {\bibinfo {volume} {96}},\ \bibinfo
  {pages} {034519} (\bibinfo {year} {2017})},\ \Eprint
  {https://arxiv.org/abs/1612.04388} {arXiv:1612.04388 [hep-lat]} \BibitemShut
  {NoStop}%
\bibitem [{\citenamefont {Berkowitz}\ \emph {et~al.}(2017)\citenamefont
  {Berkowitz}, \citenamefont {Brantley}, \citenamefont {Bouchard},
  \citenamefont {Chang}, \citenamefont {Clark}, \citenamefont {Garron},
  \citenamefont {Joo}, \citenamefont {Kurth}, \citenamefont {Monahan},
  \citenamefont {Monge-Camacho}, \citenamefont {Nicholson}, \citenamefont
  {Orginos}, \citenamefont {Rinaldi}, \citenamefont {Vranas},\ and\
  \citenamefont {Walker-Loud}}]{Berkowitz:2017gql}%
  \BibitemOpen
  \bibfield  {author} {\bibinfo {author} {\bibfnamefont {E.}~\bibnamefont
  {Berkowitz}}, \bibinfo {author} {\bibfnamefont {D.}~\bibnamefont {Brantley}},
  \bibinfo {author} {\bibfnamefont {C.}~\bibnamefont {Bouchard}}, \bibinfo
  {author} {\bibfnamefont {C.~C.}\ \bibnamefont {Chang}}, \bibinfo {author}
  {\bibfnamefont {M.~A.}\ \bibnamefont {Clark}}, \bibinfo {author}
  {\bibfnamefont {N.}~\bibnamefont {Garron}}, \bibinfo {author} {\bibfnamefont
  {B.}~\bibnamefont {Joo}}, \bibinfo {author} {\bibfnamefont {T.}~\bibnamefont
  {Kurth}}, \bibinfo {author} {\bibfnamefont {C.}~\bibnamefont {Monahan}},
  \bibinfo {author} {\bibfnamefont {H.}~\bibnamefont {Monge-Camacho}}, \bibinfo
  {author} {\bibfnamefont {A.}~\bibnamefont {Nicholson}}, \bibinfo {author}
  {\bibfnamefont {K.}~\bibnamefont {Orginos}}, \bibinfo {author} {\bibfnamefont
  {E.}~\bibnamefont {Rinaldi}}, \bibinfo {author} {\bibfnamefont
  {P.}~\bibnamefont {Vranas}},\ and\ \bibinfo {author} {\bibfnamefont
  {A.}~\bibnamefont {Walker-Loud}},\ }\href@noop {} {\bibinfo {title} {An
  accurate calculation of the nucleon axial charge with lattice {QCD}}}
  (\bibinfo {year} {2017}),\ \Eprint {https://arxiv.org/abs/1704.01114}
  {arXiv:1704.01114 [hep-lat]} \BibitemShut {NoStop}%
\bibitem [{\citenamefont {Lutz}\ \emph {et~al.}(2020)\citenamefont {Lutz},
  \citenamefont {Sauerwein},\ and\ \citenamefont {Timmermans}}]{Lutz:2020dfi}%
  \BibitemOpen
  \bibfield  {author} {\bibinfo {author} {\bibfnamefont {M.~F.~M.}\
  \bibnamefont {Lutz}}, \bibinfo {author} {\bibfnamefont {U.}~\bibnamefont
  {Sauerwein}},\ and\ \bibinfo {author} {\bibfnamefont {R.~G.~E.}\ \bibnamefont
  {Timmermans}},\ }\bibfield  {title} {\bibinfo {title} {{On the axial-vector
  form factor of the nucleon and chiral symmetry}},\ }\href
  {https://doi.org/10.1140/epjc/s10052-020-8417-5} {\bibfield  {journal}
  {\bibinfo  {journal} {Eur. Phys. J. C}\ }\textbf {\bibinfo {volume} {80}},\
  \bibinfo {pages} {844} (\bibinfo {year} {2020})},\ \Eprint
  {https://arxiv.org/abs/2003.10158} {arXiv:2003.10158 [hep-lat]} \BibitemShut
  {NoStop}%
\bibitem [{\citenamefont {Mahbub}\ \emph {et~al.}(2012)\citenamefont {Mahbub},
  \citenamefont {Kamleh}, \citenamefont {Leinweber}, \citenamefont {Moran},\
  and\ \citenamefont {Williams}}]{Mahbub:2010rm}%
  \BibitemOpen
  \bibfield  {author} {\bibinfo {author} {\bibfnamefont {M.~S.}\ \bibnamefont
  {Mahbub}}, \bibinfo {author} {\bibfnamefont {W.}~\bibnamefont {Kamleh}},
  \bibinfo {author} {\bibfnamefont {D.~B.}\ \bibnamefont {Leinweber}}, \bibinfo
  {author} {\bibfnamefont {P.~J.}\ \bibnamefont {Moran}},\ and\ \bibinfo
  {author} {\bibfnamefont {A.~G.}\ \bibnamefont {Williams}} (\bibinfo
  {collaboration} {CSSM Lattice}),\ }\bibfield  {title} {\bibinfo {title}
  {{Roper Resonance in 2+1 Flavor QCD}},\ }\href
  {https://doi.org/10.1016/j.physletb.2011.12.048} {\bibfield  {journal}
  {\bibinfo  {journal} {Phys. Lett. B}\ }\textbf {\bibinfo {volume} {707}},\
  \bibinfo {pages} {389} (\bibinfo {year} {2012})},\ \Eprint
  {https://arxiv.org/abs/1011.5724} {arXiv:1011.5724 [hep-lat]} \BibitemShut
  {NoStop}%
\bibitem [{\citenamefont {Edwards}\ \emph {et~al.}(2011)\citenamefont
  {Edwards}, \citenamefont {Dudek}, \citenamefont {Richards},\ and\
  \citenamefont {Wallace}}]{Edwards:2011jj}%
  \BibitemOpen
  \bibfield  {author} {\bibinfo {author} {\bibfnamefont {R.~G.}\ \bibnamefont
  {Edwards}}, \bibinfo {author} {\bibfnamefont {J.~J.}\ \bibnamefont {Dudek}},
  \bibinfo {author} {\bibfnamefont {D.~G.}\ \bibnamefont {Richards}},\ and\
  \bibinfo {author} {\bibfnamefont {S.~J.}\ \bibnamefont {Wallace}},\
  }\bibfield  {title} {\bibinfo {title} {{Excited state baryon spectroscopy
  from lattice QCD}},\ }\href {https://doi.org/10.1103/PhysRevD.84.074508}
  {\bibfield  {journal} {\bibinfo  {journal} {Phys. Rev. D}\ }\textbf {\bibinfo
  {volume} {84}},\ \bibinfo {pages} {074508} (\bibinfo {year} {2011})},\
  \Eprint {https://arxiv.org/abs/1104.5152} {arXiv:1104.5152 [hep-ph]}
  \BibitemShut {NoStop}%
\bibitem [{\citenamefont {Lang}\ and\ \citenamefont
  {Verduci}(2013)}]{Lang:2012db}%
  \BibitemOpen
  \bibfield  {author} {\bibinfo {author} {\bibfnamefont {C.~B.}\ \bibnamefont
  {Lang}}\ and\ \bibinfo {author} {\bibfnamefont {V.}~\bibnamefont {Verduci}},\
  }\bibfield  {title} {\bibinfo {title} {{Scattering in the \ensuremath{\pi}N
  negative parity channel in lattice QCD}},\ }\href
  {https://doi.org/10.1103/PhysRevD.87.054502} {\bibfield  {journal} {\bibinfo
  {journal} {Phys. Rev. D}\ }\textbf {\bibinfo {volume} {87}},\ \bibinfo
  {pages} {054502} (\bibinfo {year} {2013})},\ \Eprint
  {https://arxiv.org/abs/1212.5055} {arXiv:1212.5055 [hep-lat]} \BibitemShut
  {NoStop}%
\bibitem [{\citenamefont {Mahbub}\ \emph {et~al.}(2013)\citenamefont {Mahbub},
  \citenamefont {Kamleh}, \citenamefont {Leinweber}, \citenamefont {Moran},\
  and\ \citenamefont {Williams}}]{Mahbub:2013ala}%
  \BibitemOpen
  \bibfield  {author} {\bibinfo {author} {\bibfnamefont {M.~S.}\ \bibnamefont
  {Mahbub}}, \bibinfo {author} {\bibfnamefont {W.}~\bibnamefont {Kamleh}},
  \bibinfo {author} {\bibfnamefont {D.~B.}\ \bibnamefont {Leinweber}}, \bibinfo
  {author} {\bibfnamefont {P.~J.}\ \bibnamefont {Moran}},\ and\ \bibinfo
  {author} {\bibfnamefont {A.~G.}\ \bibnamefont {Williams}},\ }\bibfield
  {title} {\bibinfo {title} {{Structure and Flow of the Nucleon Eigenstates in
  Lattice QCD}},\ }\href {https://doi.org/10.1103/PhysRevD.87.094506}
  {\bibfield  {journal} {\bibinfo  {journal} {Phys. Rev. D}\ }\textbf {\bibinfo
  {volume} {87}},\ \bibinfo {pages} {094506} (\bibinfo {year} {2013})},\
  \Eprint {https://arxiv.org/abs/1302.2987} {arXiv:1302.2987 [hep-lat]}
  \BibitemShut {NoStop}%
\bibitem [{\citenamefont {Engel}\ \emph {et~al.}(2013)\citenamefont {Engel},
  \citenamefont {Lang}, \citenamefont {Mohler},\ and\ \citenamefont
  {Sch\"afer}}]{Engel:2013ig}%
  \BibitemOpen
  \bibfield  {author} {\bibinfo {author} {\bibfnamefont {G.~P.}\ \bibnamefont
  {Engel}}, \bibinfo {author} {\bibfnamefont {C.~B.}\ \bibnamefont {Lang}},
  \bibinfo {author} {\bibfnamefont {D.}~\bibnamefont {Mohler}},\ and\ \bibinfo
  {author} {\bibfnamefont {A.}~\bibnamefont {Sch\"afer}} (\bibinfo
  {collaboration} {BGR}),\ }\bibfield  {title} {\bibinfo {title} {{QCD with Two
  Light Dynamical Chirally Improved Quarks: Baryons}},\ }\href
  {https://doi.org/10.1103/PhysRevD.87.074504} {\bibfield  {journal} {\bibinfo
  {journal} {Phys. Rev. D}\ }\textbf {\bibinfo {volume} {87}},\ \bibinfo
  {pages} {074504} (\bibinfo {year} {2013})},\ \Eprint
  {https://arxiv.org/abs/1301.4318} {arXiv:1301.4318 [hep-lat]} \BibitemShut
  {NoStop}%
\bibitem [{\citenamefont {Hall}\ \emph
  {et~al.}(2013{\natexlab{c}})\citenamefont {Hall}, \citenamefont {Hsu},
  \citenamefont {Leinweber}, \citenamefont {Thomas},\ and\ \citenamefont
  {Young}}]{Hall:2013qba}%
  \BibitemOpen
  \bibfield  {author} {\bibinfo {author} {\bibfnamefont {J.~M.~M.}\
  \bibnamefont {Hall}}, \bibinfo {author} {\bibfnamefont {A.~C.~P.}\
  \bibnamefont {Hsu}}, \bibinfo {author} {\bibfnamefont {D.~B.}\ \bibnamefont
  {Leinweber}}, \bibinfo {author} {\bibfnamefont {A.~W.}\ \bibnamefont
  {Thomas}},\ and\ \bibinfo {author} {\bibfnamefont {R.~D.}\ \bibnamefont
  {Young}},\ }\bibfield  {title} {\bibinfo {title} {{Finite-volume matrix
  Hamiltonian model for a $\Delta \to N\pi$ system}},\ }\href
  {https://doi.org/10.1103/PhysRevD.87.094510} {\bibfield  {journal} {\bibinfo
  {journal} {Phys. Rev. D}\ }\textbf {\bibinfo {volume} {87}},\ \bibinfo
  {pages} {094510} (\bibinfo {year} {2013}{\natexlab{c}})},\ \Eprint
  {https://arxiv.org/abs/1303.4157} {arXiv:1303.4157 [hep-lat]} \BibitemShut
  {NoStop}%
\bibitem [{\citenamefont {Roberts}\ \emph {et~al.}(2013)\citenamefont
  {Roberts}, \citenamefont {Kamleh},\ and\ \citenamefont
  {Leinweber}}]{Roberts:2013ipa}%
  \BibitemOpen
  \bibfield  {author} {\bibinfo {author} {\bibfnamefont {D.~S.}\ \bibnamefont
  {Roberts}}, \bibinfo {author} {\bibfnamefont {W.}~\bibnamefont {Kamleh}},\
  and\ \bibinfo {author} {\bibfnamefont {D.~B.}\ \bibnamefont {Leinweber}},\
  }\bibfield  {title} {\bibinfo {title} {{Wave Function of the Roper from
  Lattice QCD}},\ }\href {https://doi.org/10.1016/j.physletb.2013.06.056}
  {\bibfield  {journal} {\bibinfo  {journal} {Phys. Lett. B}\ }\textbf
  {\bibinfo {volume} {725}},\ \bibinfo {pages} {164} (\bibinfo {year}
  {2013})},\ \Eprint {https://arxiv.org/abs/1304.0325} {arXiv:1304.0325
  [hep-lat]} \BibitemShut {NoStop}%
\bibitem [{\citenamefont {Roberts}\ \emph {et~al.}(2014)\citenamefont
  {Roberts}, \citenamefont {Kamleh},\ and\ \citenamefont
  {Leinweber}}]{Roberts:2013oea}%
  \BibitemOpen
  \bibfield  {author} {\bibinfo {author} {\bibfnamefont {D.~S.}\ \bibnamefont
  {Roberts}}, \bibinfo {author} {\bibfnamefont {W.}~\bibnamefont {Kamleh}},\
  and\ \bibinfo {author} {\bibfnamefont {D.~B.}\ \bibnamefont {Leinweber}},\
  }\bibfield  {title} {\bibinfo {title} {{Nucleon Excited State Wave Functions
  from Lattice QCD}},\ }\href {https://doi.org/10.1103/PhysRevD.89.074501}
  {\bibfield  {journal} {\bibinfo  {journal} {Phys. Rev. D}\ }\textbf {\bibinfo
  {volume} {89}},\ \bibinfo {pages} {074501} (\bibinfo {year} {2014})},\
  \Eprint {https://arxiv.org/abs/1311.6626} {arXiv:1311.6626 [hep-lat]}
  \BibitemShut {NoStop}%
\bibitem [{\citenamefont {Alexandrou}\ \emph {et~al.}(2015)\citenamefont
  {Alexandrou}, \citenamefont {Leontiou}, \citenamefont {Papanicolas},\ and\
  \citenamefont {Stiliaris}}]{Alexandrou:2014mka}%
  \BibitemOpen
  \bibfield  {author} {\bibinfo {author} {\bibfnamefont {C.}~\bibnamefont
  {Alexandrou}}, \bibinfo {author} {\bibfnamefont {T.}~\bibnamefont
  {Leontiou}}, \bibinfo {author} {\bibfnamefont {C.~N.}\ \bibnamefont
  {Papanicolas}},\ and\ \bibinfo {author} {\bibfnamefont {E.}~\bibnamefont
  {Stiliaris}},\ }\bibfield  {title} {\bibinfo {title} {{Novel analysis method
  for excited states in lattice QCD: The nucleon case}},\ }\href
  {https://doi.org/10.1103/PhysRevD.91.014506} {\bibfield  {journal} {\bibinfo
  {journal} {Phys. Rev. D}\ }\textbf {\bibinfo {volume} {91}},\ \bibinfo
  {pages} {014506} (\bibinfo {year} {2015})},\ \Eprint
  {https://arxiv.org/abs/1411.6765} {arXiv:1411.6765 [hep-lat]} \BibitemShut
  {NoStop}%
\bibitem [{\citenamefont {Liu}\ \emph {et~al.}(2016)\citenamefont {Liu},
  \citenamefont {Kamleh}, \citenamefont {Leinweber}, \citenamefont {Stokes},
  \citenamefont {Thomas},\ and\ \citenamefont {Wu}}]{Liu:2015ktc}%
  \BibitemOpen
  \bibfield  {author} {\bibinfo {author} {\bibfnamefont {Z.-W.}\ \bibnamefont
  {Liu}}, \bibinfo {author} {\bibfnamefont {W.}~\bibnamefont {Kamleh}},
  \bibinfo {author} {\bibfnamefont {D.~B.}\ \bibnamefont {Leinweber}}, \bibinfo
  {author} {\bibfnamefont {F.~M.}\ \bibnamefont {Stokes}}, \bibinfo {author}
  {\bibfnamefont {A.~W.}\ \bibnamefont {Thomas}},\ and\ \bibinfo {author}
  {\bibfnamefont {J.-J.}\ \bibnamefont {Wu}},\ }\bibfield  {title} {\bibinfo
  {title} {{Hamiltonian effective field theory study of the
  $\mathbf{N^*(1535)}$ resonance in lattice QCD}},\ }\href
  {https://doi.org/10.1103/PhysRevLett.116.082004} {\bibfield  {journal}
  {\bibinfo  {journal} {Phys. Rev. Lett.}\ }\textbf {\bibinfo {volume} {116}},\
  \bibinfo {pages} {082004} (\bibinfo {year} {2016})},\ \Eprint
  {https://arxiv.org/abs/1512.00140} {arXiv:1512.00140 [hep-lat]} \BibitemShut
  {NoStop}%
\bibitem [{\citenamefont {Kiratidis}\ \emph {et~al.}(2015)\citenamefont
  {Kiratidis}, \citenamefont {Kamleh}, \citenamefont {Leinweber},\ and\
  \citenamefont {Owen}}]{Kiratidis:2015vpa}%
  \BibitemOpen
  \bibfield  {author} {\bibinfo {author} {\bibfnamefont {A.~L.}\ \bibnamefont
  {Kiratidis}}, \bibinfo {author} {\bibfnamefont {W.}~\bibnamefont {Kamleh}},
  \bibinfo {author} {\bibfnamefont {D.~B.}\ \bibnamefont {Leinweber}},\ and\
  \bibinfo {author} {\bibfnamefont {B.~J.}\ \bibnamefont {Owen}},\ }\bibfield
  {title} {\bibinfo {title} {{Lattice baryon spectroscopy with multi-particle
  interpolators}},\ }\href {https://doi.org/10.1103/PhysRevD.91.094509}
  {\bibfield  {journal} {\bibinfo  {journal} {Phys. Rev. D}\ }\textbf {\bibinfo
  {volume} {91}},\ \bibinfo {pages} {094509} (\bibinfo {year} {2015})},\
  \Eprint {https://arxiv.org/abs/1501.07667} {arXiv:1501.07667 [hep-lat]}
  \BibitemShut {NoStop}%
\bibitem [{\citenamefont {Leinweber}\ \emph {et~al.}(2016)\citenamefont
  {Leinweber}, \citenamefont {Kamleh}, \citenamefont {Kiratidis}, \citenamefont
  {Liu}, \citenamefont {Mahbub}, \citenamefont {Roberts}, \citenamefont
  {Stokes}, \citenamefont {Thomas},\ and\ \citenamefont
  {Wu}}]{Leinweber:2015kyz}%
  \BibitemOpen
  \bibfield  {author} {\bibinfo {author} {\bibfnamefont {D.}~\bibnamefont
  {Leinweber}}, \bibinfo {author} {\bibfnamefont {W.}~\bibnamefont {Kamleh}},
  \bibinfo {author} {\bibfnamefont {A.}~\bibnamefont {Kiratidis}}, \bibinfo
  {author} {\bibfnamefont {Z.-W.}\ \bibnamefont {Liu}}, \bibinfo {author}
  {\bibfnamefont {S.}~\bibnamefont {Mahbub}}, \bibinfo {author} {\bibfnamefont
  {D.}~\bibnamefont {Roberts}}, \bibinfo {author} {\bibfnamefont
  {F.}~\bibnamefont {Stokes}}, \bibinfo {author} {\bibfnamefont {A.~W.}\
  \bibnamefont {Thomas}},\ and\ \bibinfo {author} {\bibfnamefont
  {J.}~\bibnamefont {Wu}},\ }\bibfield  {title} {\bibinfo {title} {{N*
  Spectroscopy from Lattice QCD: The Roper Explained}},\ }\href
  {https://doi.org/10.7566/JPSCP.10.010011} {\bibfield  {journal} {\bibinfo
  {journal} {JPS Conf. Proc.}\ }\textbf {\bibinfo {volume} {10}},\ \bibinfo
  {pages} {010011} (\bibinfo {year} {2016})},\ \Eprint
  {https://arxiv.org/abs/1511.09146} {arXiv:1511.09146 [hep-lat]} \BibitemShut
  {NoStop}%
\bibitem [{\citenamefont {Stokes}\ \emph {et~al.}(2015)\citenamefont {Stokes},
  \citenamefont {Kamleh}, \citenamefont {Leinweber}, \citenamefont {Mahbub},
  \citenamefont {Menadue},\ and\ \citenamefont {Owen}}]{Stokes:2013fgw}%
  \BibitemOpen
  \bibfield  {author} {\bibinfo {author} {\bibfnamefont {F.~M.}\ \bibnamefont
  {Stokes}}, \bibinfo {author} {\bibfnamefont {W.}~\bibnamefont {Kamleh}},
  \bibinfo {author} {\bibfnamefont {D.~B.}\ \bibnamefont {Leinweber}}, \bibinfo
  {author} {\bibfnamefont {M.~S.}\ \bibnamefont {Mahbub}}, \bibinfo {author}
  {\bibfnamefont {B.~J.}\ \bibnamefont {Menadue}},\ and\ \bibinfo {author}
  {\bibfnamefont {B.~J.}\ \bibnamefont {Owen}},\ }\bibfield  {title} {\bibinfo
  {title} {{Parity-expanded variational analysis for nonzero momentum}},\
  }\href {https://doi.org/10.1103/PhysRevD.92.114506} {\bibfield  {journal}
  {\bibinfo  {journal} {Phys. Rev. D}\ }\textbf {\bibinfo {volume} {92}},\
  \bibinfo {pages} {114506} (\bibinfo {year} {2015})},\ \Eprint
  {https://arxiv.org/abs/1302.4152} {arXiv:1302.4152 [hep-lat]} \BibitemShut
  {NoStop}%
\bibitem [{\citenamefont {Kiratidis}\ \emph {et~al.}(2017)\citenamefont
  {Kiratidis}, \citenamefont {Kamleh}, \citenamefont {Leinweber}, \citenamefont
  {Liu}, \citenamefont {Stokes},\ and\ \citenamefont
  {Thomas}}]{Kiratidis:2016hda}%
  \BibitemOpen
  \bibfield  {author} {\bibinfo {author} {\bibfnamefont {A.~L.}\ \bibnamefont
  {Kiratidis}}, \bibinfo {author} {\bibfnamefont {W.}~\bibnamefont {Kamleh}},
  \bibinfo {author} {\bibfnamefont {D.~B.}\ \bibnamefont {Leinweber}}, \bibinfo
  {author} {\bibfnamefont {Z.-W.}\ \bibnamefont {Liu}}, \bibinfo {author}
  {\bibfnamefont {F.~M.}\ \bibnamefont {Stokes}},\ and\ \bibinfo {author}
  {\bibfnamefont {A.~W.}\ \bibnamefont {Thomas}},\ }\bibfield  {title}
  {\bibinfo {title} {{Search for low-lying lattice QCD eigenstates in the Roper
  regime}},\ }\href {https://doi.org/10.1103/PhysRevD.95.074507} {\bibfield
  {journal} {\bibinfo  {journal} {Phys. Rev. D}\ }\textbf {\bibinfo {volume}
  {95}},\ \bibinfo {pages} {074507} (\bibinfo {year} {2017})},\ \Eprint
  {https://arxiv.org/abs/1608.03051} {arXiv:1608.03051 [hep-lat]} \BibitemShut
  {NoStop}%
\bibitem [{\citenamefont {Liu}\ \emph {et~al.}(2017)\citenamefont {Liu},
  \citenamefont {Kamleh}, \citenamefont {Leinweber}, \citenamefont {Stokes},
  \citenamefont {Thomas},\ and\ \citenamefont {Wu}}]{Liu:2016uzk}%
  \BibitemOpen
  \bibfield  {author} {\bibinfo {author} {\bibfnamefont {Z.-W.}\ \bibnamefont
  {Liu}}, \bibinfo {author} {\bibfnamefont {W.}~\bibnamefont {Kamleh}},
  \bibinfo {author} {\bibfnamefont {D.~B.}\ \bibnamefont {Leinweber}}, \bibinfo
  {author} {\bibfnamefont {F.~M.}\ \bibnamefont {Stokes}}, \bibinfo {author}
  {\bibfnamefont {A.~W.}\ \bibnamefont {Thomas}},\ and\ \bibinfo {author}
  {\bibfnamefont {J.-J.}\ \bibnamefont {Wu}},\ }\bibfield  {title} {\bibinfo
  {title} {{Hamiltonian effective field theory study of the
  $\mathbf{N^*(1440)}$ resonance in lattice QCD}},\ }\href
  {https://doi.org/10.1103/PhysRevD.95.034034} {\bibfield  {journal} {\bibinfo
  {journal} {Phys. Rev. D}\ }\textbf {\bibinfo {volume} {95}},\ \bibinfo
  {pages} {034034} (\bibinfo {year} {2017})},\ \Eprint
  {https://arxiv.org/abs/1607.04536} {arXiv:1607.04536 [nucl-th]} \BibitemShut
  {NoStop}%
\bibitem [{\citenamefont {Wu}\ \emph {et~al.}(2017)\citenamefont {Wu},
  \citenamefont {Kamano}, \citenamefont {Lee}, \citenamefont {Leinweber},\ and\
  \citenamefont {Thomas}}]{Wu:2016ixr}%
  \BibitemOpen
  \bibfield  {author} {\bibinfo {author} {\bibfnamefont {J.-J.}\ \bibnamefont
  {Wu}}, \bibinfo {author} {\bibfnamefont {H.}~\bibnamefont {Kamano}}, \bibinfo
  {author} {\bibfnamefont {T.~S.~H.}\ \bibnamefont {Lee}}, \bibinfo {author}
  {\bibfnamefont {D.~B.}\ \bibnamefont {Leinweber}},\ and\ \bibinfo {author}
  {\bibfnamefont {A.~W.}\ \bibnamefont {Thomas}},\ }\bibfield  {title}
  {\bibinfo {title} {{Nucleon resonance structure in the finite volume of
  lattice QCD}},\ }\href {https://doi.org/10.1103/PhysRevD.95.114507}
  {\bibfield  {journal} {\bibinfo  {journal} {Phys. Rev. D}\ }\textbf {\bibinfo
  {volume} {95}},\ \bibinfo {pages} {114507} (\bibinfo {year} {2017})},\
  \Eprint {https://arxiv.org/abs/1611.05970} {arXiv:1611.05970 [hep-lat]}
  \BibitemShut {NoStop}%
\bibitem [{\citenamefont {Lang}\ \emph {et~al.}(2017)\citenamefont {Lang},
  \citenamefont {Leskovec}, \citenamefont {Padmanath},\ and\ \citenamefont
  {Prelovsek}}]{Lang:2016hnn}%
  \BibitemOpen
  \bibfield  {author} {\bibinfo {author} {\bibfnamefont {C.~B.}\ \bibnamefont
  {Lang}}, \bibinfo {author} {\bibfnamefont {L.}~\bibnamefont {Leskovec}},
  \bibinfo {author} {\bibfnamefont {M.}~\bibnamefont {Padmanath}},\ and\
  \bibinfo {author} {\bibfnamefont {S.}~\bibnamefont {Prelovsek}},\ }\bibfield
  {title} {\bibinfo {title} {{Pion-nucleon scattering in the Roper channel from
  lattice QCD}},\ }\href {https://doi.org/10.1103/PhysRevD.95.014510}
  {\bibfield  {journal} {\bibinfo  {journal} {Phys. Rev. D}\ }\textbf {\bibinfo
  {volume} {95}},\ \bibinfo {pages} {014510} (\bibinfo {year} {2017})},\
  \Eprint {https://arxiv.org/abs/1610.01422} {arXiv:1610.01422 [hep-lat]}
  \BibitemShut {NoStop}%
\bibitem [{\citenamefont {Wu}\ \emph {et~al.}(2018)\citenamefont {Wu},
  \citenamefont {Leinweber}, \citenamefont {Liu},\ and\ \citenamefont
  {Thomas}}]{Wu:2017qve}%
  \BibitemOpen
  \bibfield  {author} {\bibinfo {author} {\bibfnamefont {J.-j.}\ \bibnamefont
  {Wu}}, \bibinfo {author} {\bibfnamefont {D.~B.}\ \bibnamefont {Leinweber}},
  \bibinfo {author} {\bibfnamefont {Z.-w.}\ \bibnamefont {Liu}},\ and\ \bibinfo
  {author} {\bibfnamefont {A.~W.}\ \bibnamefont {Thomas}},\ }\bibfield  {title}
  {\bibinfo {title} {{Structure of the Roper Resonance from Lattice QCD
  Constraints}},\ }\href {https://doi.org/10.1103/PhysRevD.97.094509}
  {\bibfield  {journal} {\bibinfo  {journal} {Phys. Rev. D}\ }\textbf {\bibinfo
  {volume} {97}},\ \bibinfo {pages} {094509} (\bibinfo {year} {2018})},\
  \Eprint {https://arxiv.org/abs/1703.10715} {arXiv:1703.10715 [nucl-th]}
  \BibitemShut {NoStop}%
\bibitem [{\citenamefont {Andersen}\ \emph {et~al.}(2018)\citenamefont
  {Andersen}, \citenamefont {Bulava}, \citenamefont {H\"orz},\ and\
  \citenamefont {Morningstar}}]{Andersen:2017una}%
  \BibitemOpen
  \bibfield  {author} {\bibinfo {author} {\bibfnamefont {C.~W.}\ \bibnamefont
  {Andersen}}, \bibinfo {author} {\bibfnamefont {J.}~\bibnamefont {Bulava}},
  \bibinfo {author} {\bibfnamefont {B.}~\bibnamefont {H\"orz}},\ and\ \bibinfo
  {author} {\bibfnamefont {C.}~\bibnamefont {Morningstar}},\ }\bibfield
  {title} {\bibinfo {title} {{Elastic $I=3/2 p$-wave nucleon-pion scattering
  amplitude and the $\Delta$(1232) resonance from N$_f$=2+1 lattice QCD}},\
  }\href {https://doi.org/10.1103/PhysRevD.97.014506} {\bibfield  {journal}
  {\bibinfo  {journal} {Phys. Rev. D}\ }\textbf {\bibinfo {volume} {97}},\
  \bibinfo {pages} {014506} (\bibinfo {year} {2018})},\ \Eprint
  {https://arxiv.org/abs/1710.01557} {arXiv:1710.01557 [hep-lat]} \BibitemShut
  {NoStop}%
\bibitem [{\citenamefont {Stokes}\ \emph {et~al.}(2020)\citenamefont {Stokes},
  \citenamefont {Kamleh},\ and\ \citenamefont {Leinweber}}]{Stokes:2019zdd}%
  \BibitemOpen
  \bibfield  {author} {\bibinfo {author} {\bibfnamefont {F.~M.}\ \bibnamefont
  {Stokes}}, \bibinfo {author} {\bibfnamefont {W.}~\bibnamefont {Kamleh}},\
  and\ \bibinfo {author} {\bibfnamefont {D.~B.}\ \bibnamefont {Leinweber}},\
  }\bibfield  {title} {\bibinfo {title} {{Elastic Form Factors of Nucleon
  Excitations in Lattice QCD}},\ }\href
  {https://doi.org/10.1103/PhysRevD.102.014507} {\bibfield  {journal} {\bibinfo
   {journal} {Phys. Rev. D}\ }\textbf {\bibinfo {volume} {102}},\ \bibinfo
  {pages} {014507} (\bibinfo {year} {2020})},\ \Eprint
  {https://arxiv.org/abs/1907.00177} {arXiv:1907.00177 [hep-lat]} \BibitemShut
  {NoStop}%
\bibitem [{\citenamefont {Stokes}\ \emph {et~al.}(2019)\citenamefont {Stokes},
  \citenamefont {Kamleh},\ and\ \citenamefont {Leinweber}}]{Stokes:2019yiz}%
  \BibitemOpen
  \bibfield  {author} {\bibinfo {author} {\bibfnamefont {F.~M.}\ \bibnamefont
  {Stokes}}, \bibinfo {author} {\bibfnamefont {W.}~\bibnamefont {Kamleh}},\
  and\ \bibinfo {author} {\bibfnamefont {D.~B.}\ \bibnamefont {Leinweber}},\
  }\bibfield  {title} {\bibinfo {title} {{Structure and transitions of nucleon
  excitations via parity-expanded variational analysis}},\ }\href
  {https://doi.org/10.22323/1.363.0182} {\bibfield  {journal} {\bibinfo
  {journal} {PoS}\ }\textbf {\bibinfo {volume} {LATTICE2019}},\ \bibinfo
  {pages} {182} (\bibinfo {year} {2019})},\ \Eprint
  {https://arxiv.org/abs/2001.07919} {arXiv:2001.07919 [hep-lat]} \BibitemShut
  {NoStop}%
\bibitem [{\citenamefont {Virgili}\ \emph {et~al.}(2020)\citenamefont
  {Virgili}, \citenamefont {Kamleh},\ and\ \citenamefont
  {Leinweber}}]{Virgili:2019shg}%
  \BibitemOpen
  \bibfield  {author} {\bibinfo {author} {\bibfnamefont {A.}~\bibnamefont
  {Virgili}}, \bibinfo {author} {\bibfnamefont {W.}~\bibnamefont {Kamleh}},\
  and\ \bibinfo {author} {\bibfnamefont {D.}~\bibnamefont {Leinweber}},\
  }\bibfield  {title} {\bibinfo {title} {{Role of chiral symmetry in the
  nucleon excitation spectrum}},\ }\href
  {https://doi.org/10.1103/PhysRevD.101.074504} {\bibfield  {journal} {\bibinfo
   {journal} {Phys. Rev. D}\ }\textbf {\bibinfo {volume} {101}},\ \bibinfo
  {pages} {074504} (\bibinfo {year} {2020})},\ \Eprint
  {https://arxiv.org/abs/1910.13782} {arXiv:1910.13782 [hep-lat]} \BibitemShut
  {NoStop}%
\bibitem [{\citenamefont {Khan}\ \emph {et~al.}(2021)\citenamefont {Khan},
  \citenamefont {Richards},\ and\ \citenamefont {Winter}}]{Khan:2020ahz}%
  \BibitemOpen
  \bibfield  {author} {\bibinfo {author} {\bibfnamefont {T.}~\bibnamefont
  {Khan}}, \bibinfo {author} {\bibfnamefont {D.}~\bibnamefont {Richards}},\
  and\ \bibinfo {author} {\bibfnamefont {F.}~\bibnamefont {Winter}},\
  }\bibfield  {title} {\bibinfo {title} {{Positive-parity baryon spectrum and
  the role of hybrid baryons}},\ }\href
  {https://doi.org/10.1103/PhysRevD.104.034503} {\bibfield  {journal} {\bibinfo
   {journal} {Phys. Rev. D}\ }\textbf {\bibinfo {volume} {104}},\ \bibinfo
  {pages} {034503} (\bibinfo {year} {2021})},\ \Eprint
  {https://arxiv.org/abs/2010.03052} {arXiv:2010.03052 [hep-lat]} \BibitemShut
  {NoStop}%
\bibitem [{\citenamefont {Morningstar}\ \emph {et~al.}(2022)\citenamefont
  {Morningstar}, \citenamefont {Bulava}, \citenamefont {Hanlon}, \citenamefont
  {H\"orz}, \citenamefont {Mohler}, \citenamefont {Nicholson}, \citenamefont
  {Skinner},\ and\ \citenamefont {Walker-Loud}}]{Morningstar:2021ewk}%
  \BibitemOpen
  \bibfield  {author} {\bibinfo {author} {\bibfnamefont {C.}~\bibnamefont
  {Morningstar}}, \bibinfo {author} {\bibfnamefont {J.}~\bibnamefont {Bulava}},
  \bibinfo {author} {\bibfnamefont {A.~D.}\ \bibnamefont {Hanlon}}, \bibinfo
  {author} {\bibfnamefont {B.}~\bibnamefont {H\"orz}}, \bibinfo {author}
  {\bibfnamefont {D.}~\bibnamefont {Mohler}}, \bibinfo {author} {\bibfnamefont
  {A.}~\bibnamefont {Nicholson}}, \bibinfo {author} {\bibfnamefont
  {S.}~\bibnamefont {Skinner}},\ and\ \bibinfo {author} {\bibfnamefont
  {A.}~\bibnamefont {Walker-Loud}},\ }\bibfield  {title} {\bibinfo {title}
  {{Progress on Meson-Baryon Scattering}},\ }\href
  {https://doi.org/10.22323/1.396.0170} {\bibfield  {journal} {\bibinfo
  {journal} {PoS}\ }\textbf {\bibinfo {volume} {LATTICE2021}},\ \bibinfo
  {pages} {170} (\bibinfo {year} {2022})},\ \Eprint
  {https://arxiv.org/abs/2111.07755} {arXiv:2111.07755 [hep-lat]} \BibitemShut
  {NoStop}%
\bibitem [{\citenamefont {Abell}\ \emph {et~al.}(2022)\citenamefont {Abell},
  \citenamefont {Leinweber}, \citenamefont {Thomas},\ and\ \citenamefont
  {Wu}}]{Abell:2021awi}%
  \BibitemOpen
  \bibfield  {author} {\bibinfo {author} {\bibfnamefont {C.~D.}\ \bibnamefont
  {Abell}}, \bibinfo {author} {\bibfnamefont {D.~B.}\ \bibnamefont
  {Leinweber}}, \bibinfo {author} {\bibfnamefont {A.~W.}\ \bibnamefont
  {Thomas}},\ and\ \bibinfo {author} {\bibfnamefont {J.-J.}\ \bibnamefont
  {Wu}},\ }\bibfield  {title} {\bibinfo {title} {{Regularization in
  nonperturbative extensions of effective field theory}},\ }\href
  {https://doi.org/10.1103/PhysRevD.106.034506} {\bibfield  {journal} {\bibinfo
   {journal} {Phys. Rev. D}\ }\textbf {\bibinfo {volume} {106}},\ \bibinfo
  {pages} {034506} (\bibinfo {year} {2022})},\ \Eprint
  {https://arxiv.org/abs/2110.14113} {arXiv:2110.14113 [hep-lat]} \BibitemShut
  {NoStop}%
\bibitem [{\citenamefont {Bulava}\ \emph
  {et~al.}(2023{\natexlab{a}})\citenamefont {Bulava}, \citenamefont {Hanlon},
  \citenamefont {H\"orz}, \citenamefont {Morningstar}, \citenamefont
  {Nicholson}, \citenamefont {Romero-L\'opez}, \citenamefont {Skinner},
  \citenamefont {Vranas},\ and\ \citenamefont {Walker-Loud}}]{Bulava:2022vpq}%
  \BibitemOpen
  \bibfield  {author} {\bibinfo {author} {\bibfnamefont {J.}~\bibnamefont
  {Bulava}}, \bibinfo {author} {\bibfnamefont {A.~D.}\ \bibnamefont {Hanlon}},
  \bibinfo {author} {\bibfnamefont {B.}~\bibnamefont {H\"orz}}, \bibinfo
  {author} {\bibfnamefont {C.}~\bibnamefont {Morningstar}}, \bibinfo {author}
  {\bibfnamefont {A.}~\bibnamefont {Nicholson}}, \bibinfo {author}
  {\bibfnamefont {F.}~\bibnamefont {Romero-L\'opez}}, \bibinfo {author}
  {\bibfnamefont {S.}~\bibnamefont {Skinner}}, \bibinfo {author} {\bibfnamefont
  {P.}~\bibnamefont {Vranas}},\ and\ \bibinfo {author} {\bibfnamefont
  {A.}~\bibnamefont {Walker-Loud}},\ }\bibfield  {title} {\bibinfo {title}
  {{Elastic nucleon-pion scattering at m\ensuremath{\pi}=200 MeV from lattice
  QCD}},\ }\href {https://doi.org/10.1016/j.nuclphysb.2023.116105} {\bibfield
  {journal} {\bibinfo  {journal} {Nucl. Phys. B}\ }\textbf {\bibinfo {volume}
  {987}},\ \bibinfo {pages} {116105} (\bibinfo {year} {2023}{\natexlab{a}})},\
  \Eprint {https://arxiv.org/abs/2208.03867} {arXiv:2208.03867 [hep-lat]}
  \BibitemShut {NoStop}%
\bibitem [{\citenamefont {Bulava}\ \emph
  {et~al.}(2023{\natexlab{b}})\citenamefont {Bulava} \emph
  {et~al.}}]{Bulava:2023uma}%
  \BibitemOpen
  \bibfield  {author} {\bibinfo {author} {\bibfnamefont {J.}~\bibnamefont
  {Bulava}} \emph {et~al.},\ }\bibfield  {title} {\bibinfo {title} {{Low-lying
  baryon resonances from lattice QCD}}\ }(\bibinfo {year} {2023})\ \Eprint
  {https://arxiv.org/abs/2310.08375} {arXiv:2310.08375 [hep-lat]} \BibitemShut
  {NoStop}%
\bibitem [{\citenamefont {Abell}\ \emph
  {et~al.}(2023{\natexlab{a}})\citenamefont {Abell}, \citenamefont {Leinweber},
  \citenamefont {Thomas},\ and\ \citenamefont {Wu}}]{Abell:2023qgj}%
  \BibitemOpen
  \bibfield  {author} {\bibinfo {author} {\bibfnamefont {C.~D.}\ \bibnamefont
  {Abell}}, \bibinfo {author} {\bibfnamefont {D.~B.}\ \bibnamefont
  {Leinweber}}, \bibinfo {author} {\bibfnamefont {A.~W.}\ \bibnamefont
  {Thomas}},\ and\ \bibinfo {author} {\bibfnamefont {J.-J.}\ \bibnamefont
  {Wu}},\ }\href@noop {} {\bibinfo {title} {Effects of multiple single-particle
  basis states in scattering systems}} (\bibinfo {year} {2023}{\natexlab{a}}),\
  \Eprint {https://arxiv.org/abs/2305.18790} {arXiv:2305.18790 [nucl-th]}
  \BibitemShut {NoStop}%
\bibitem [{\citenamefont {Abell}\ \emph
  {et~al.}(2023{\natexlab{b}})\citenamefont {Abell}, \citenamefont {Leinweber},
  \citenamefont {Liu}, \citenamefont {Thomas},\ and\ \citenamefont
  {Wu}}]{Abell:2023nex}%
  \BibitemOpen
  \bibfield  {author} {\bibinfo {author} {\bibfnamefont {C.~D.}\ \bibnamefont
  {Abell}}, \bibinfo {author} {\bibfnamefont {D.~B.}\ \bibnamefont
  {Leinweber}}, \bibinfo {author} {\bibfnamefont {Z.-W.}\ \bibnamefont {Liu}},
  \bibinfo {author} {\bibfnamefont {A.~W.}\ \bibnamefont {Thomas}},\ and\
  \bibinfo {author} {\bibfnamefont {J.-J.}\ \bibnamefont {Wu}},\ }\href@noop {}
  {\bibinfo {title} {Low-lying odd-parity nucleon resonances as quark-model
  like states}} (\bibinfo {year} {2023}{\natexlab{b}}),\ \Eprint
  {https://arxiv.org/abs/2306.00337} {arXiv:2306.00337 [hep-lat]} \BibitemShut
  {NoStop}%
\bibitem [{\citenamefont {Wang}\ and\ \citenamefont
  {Thomas}(2010)}]{Wang:2010hp}%
  \BibitemOpen
  \bibfield  {author} {\bibinfo {author} {\bibfnamefont {P.}~\bibnamefont
  {Wang}}\ and\ \bibinfo {author} {\bibfnamefont {A.~W.}\ \bibnamefont
  {Thomas}},\ }\bibfield  {title} {\bibinfo {title} {{The First Moments of
  Nucleon Generalized Parton Distributions}},\ }\href
  {https://doi.org/10.1103/PhysRevD.81.114015} {\bibfield  {journal} {\bibinfo
  {journal} {Phys. Rev. D}\ }\textbf {\bibinfo {volume} {81}},\ \bibinfo
  {pages} {114015} (\bibinfo {year} {2010})},\ \Eprint
  {https://arxiv.org/abs/1003.0957} {arXiv:1003.0957 [hep-ph]} \BibitemShut
  {NoStop}%
\bibitem [{\citenamefont {Scapellato}\ \emph {et~al.}(2022)\citenamefont
  {Scapellato}, \citenamefont {Alexandrou}, \citenamefont {Cichy},
  \citenamefont {Constantinou}, \citenamefont {Hadjiyiannakou}, \citenamefont
  {Jansen},\ and\ \citenamefont {Steffens}}]{Scapellato:2021uke}%
  \BibitemOpen
  \bibfield  {author} {\bibinfo {author} {\bibfnamefont {A.}~\bibnamefont
  {Scapellato}}, \bibinfo {author} {\bibfnamefont {C.}~\bibnamefont
  {Alexandrou}}, \bibinfo {author} {\bibfnamefont {K.}~\bibnamefont {Cichy}},
  \bibinfo {author} {\bibfnamefont {M.}~\bibnamefont {Constantinou}}, \bibinfo
  {author} {\bibfnamefont {K.}~\bibnamefont {Hadjiyiannakou}}, \bibinfo
  {author} {\bibfnamefont {K.}~\bibnamefont {Jansen}},\ and\ \bibinfo {author}
  {\bibfnamefont {F.}~\bibnamefont {Steffens}},\ }\bibfield  {title} {\bibinfo
  {title} {{Generalized parton distributions of the proton from lattice QCD}},\
  }\href {https://doi.org/10.22323/1.396.0129} {\bibfield  {journal} {\bibinfo
  {journal} {PoS}\ }\textbf {\bibinfo {volume} {LATTICE2021}},\ \bibinfo
  {pages} {129} (\bibinfo {year} {2022})},\ \Eprint
  {https://arxiv.org/abs/2111.03226} {arXiv:2111.03226 [hep-lat]} \BibitemShut
  {NoStop}%
\bibitem [{\citenamefont {He}\ \emph {et~al.}(2022)\citenamefont {He},
  \citenamefont {Ji}, \citenamefont {Melnitchouk}, \citenamefont {Thomas},\
  and\ \citenamefont {Wang}}]{He:2022leb}%
  \BibitemOpen
  \bibfield  {author} {\bibinfo {author} {\bibfnamefont {F.}~\bibnamefont
  {He}}, \bibinfo {author} {\bibfnamefont {C.-R.}\ \bibnamefont {Ji}}, \bibinfo
  {author} {\bibfnamefont {W.}~\bibnamefont {Melnitchouk}}, \bibinfo {author}
  {\bibfnamefont {A.~W.}\ \bibnamefont {Thomas}},\ and\ \bibinfo {author}
  {\bibfnamefont {P.}~\bibnamefont {Wang}},\ }\bibfield  {title} {\bibinfo
  {title} {{Generalized parton distributions of sea quarks in the proton from
  nonlocal chiral effective theory}},\ }\href
  {https://doi.org/10.1103/PhysRevD.106.054006} {\bibfield  {journal} {\bibinfo
   {journal} {Phys. Rev. D}\ }\textbf {\bibinfo {volume} {106}},\ \bibinfo
  {pages} {054006} (\bibinfo {year} {2022})},\ \Eprint
  {https://arxiv.org/abs/2202.00266} {arXiv:2202.00266 [hep-ph]} \BibitemShut
  {NoStop}%
\bibitem [{\citenamefont {Lin}(2023)}]{Lin:2023gxz}%
  \BibitemOpen
  \bibfield  {author} {\bibinfo {author} {\bibfnamefont {H.-W.}\ \bibnamefont
  {Lin}},\ }\bibfield  {title} {\bibinfo {title} {{Pion valence-quark
  generalized parton distribution at physical pion mass}},\ }\href
  {https://doi.org/10.1016/j.physletb.2023.138181} {\bibfield  {journal}
  {\bibinfo  {journal} {Phys. Lett. B}\ }\textbf {\bibinfo {volume} {846}},\
  \bibinfo {pages} {138181} (\bibinfo {year} {2023})}\BibitemShut {NoStop}%
\bibitem [{\citenamefont {Copeland}\ \emph {et~al.}(2021)\citenamefont
  {Copeland}, \citenamefont {Ji},\ and\ \citenamefont
  {Melnitchouk}}]{LagrangianCopeland}%
  \BibitemOpen
  \bibfield  {author} {\bibinfo {author} {\bibfnamefont {P.~M.}\ \bibnamefont
  {Copeland}}, \bibinfo {author} {\bibfnamefont {C.-R.}\ \bibnamefont {Ji}},\
  and\ \bibinfo {author} {\bibfnamefont {W.}~\bibnamefont {Melnitchouk}},\
  }\bibfield  {title} {\bibinfo {title} {Octet and decuplet baryon
  self-energies in relativistic su(3) chiral effective theory},\ }\href
  {https://doi.org/10.1103/PhysRevD.103.094019} {\bibfield  {journal} {\bibinfo
   {journal} {Phys. Rev. D}\ }\textbf {\bibinfo {volume} {103}},\ \bibinfo
  {pages} {094019} (\bibinfo {year} {2021})}\BibitemShut {NoStop}%
\bibitem [{\citenamefont {Jenkins}\ and\ \citenamefont
  {Manohar}(1991)}]{LagrangianJenkins}%
  \BibitemOpen
  \bibfield  {author} {\bibinfo {author} {\bibfnamefont {E.}~\bibnamefont
  {Jenkins}}\ and\ \bibinfo {author} {\bibfnamefont {A.~V.}\ \bibnamefont
  {Manohar}},\ }\bibfield  {title} {\bibinfo {title} {Chiral corrections to the
  baryon axial currents},\ }\href
  {https://doi.org/https://doi.org/10.1016/0370-2693(91)90840-M} {\bibfield
  {journal} {\bibinfo  {journal} {Physics Letters B}\ }\textbf {\bibinfo
  {volume} {259}},\ \bibinfo {pages} {353} (\bibinfo {year}
  {1991})}\BibitemShut {NoStop}%
\bibitem [{\citenamefont {Lutz}\ and\ \citenamefont
  {Kolomeitsev}(2002)}]{LagrangianLutz}%
  \BibitemOpen
  \bibfield  {author} {\bibinfo {author} {\bibfnamefont {M.}~\bibnamefont
  {Lutz}}\ and\ \bibinfo {author} {\bibfnamefont {E.}~\bibnamefont
  {Kolomeitsev}},\ }\bibfield  {title} {\bibinfo {title} {Relativistic chiral
  su(3) symmetry, large-nc sum rules and meson–baryon scattering},\ }\href
  {https://doi.org/https://doi.org/10.1016/S0375-9474(01)01312-4} {\bibfield
  {journal} {\bibinfo  {journal} {Nuclear Physics A}\ }\textbf {\bibinfo
  {volume} {700}},\ \bibinfo {pages} {193} (\bibinfo {year}
  {2002})}\BibitemShut {NoStop}%
\bibitem [{\citenamefont {Ledwig}\ \emph {et~al.}(2014)\citenamefont {Ledwig},
  \citenamefont {Camalich}, \citenamefont {Geng},\ and\ \citenamefont
  {Vacas}}]{LagrangianLedwig}%
  \BibitemOpen
  \bibfield  {author} {\bibinfo {author} {\bibfnamefont {T.}~\bibnamefont
  {Ledwig}}, \bibinfo {author} {\bibfnamefont {J.~M.}\ \bibnamefont
  {Camalich}}, \bibinfo {author} {\bibfnamefont {L.~S.}\ \bibnamefont {Geng}},\
  and\ \bibinfo {author} {\bibfnamefont {M.~J.~V.}\ \bibnamefont {Vacas}},\
  }\bibfield  {title} {\bibinfo {title} {Octet-baryon axial-vector charges and
  su(3)-breaking effects in the semileptonic hyperon decays},\ }\href
  {https://doi.org/10.1103/PhysRevD.90.054502} {\bibfield  {journal} {\bibinfo
  {journal} {Phys. Rev. D}\ }\textbf {\bibinfo {volume} {90}},\ \bibinfo
  {pages} {054502} (\bibinfo {year} {2014})}\BibitemShut {NoStop}%
\bibitem [{\citenamefont {Oller}\ \emph {et~al.}(2006)\citenamefont {Oller},
  \citenamefont {Verbeni},\ and\ \citenamefont {Prades}}]{2ndLagrangianOller}%
  \BibitemOpen
  \bibfield  {author} {\bibinfo {author} {\bibfnamefont {J.~A.}\ \bibnamefont
  {Oller}}, \bibinfo {author} {\bibfnamefont {M.}~\bibnamefont {Verbeni}},\
  and\ \bibinfo {author} {\bibfnamefont {J.}~\bibnamefont {Prades}},\
  }\bibfield  {title} {\bibinfo {title} {Meson-baryon effective chiral
  lagrangians to $\mathcal{O}(q^3)$},\ }\href
  {https://doi.org/10.1088/1126-6708/2006/09/079} {\bibfield  {journal}
  {\bibinfo  {journal} {Journal of High Energy Physics}\ }\textbf {\bibinfo
  {volume} {2006}},\ \bibinfo {pages} {079} (\bibinfo {year}
  {2006})}\BibitemShut {NoStop}%
\bibitem [{\citenamefont {Gasser}\ \emph {et~al.}(1988)\citenamefont {Gasser},
  \citenamefont {Sainio},\ and\ \citenamefont
  {Švarc}}]{SU2NLOLagrangianGasser}%
  \BibitemOpen
  \bibfield  {author} {\bibinfo {author} {\bibfnamefont {J.}~\bibnamefont
  {Gasser}}, \bibinfo {author} {\bibfnamefont {M.}~\bibnamefont {Sainio}},\
  and\ \bibinfo {author} {\bibfnamefont {A.}~\bibnamefont {Švarc}},\
  }\bibfield  {title} {\bibinfo {title} {Nucleons with chiral loops},\ }\href
  {https://doi.org/https://doi.org/10.1016/0550-3213(88)90108-3} {\bibfield
  {journal} {\bibinfo  {journal} {Nuclear Physics B}\ }\textbf {\bibinfo
  {volume} {307}},\ \bibinfo {pages} {779} (\bibinfo {year}
  {1988})}\BibitemShut {NoStop}%
\bibitem [{\citenamefont {Fettes}\ \emph {et~al.}(2000)\citenamefont {Fettes},
  \citenamefont {Meißner}, \citenamefont {Mojžiš},\ and\ \citenamefont
  {Steininger}}]{SU2NLOLagrangianFettes}%
  \BibitemOpen
  \bibfield  {author} {\bibinfo {author} {\bibfnamefont {N.}~\bibnamefont
  {Fettes}}, \bibinfo {author} {\bibfnamefont {U.-G.}\ \bibnamefont
  {Meißner}}, \bibinfo {author} {\bibfnamefont {M.}~\bibnamefont {Mojžiš}},\
  and\ \bibinfo {author} {\bibfnamefont {S.}~\bibnamefont {Steininger}},\
  }\bibfield  {title} {\bibinfo {title} {The chiral effective pion-nucleon
  lagrangian of order p4},\ }\href
  {https://doi.org/https://doi.org/10.1006/aphy.2000.6059} {\bibfield
  {journal} {\bibinfo  {journal} {Annals of Physics}\ }\textbf {\bibinfo
  {volume} {283}},\ \bibinfo {pages} {273} (\bibinfo {year}
  {2000})}\BibitemShut {NoStop}%
\bibitem [{\citenamefont {Fuchs}\ \emph {et~al.}(2003)\citenamefont {Fuchs},
  \citenamefont {Gegelia}, \citenamefont {Japaridze},\ and\ \citenamefont
  {Scherer}}]{EOMSFuchs}%
  \BibitemOpen
  \bibfield  {author} {\bibinfo {author} {\bibfnamefont {T.}~\bibnamefont
  {Fuchs}}, \bibinfo {author} {\bibfnamefont {J.}~\bibnamefont {Gegelia}},
  \bibinfo {author} {\bibfnamefont {G.}~\bibnamefont {Japaridze}},\ and\
  \bibinfo {author} {\bibfnamefont {S.}~\bibnamefont {Scherer}},\ }\bibfield
  {title} {\bibinfo {title} {Renormalization of relativistic baryon chiral
  perturbation theory and power counting},\ }\href
  {https://doi.org/10.1103/PhysRevD.68.056005} {\bibfield  {journal} {\bibinfo
  {journal} {Phys. Rev. D}\ }\textbf {\bibinfo {volume} {68}},\ \bibinfo
  {pages} {056005} (\bibinfo {year} {2003})}\BibitemShut {NoStop}%
\bibitem [{\citenamefont {Armstrong}\ and\ \citenamefont
  {McKeown}(2012)}]{Armstrong:2012bi}%
  \BibitemOpen
  \bibfield  {author} {\bibinfo {author} {\bibfnamefont {D.~S.}\ \bibnamefont
  {Armstrong}}\ and\ \bibinfo {author} {\bibfnamefont {R.~D.}\ \bibnamefont
  {McKeown}},\ }\bibfield  {title} {\bibinfo {title} {{Parity-Violating
  Electron Scattering and the Electric and Magnetic Strange Form Factors of the
  Nucleon}},\ }\href {https://doi.org/10.1146/annurev-nucl-102010-130419}
  {\bibfield  {journal} {\bibinfo  {journal} {Ann. Rev. Nucl. Part. Sci.}\
  }\textbf {\bibinfo {volume} {62}},\ \bibinfo {pages} {337} (\bibinfo {year}
  {2012})},\ \Eprint {https://arxiv.org/abs/1207.5238} {arXiv:1207.5238
  [nucl-ex]} \BibitemShut {NoStop}%
\bibitem [{\citenamefont {Aniol}\ \emph {et~al.}(1999)\citenamefont {Aniol}
  \emph {et~al.}}]{HAPPEX:1998epc}%
  \BibitemOpen
  \bibfield  {author} {\bibinfo {author} {\bibfnamefont {K.~A.}\ \bibnamefont
  {Aniol}} \emph {et~al.} (\bibinfo {collaboration} {HAPPEX}),\ }\bibfield
  {title} {\bibinfo {title} {{Measurement of the neutral weak form-factors of
  the proton}},\ }\href {https://doi.org/10.1103/PhysRevLett.82.1096}
  {\bibfield  {journal} {\bibinfo  {journal} {Phys. Rev. Lett.}\ }\textbf
  {\bibinfo {volume} {82}},\ \bibinfo {pages} {1096} (\bibinfo {year}
  {1999})},\ \Eprint {https://arxiv.org/abs/nucl-ex/9810012}
  {arXiv:nucl-ex/9810012} \BibitemShut {NoStop}%
\bibitem [{\citenamefont {Androic}\ \emph {et~al.}(2010)\citenamefont {Androic}
  \emph {et~al.}}]{G0:2009wvv}%
  \BibitemOpen
  \bibfield  {author} {\bibinfo {author} {\bibfnamefont {D.}~\bibnamefont
  {Androic}} \emph {et~al.} (\bibinfo {collaboration} {G0}),\ }\bibfield
  {title} {\bibinfo {title} {{Strange Quark Contributions to Parity-Violating
  Asymmetries in the Backward Angle G0 Electron Scattering Experiment}},\
  }\href {https://doi.org/10.1103/PhysRevLett.104.012001} {\bibfield  {journal}
  {\bibinfo  {journal} {Phys. Rev. Lett.}\ }\textbf {\bibinfo {volume} {104}},\
  \bibinfo {pages} {012001} (\bibinfo {year} {2010})},\ \Eprint
  {https://arxiv.org/abs/0909.5107} {arXiv:0909.5107 [nucl-ex]} \BibitemShut
  {NoStop}%
\bibitem [{\citenamefont {Thomas}(1983)}]{Thomas:1983fh}%
  \BibitemOpen
  \bibfield  {author} {\bibinfo {author} {\bibfnamefont {A.~W.}\ \bibnamefont
  {Thomas}},\ }\bibfield  {title} {\bibinfo {title} {{A Limit on the Pionic
  Component of the Nucleon Through SU(3) Flavor Breaking in the Sea}},\ }\href
  {https://doi.org/10.1016/0370-2693(83)90026-6} {\bibfield  {journal}
  {\bibinfo  {journal} {Phys. Lett. B}\ }\textbf {\bibinfo {volume} {126}},\
  \bibinfo {pages} {97} (\bibinfo {year} {1983})}\BibitemShut {NoStop}%
\bibitem [{\citenamefont {Thomas}(1984)}]{Thomas:1982kv}%
  \BibitemOpen
  \bibfield  {author} {\bibinfo {author} {\bibfnamefont {A.~W.}\ \bibnamefont
  {Thomas}},\ }\bibfield  {title} {\bibinfo {title} {{Chiral Symmetry and the
  Bag Model: A New Starting Point for Nuclear Physics}},\ }\href
  {https://doi.org/10.1007/978-1-4613-9892-9_1} {\bibfield  {journal} {\bibinfo
   {journal} {Adv. Nucl. Phys.}\ }\textbf {\bibinfo {volume} {13}},\ \bibinfo
  {pages} {1} (\bibinfo {year} {1984})}\BibitemShut {NoStop}%
\bibitem [{\citenamefont {Thomas}(2003)}]{Thomas:2002sj}%
  \BibitemOpen
  \bibfield  {author} {\bibinfo {author} {\bibfnamefont {A.~W.}\ \bibnamefont
  {Thomas}},\ }\bibfield  {title} {\bibinfo {title} {{Chiral extrapolation of
  hadronic observables}},\ }\href
  {https://doi.org/10.1016/S0920-5632(03)01492-0} {\bibfield  {journal}
  {\bibinfo  {journal} {Nucl. Phys. B Proc. Suppl.}\ }\textbf {\bibinfo
  {volume} {119}},\ \bibinfo {pages} {50} (\bibinfo {year} {2003})},\ \Eprint
  {https://arxiv.org/abs/hep-lat/0208023} {arXiv:hep-lat/0208023} \BibitemShut
  {NoStop}%
\bibitem [{\citenamefont {Donoghue}\ \emph {et~al.}(1999)\citenamefont
  {Donoghue}, \citenamefont {Holstein},\ and\ \citenamefont
  {Borasoy}}]{Donoghue:1998bs}%
  \BibitemOpen
  \bibfield  {author} {\bibinfo {author} {\bibfnamefont {J.~F.}\ \bibnamefont
  {Donoghue}}, \bibinfo {author} {\bibfnamefont {B.~R.}\ \bibnamefont
  {Holstein}},\ and\ \bibinfo {author} {\bibfnamefont {B.}~\bibnamefont
  {Borasoy}},\ }\bibfield  {title} {\bibinfo {title} {{SU(3) baryon chiral
  perturbation theory and long distance regularization}},\ }\href
  {https://doi.org/10.1103/PhysRevD.59.036002} {\bibfield  {journal} {\bibinfo
  {journal} {Phys. Rev. D}\ }\textbf {\bibinfo {volume} {59}},\ \bibinfo
  {pages} {036002} (\bibinfo {year} {1999})},\ \Eprint
  {https://arxiv.org/abs/hep-ph/9804281} {arXiv:hep-ph/9804281} \BibitemShut
  {NoStop}%
\bibitem [{\citenamefont {Hall}\ \emph {et~al.}(2011)\citenamefont {Hall},
  \citenamefont {Lee}, \citenamefont {Leinweber}, \citenamefont {Liu},
  \citenamefont {Mathur}, \citenamefont {Young},\ and\ \citenamefont
  {Zhang}}]{Hall:2011en}%
  \BibitemOpen
  \bibfield  {author} {\bibinfo {author} {\bibfnamefont {J.~M.~M.}\
  \bibnamefont {Hall}}, \bibinfo {author} {\bibfnamefont {F.~X.}\ \bibnamefont
  {Lee}}, \bibinfo {author} {\bibfnamefont {D.~B.}\ \bibnamefont {Leinweber}},
  \bibinfo {author} {\bibfnamefont {K.~F.}\ \bibnamefont {Liu}}, \bibinfo
  {author} {\bibfnamefont {N.}~\bibnamefont {Mathur}}, \bibinfo {author}
  {\bibfnamefont {R.~D.}\ \bibnamefont {Young}},\ and\ \bibinfo {author}
  {\bibfnamefont {J.~B.}\ \bibnamefont {Zhang}},\ }\bibfield  {title} {\bibinfo
  {title} {{Chiral extrapolation beyond the power-counting regime}},\ }\href
  {https://doi.org/10.1103/PhysRevD.84.114011} {\bibfield  {journal} {\bibinfo
  {journal} {Phys. Rev. D}\ }\textbf {\bibinfo {volume} {84}},\ \bibinfo
  {pages} {114011} (\bibinfo {year} {2011})},\ \Eprint
  {https://arxiv.org/abs/1101.4411} {arXiv:1101.4411 [hep-lat]} \BibitemShut
  {NoStop}%
\bibitem [{\citenamefont {McGovern}\ and\ \citenamefont
  {Birse}(2006)}]{McGovern:2006fm}%
  \BibitemOpen
  \bibfield  {author} {\bibinfo {author} {\bibfnamefont {J.~A.}\ \bibnamefont
  {McGovern}}\ and\ \bibinfo {author} {\bibfnamefont {M.~C.}\ \bibnamefont
  {Birse}},\ }\bibfield  {title} {\bibinfo {title} {{Convergence of the chiral
  expansion for the nucleon mass}},\ }\href
  {https://doi.org/10.1103/PhysRevD.74.097501} {\bibfield  {journal} {\bibinfo
  {journal} {Phys. Rev. D}\ }\textbf {\bibinfo {volume} {74}},\ \bibinfo
  {pages} {097501} (\bibinfo {year} {2006})},\ \Eprint
  {https://arxiv.org/abs/hep-lat/0608002} {arXiv:hep-lat/0608002} \BibitemShut
  {NoStop}%
\bibitem [{\citenamefont {Alarcon}\ \emph {et~al.}(2011)\citenamefont
  {Alarcon}, \citenamefont {Martin~Camalich}, \citenamefont {Oller},\ and\
  \citenamefont {Alvarez-Ruso}}]{Alarcon:2011kh}%
  \BibitemOpen
  \bibfield  {author} {\bibinfo {author} {\bibfnamefont {J.~M.}\ \bibnamefont
  {Alarcon}}, \bibinfo {author} {\bibfnamefont {J.}~\bibnamefont
  {Martin~Camalich}}, \bibinfo {author} {\bibfnamefont {J.~A.}\ \bibnamefont
  {Oller}},\ and\ \bibinfo {author} {\bibfnamefont {L.}~\bibnamefont
  {Alvarez-Ruso}},\ }\bibfield  {title} {\bibinfo {title} {{{$\pi N$}
  scattering in relativistic baryon chiral perturbation theory revisited}},\
  }\href {https://doi.org/10.1103/PhysRevC.83.055205} {\bibfield  {journal}
  {\bibinfo  {journal} {Phys. Rev. C}\ }\textbf {\bibinfo {volume} {83}},\
  \bibinfo {pages} {055205} (\bibinfo {year} {2011})},\ \bibinfo {note}
  {[Erratum: Phys.Rev.C 87, 059901 (2013)]},\ \Eprint
  {https://arxiv.org/abs/1102.1537} {arXiv:1102.1537 [nucl-th]} \BibitemShut
  {NoStop}%
\bibitem [{\citenamefont {Frink}\ \emph {et~al.}(2005)\citenamefont {Frink},
  \citenamefont {Meissner},\ and\ \citenamefont {Scheller}}]{Frink:2005ru}%
  \BibitemOpen
  \bibfield  {author} {\bibinfo {author} {\bibfnamefont {M.}~\bibnamefont
  {Frink}}, \bibinfo {author} {\bibfnamefont {U.-G.}\ \bibnamefont
  {Meissner}},\ and\ \bibinfo {author} {\bibfnamefont {I.}~\bibnamefont
  {Scheller}},\ }\bibfield  {title} {\bibinfo {title} {{Baryon masses, chiral
  extrapolations, and all that}},\ }\href
  {https://doi.org/10.1140/epja/i2005-10063-9} {\bibfield  {journal} {\bibinfo
  {journal} {Eur. Phys. J. A}\ }\textbf {\bibinfo {volume} {24}},\ \bibinfo
  {pages} {395} (\bibinfo {year} {2005})},\ \Eprint
  {https://arxiv.org/abs/hep-lat/0501024} {arXiv:hep-lat/0501024} \BibitemShut
  {NoStop}%
\bibitem [{\citenamefont {Djukanovic}\ \emph {et~al.}(2005)\citenamefont
  {Djukanovic}, \citenamefont {Schindler}, \citenamefont {Gegelia},\ and\
  \citenamefont {Scherer}}]{ModpropDjukanovic}%
  \BibitemOpen
  \bibfield  {author} {\bibinfo {author} {\bibfnamefont {D.}~\bibnamefont
  {Djukanovic}}, \bibinfo {author} {\bibfnamefont {M.~R.}\ \bibnamefont
  {Schindler}}, \bibinfo {author} {\bibfnamefont {J.}~\bibnamefont {Gegelia}},\
  and\ \bibinfo {author} {\bibfnamefont {S.}~\bibnamefont {Scherer}},\
  }\bibfield  {title} {\bibinfo {title} {Improving the ultraviolet behavior in
  baryon chiral perturbation theory},\ }\href
  {https://doi.org/10.1103/PhysRevD.72.045002} {\bibfield  {journal} {\bibinfo
  {journal} {Phys. Rev. D}\ }\textbf {\bibinfo {volume} {72}},\ \bibinfo
  {pages} {045002} (\bibinfo {year} {2005})}\BibitemShut {NoStop}%
\bibitem [{\citenamefont {Frink}\ and\ \citenamefont
  {Meißner}(2004)}]{c1c2c3Frink}%
  \BibitemOpen
  \bibfield  {author} {\bibinfo {author} {\bibfnamefont {M.}~\bibnamefont
  {Frink}}\ and\ \bibinfo {author} {\bibfnamefont {U.-G.}\ \bibnamefont
  {Meißner}},\ }\bibfield  {title} {\bibinfo {title} {Chiral extrapolations of
  baryon masses for unquenched three-flavor lattice simulations},\ }\href
  {https://doi.org/10.1088/1126-6708/2004/07/028} {\bibfield  {journal}
  {\bibinfo  {journal} {Journal of High Energy Physics}\ }\textbf {\bibinfo
  {volume} {2004}},\ \bibinfo {pages} {028} (\bibinfo {year}
  {2004})}\BibitemShut {NoStop}%
\bibitem [{\citenamefont {Siemens}\ \emph {et~al.}(2017)\citenamefont
  {Siemens}, \citenamefont {de~Elvira}, \citenamefont {Epelbaum}, \citenamefont
  {Hoferichter}, \citenamefont {Krebs}, \citenamefont {Kubis},\ and\
  \citenamefont {Mei{\ss}ner}}]{LECsSiemens:2017}%
  \BibitemOpen
  \bibfield  {author} {\bibinfo {author} {\bibfnamefont {D.}~\bibnamefont
  {Siemens}}, \bibinfo {author} {\bibfnamefont {J.~R.}\ \bibnamefont
  {de~Elvira}}, \bibinfo {author} {\bibfnamefont {E.}~\bibnamefont {Epelbaum}},
  \bibinfo {author} {\bibfnamefont {M.}~\bibnamefont {Hoferichter}}, \bibinfo
  {author} {\bibfnamefont {H.}~\bibnamefont {Krebs}}, \bibinfo {author}
  {\bibfnamefont {B.}~\bibnamefont {Kubis}},\ and\ \bibinfo {author}
  {\bibfnamefont {U.-G.}\ \bibnamefont {Mei{\ss}ner}},\ }\bibfield  {title}
  {\bibinfo {title} {Reconciling threshold and subthreshold expansions for
  pion--nucleon scattering},\ }\href@noop {} {\bibfield  {journal} {\bibinfo
  {journal} {Physics Letters B}\ }\textbf {\bibinfo {volume} {770}},\ \bibinfo
  {pages} {27} (\bibinfo {year} {2017})}\BibitemShut {NoStop}%
\bibitem [{\citenamefont {Menadue}\ \emph {et~al.}(2012)\citenamefont
  {Menadue}, \citenamefont {Kamleh}, \citenamefont {Leinweber},\ and\
  \citenamefont {Mahbub}}]{Menadue:2011pd}%
  \BibitemOpen
  \bibfield  {author} {\bibinfo {author} {\bibfnamefont {B.~J.}\ \bibnamefont
  {Menadue}}, \bibinfo {author} {\bibfnamefont {W.}~\bibnamefont {Kamleh}},
  \bibinfo {author} {\bibfnamefont {D.~B.}\ \bibnamefont {Leinweber}},\ and\
  \bibinfo {author} {\bibfnamefont {M.~S.}\ \bibnamefont {Mahbub}},\ }\bibfield
   {title} {\bibinfo {title} {{Isolating the $\Lambda(1405)$ in Lattice QCD}},\
  }\href {https://doi.org/10.1103/PhysRevLett.108.112001} {\bibfield  {journal}
  {\bibinfo  {journal} {Phys. Rev. Lett.}\ }\textbf {\bibinfo {volume} {108}},\
  \bibinfo {pages} {112001} (\bibinfo {year} {2012})},\ \Eprint
  {https://arxiv.org/abs/1109.6716} {arXiv:1109.6716 [hep-lat]} \BibitemShut
  {NoStop}%
\bibitem [{\citenamefont {L{\"u}scher}(2010)}]{Luscher:2010iy}%
  \BibitemOpen
  \bibfield  {author} {\bibinfo {author} {\bibfnamefont {M.}~\bibnamefont
  {L{\"u}scher}},\ }\bibfield  {title} {\bibinfo {title} {Properties and uses
  of the wilson flow in lattice {QCD}},\ }\href
  {https://doi.org/10.1007/JHEP08(2010)071} {\bibfield  {journal} {\bibinfo
  {journal} {Journal of High Energy Physics}\ }\textbf {\bibinfo {volume}
  {2010}},\ \bibinfo {pages} {71} (\bibinfo {year} {2010})}\BibitemShut
  {NoStop}%
\bibitem [{\citenamefont {Steininger}\ \emph {et~al.}(1998)\citenamefont
  {Steininger}, \citenamefont {Meißner},\ and\ \citenamefont
  {Fettes}}]{SvenSteininger_1998}%
  \BibitemOpen
  \bibfield  {author} {\bibinfo {author} {\bibfnamefont {S.}~\bibnamefont
  {Steininger}}, \bibinfo {author} {\bibfnamefont {U.-G.}\ \bibnamefont
  {Meißner}},\ and\ \bibinfo {author} {\bibfnamefont {N.}~\bibnamefont
  {Fettes}},\ }\bibfield  {title} {\bibinfo {title} {On wave function
  renormalization and related aspects in heavy fermion effective field
  theories},\ }\href {https://doi.org/10.1088/1126-6708/1998/09/008} {\bibfield
   {journal} {\bibinfo  {journal} {Journal of High Energy Physics}\ }\textbf
  {\bibinfo {volume} {1998}},\ \bibinfo {pages} {008} (\bibinfo {year}
  {1998})}\BibitemShut {NoStop}%
\bibitem [{\citenamefont {Becher}\ and\ \citenamefont
  {Leutwyler}(1999)}]{SU2chiPTmassexpBecher}%
  \BibitemOpen
  \bibfield  {author} {\bibinfo {author} {\bibfnamefont {T.}~\bibnamefont
  {Becher}}\ and\ \bibinfo {author} {\bibfnamefont {H.}~\bibnamefont
  {Leutwyler}},\ }\bibfield  {title} {\bibinfo {title} {Baryon chiral
  perturbation theory in manifestly lorentz invariant form},\ }\href
  {https://doi.org/10.1007/PL00021673} {\bibfield  {journal} {\bibinfo
  {journal} {The European Physical Journal C - Particles and Fields}\ }\textbf
  {\bibinfo {volume} {9}},\ \bibinfo {pages} {643} (\bibinfo {year}
  {1999})}\BibitemShut {NoStop}%
\bibitem [{\citenamefont {Schindler}\ \emph {et~al.}(2007)\citenamefont
  {Schindler}, \citenamefont {Djukanovic}, \citenamefont {Gegelia},\ and\
  \citenamefont {Scherer}}]{SU2chiPTmassexpSchindler}%
  \BibitemOpen
  \bibfield  {author} {\bibinfo {author} {\bibfnamefont {M.~R.}\ \bibnamefont
  {Schindler}}, \bibinfo {author} {\bibfnamefont {D.}~\bibnamefont
  {Djukanovic}}, \bibinfo {author} {\bibfnamefont {J.}~\bibnamefont
  {Gegelia}},\ and\ \bibinfo {author} {\bibfnamefont {S.}~\bibnamefont
  {Scherer}},\ }\bibfield  {title} {\bibinfo {title} {Chiral expansion of the
  nucleon mass to order $\mathcal{O}(q^6)$},\ }\href
  {https://doi.org/https://doi.org/10.1016/j.physletb.2007.04.034} {\bibfield
  {journal} {\bibinfo  {journal} {Physics Letters B}\ }\textbf {\bibinfo
  {volume} {649}},\ \bibinfo {pages} {390} (\bibinfo {year}
  {2007})}\BibitemShut {NoStop}%
\bibitem [{\citenamefont {Hoferichter}\ \emph {et~al.}(2023)\citenamefont
  {Hoferichter}, \citenamefont {{Ruiz de Elvira}}, \citenamefont {Kubis},\ and\
  \citenamefont {Meißner}}]{Hoferichter:2023}%
  \BibitemOpen
  \bibfield  {author} {\bibinfo {author} {\bibfnamefont {M.}~\bibnamefont
  {Hoferichter}}, \bibinfo {author} {\bibfnamefont {J.}~\bibnamefont {{Ruiz de
  Elvira}}}, \bibinfo {author} {\bibfnamefont {B.}~\bibnamefont {Kubis}},\ and\
  \bibinfo {author} {\bibfnamefont {U.-G.}\ \bibnamefont {Meißner}},\
  }\bibfield  {title} {\bibinfo {title} {On the role of isospin violation in
  the pion–nucleon $\sigma$-term},\ }\href
  {https://doi.org/https://doi.org/10.1016/j.physletb.2023.138001} {\bibfield
  {journal} {\bibinfo  {journal} {Physics Letters B}\ }\textbf {\bibinfo
  {volume} {843}},\ \bibinfo {pages} {138001} (\bibinfo {year}
  {2023})}\BibitemShut {NoStop}%
\bibitem [{\citenamefont {Ruiz~de Elvira}\ \emph {et~al.}(2018)\citenamefont
  {Ruiz~de Elvira}, \citenamefont {Hoferichter}, \citenamefont {Kubis},\ and\
  \citenamefont {Mei\ss{}ner}}]{RuizdeElvira:2017stg}%
  \BibitemOpen
  \bibfield  {author} {\bibinfo {author} {\bibfnamefont {J.}~\bibnamefont
  {Ruiz~de Elvira}}, \bibinfo {author} {\bibfnamefont {M.}~\bibnamefont
  {Hoferichter}}, \bibinfo {author} {\bibfnamefont {B.}~\bibnamefont {Kubis}},\
  and\ \bibinfo {author} {\bibfnamefont {U.-G.}\ \bibnamefont {Mei\ss{}ner}},\
  }\bibfield  {title} {\bibinfo {title} {{Extracting the $\sigma$-term from
  low-energy pion-nucleon scattering}},\ }\href
  {https://doi.org/10.1088/1361-6471/aa9422} {\bibfield  {journal} {\bibinfo
  {journal} {J. Phys. G}\ }\textbf {\bibinfo {volume} {45}},\ \bibinfo {pages}
  {024001} (\bibinfo {year} {2018})},\ \Eprint
  {https://arxiv.org/abs/1706.01465} {arXiv:1706.01465 [hep-ph]} \BibitemShut
  {NoStop}%
\bibitem [{\citenamefont {Alarc\'on}\ \emph {et~al.}(2012)\citenamefont
  {Alarc\'on}, \citenamefont {Camalich},\ and\ \citenamefont
  {Oller}}]{PhysRevD.85.051503}%
  \BibitemOpen
  \bibfield  {author} {\bibinfo {author} {\bibfnamefont {J.~M.}\ \bibnamefont
  {Alarc\'on}}, \bibinfo {author} {\bibfnamefont {J.~M.}\ \bibnamefont
  {Camalich}},\ and\ \bibinfo {author} {\bibfnamefont {J.~A.}\ \bibnamefont
  {Oller}},\ }\bibfield  {title} {\bibinfo {title} {Chiral representation of
  the {$\ensuremath{\pi}N$} scattering amplitude and the pion-nucleon sigma
  term},\ }\href {https://doi.org/10.1103/PhysRevD.85.051503} {\bibfield
  {journal} {\bibinfo  {journal} {Phys. Rev. D}\ }\textbf {\bibinfo {volume}
  {85}},\ \bibinfo {pages} {051503} (\bibinfo {year} {2012})}\BibitemShut
  {NoStop}%
\bibitem [{\citenamefont {Yang}\ \emph {et~al.}(2016)\citenamefont {Yang},
  \citenamefont {Alexandru}, \citenamefont {Draper}, \citenamefont {Liang},\
  and\ \citenamefont {Liu}}]{Yang:2015uis}%
  \BibitemOpen
  \bibfield  {author} {\bibinfo {author} {\bibfnamefont {Y.-B.}\ \bibnamefont
  {Yang}}, \bibinfo {author} {\bibfnamefont {A.}~\bibnamefont {Alexandru}},
  \bibinfo {author} {\bibfnamefont {T.}~\bibnamefont {Draper}}, \bibinfo
  {author} {\bibfnamefont {J.}~\bibnamefont {Liang}},\ and\ \bibinfo {author}
  {\bibfnamefont {K.-F.}\ \bibnamefont {Liu}} (\bibinfo {collaboration}
  {xQCD}),\ }\bibfield  {title} {\bibinfo {title} {{$\pi N$ and strangeness
  sigma terms at the physical point with chiral fermions}},\ }\href
  {https://doi.org/10.1103/PhysRevD.94.054503} {\bibfield  {journal} {\bibinfo
  {journal} {Phys. Rev. D}\ }\textbf {\bibinfo {volume} {94}},\ \bibinfo
  {pages} {054503} (\bibinfo {year} {2016})},\ \Eprint
  {https://arxiv.org/abs/1511.09089} {arXiv:1511.09089 [hep-lat]} \BibitemShut
  {NoStop}%
\bibitem [{\citenamefont {Alexandrou}\ \emph {et~al.}(2020)\citenamefont
  {Alexandrou}, \citenamefont {Bacchio}, \citenamefont {Constantinou},
  \citenamefont {Finkenrath}, \citenamefont {Hadjiyiannakou}, \citenamefont
  {Jansen}, \citenamefont {Koutsou},\ and\ \citenamefont {Vaquero
  Aviles-Casco}}]{Alexandrou:2019brg}%
  \BibitemOpen
  \bibfield  {author} {\bibinfo {author} {\bibfnamefont {C.}~\bibnamefont
  {Alexandrou}}, \bibinfo {author} {\bibfnamefont {S.}~\bibnamefont {Bacchio}},
  \bibinfo {author} {\bibfnamefont {M.}~\bibnamefont {Constantinou}}, \bibinfo
  {author} {\bibfnamefont {J.}~\bibnamefont {Finkenrath}}, \bibinfo {author}
  {\bibfnamefont {K.}~\bibnamefont {Hadjiyiannakou}}, \bibinfo {author}
  {\bibfnamefont {K.}~\bibnamefont {Jansen}}, \bibinfo {author} {\bibfnamefont
  {G.}~\bibnamefont {Koutsou}},\ and\ \bibinfo {author} {\bibfnamefont
  {A.}~\bibnamefont {Vaquero Aviles-Casco}},\ }\bibfield  {title} {\bibinfo
  {title} {{Nucleon axial, tensor, and scalar charges and $\sigma$-terms in
  lattice QCD}},\ }\href {https://doi.org/10.1103/PhysRevD.102.054517}
  {\bibfield  {journal} {\bibinfo  {journal} {Phys. Rev. D}\ }\textbf {\bibinfo
  {volume} {102}},\ \bibinfo {pages} {054517} (\bibinfo {year} {2020})},\
  \Eprint {https://arxiv.org/abs/1909.00485} {arXiv:1909.00485 [hep-lat]}
  \BibitemShut {NoStop}%
\bibitem [{\citenamefont {Ericson}(1987)}]{Ericson:1987uf}%
  \BibitemOpen
  \bibfield  {author} {\bibinfo {author} {\bibfnamefont {T.~E.~O.}\
  \bibnamefont {Ericson}},\ }\bibfield  {title} {\bibinfo {title} {{A New
  Interpretation of the $\pi N \sigma$ - Term}},\ }\href
  {https://doi.org/10.1016/0370-2693(87)91180-4} {\bibfield  {journal}
  {\bibinfo  {journal} {Phys. Lett. B}\ }\textbf {\bibinfo {volume} {195}},\
  \bibinfo {pages} {116} (\bibinfo {year} {1987})}\BibitemShut {NoStop}%
\bibitem [{\citenamefont {Scherer}\ and\ \citenamefont
  {Schindler}(2011)}]{chiPTtextbookScherer}%
  \BibitemOpen
  \bibfield  {author} {\bibinfo {author} {\bibfnamefont {S.}~\bibnamefont
  {Scherer}}\ and\ \bibinfo {author} {\bibfnamefont {M.~R.}\ \bibnamefont
  {Schindler}},\ }\href@noop {} {\emph {\bibinfo {title} {A primer for chiral
  perturbation theory}}},\ Vol.\ \bibinfo {volume} {830}\ (\bibinfo
  {publisher} {Springer Science \& Business Media},\ \bibinfo {year}
  {2011})\BibitemShut {NoStop}%
\end{thebibliography}%

\end{document}